\documentclass[12pt]{article}

\usepackage[a4paper,pdftex]{geometry}
\usepackage[T1]{fontenc}
\usepackage[english]{babel}
\usepackage[authoryear]{natbib}
\usepackage[small]{titlesec}
\usepackage[justification=centering]{caption}
\usepackage[final]{pdfpages}
\usepackage{soul}

\usepackage{array,amsmath,amsthm,amssymb,graphicx,enumerate,booktabs,bigstrut,rotating,multirow,float,etoolbox,lmodern,comment,listings,qtree,a4wide,moresize,bbm,subcaption,tikz,pdfescape,mathtools,flafter,enumitem,setspace,ragged2e,colortbl,lineno,lscape,pdflscape,stackengine,xcolor}

\usetikzlibrary{shapes.geometric, arrows, positioning, decorations.pathreplacing}

\usepackage{datatool}
\DTLsetseparator{,}
\DTLloaddb[keys={key}]{est}{results/kv_store.csv}
\newcommand{\getval}[1]{\DTLfetch{est}{key}{#1}{value}}

\DeclareMathOperator{\E}{\mathbb{E}}

\newcolumntype{L}[1]{>{\raggedright\let\newline\\\arraybackslash\hspace{0pt}}m{#1}}
\newcolumntype{C}[1]{>{\centering\let\newline\\\arraybackslash\hspace{0pt}}m{#1}}
\newcolumntype{R}[1]{>{\raggedleft\let\newline\\\arraybackslash\hspace{0pt}}m{#1}}

\usepackage{hyperref}
\hypersetup{colorlinks,linkcolor={red},citecolor={blue},urlcolor={blue}}

\makeatletter
\renewcommand\subsubsection{\@startsection{subsubsection}{3}{\z@}%
	{-3.25ex\@plus -1ex \@minus -.2ex}%
	{-1.5ex \@plus -.2ex}%
	{\normalfont\normalsize\bfseries}
}
\def\@biblabel#1{\hspace*{-\labelsep}}

\makeatother

\def\sym#1{\ifmmode^{#1}\else\(^{#1}\)\fi}

\pdfpageheight\paperheight
\pdfpagewidth\paperwidth

\definecolor{blue_lab}{RGB}{0,114,178}
\definecolor{red_lab}{RGB}{213,94,0}
\definecolor{yellow_lab}{RGB}{240,228,66}
\definecolor{green_lab}{RGB}{0,158,115}

\newtheorem{definition}{Definition}

\begin{document}
\begin{filecontents*}{results/kv_store.csv}
key,value,timestamp,label
n_workers_paired,880,,"N workers (worker.dta + worker2.dta), all paired"
n_worker_pairs,440,,N worker pairs
n_mba_started,338,,N MBA who started the experiment
n_mba_excluded,67,,N MBA excluded: quit before first redistribution
n_dyson_recruited,167,,N Dyson undergraduate spectators matched
worker_bonus_mean,2.86,,Mean worker bonus = realized final earnings
worker_comp_mean,4.36,,Mean worker total comp = 1.50 fee + bonus
n_mba,271,,N MBA spectators
mba_pct_male,55,,Pct male
mba_pct_born_usa,56,,Pct US-born
mba_pct_stay_usa,94,,Pct plan to stay in US
mba_pct_voted,69,,Pct voted recent elections
mba_pct_private,84,,Pct prefer private sector
mba_pct_manager,91,,Pct aspire managerial
mba_pct_hardwork,81,,Pct hard work better life
mba_pct_age_2431,83,,"Pct MBA aged 24-31 (age_grp 2 or 3, non-missing)"
gini_b0,0.420,,Mean Gini luck condition (beta0)
gini_b1,0.240,,Performance effect on Gini (beta1)
gini_b2,0.188,,Efficiency-cost effect on Gini (beta2)
gini_b1_pct,57,,Performance effect pct of luck Gini
gini_b2_pct,45,,Efficiency effect pct of luck Gini
gini_elast,0.17,,Implied arc elasticity wrt efficiency cost
gini_b1_fd,0.191,,"Performance effect on Gini, first decision"
gini_b1_fd_pct,44,,First-decision performance effect pct of luck Gini
gini_b2_fd,0.21,,"Efficiency effect on Gini, first decision (2dp, intro)"
gini_mean_luck,0.43,,"Mean Gini luck condition (raw, first decision)"
gini_mean_perf,0.62,,"Mean Gini performance condition (raw, first decision)"
gini_mean_eff,0.64,,"Mean Gini efficiency condition (raw, first decision)"
gini_eff_diff,0.21,,Efficiency minus luck mean Gini
desc_gini_overall,0.56,,Overall mean implemented Gini
desc_redist_dollars,1.30,,Mean dollars received by loser (net of efficiency cost)
desc_pct_perfineq,34.8,,Pct of decisions at perfect inequality
desc_pct_perfeq,22.2,,Pct of decisions at perfect equality
cmp_mba_b0,0.425,,"MBA luck-condition Gini, comparison spec"
cmp_mba_perf,0.184,,"MBA performance effect, comparison spec"
cmp_mba_eff,0.211,,"MBA efficiency-cost effect, comparison spec"
cmp_mba_perfgini,0.609,,MBA performance-condition Gini
cmp_mba_b0_2dp,0.43,,"MBA luck Gini, comparison spec (2dp)"
cmp_mba_perfgini_2dp,0.61,,MBA performance-condition Gini (2dp)
cohn_top5_luck,0.271,,Cohn top-5% luck-condition Gini
mba_vs_cohn_pct,36,,Cohn top5 luck Gini pct smaller than MBA
mba_luck_gini_fd,0.432,,"MBA luck Gini, no controls, first decision"
almas_hi_eff,0.174,,Almas income>100k efficiency effect
pool_luck_genpop,0.292,,"DL pooled luck-condition Gini, general pop"
pool_perf_genpop,0.279,,"DL pooled performance effect, general pop"
pool_perfgini_genpop,0.571,,"DL pooled performance-condition Gini, general pop"
pool_luck_genpop_2dp,0.29,,"DL pooled luck-condition Gini, general pop (2dp, intro)"
pool_perfgini_genpop_2dp,0.57,,"DL pooled performance-condition Gini, general pop (2dp, intro)"
p_mba_vs_pool_luck,0.138,,p: MBA vs pooled luck-condition Gini level
p_mba_vs_pool_perf,0.136,,p: MBA vs pooled performance treatment effect
p_mba_vs_almas_eff,0.004,,p: MBA vs Almas efficiency-cost effect
p_mba_vs_almas_luck,0.410,,p: MBA vs almas luck-condition Gini level
p_mba_vs_cohn_luck,0.000,,p: MBA vs cohn luck-condition Gini level
p_mba_vs_sce_luck,0.663,,p: MBA vs sce luck-condition Gini level
p_mba_vs_hs_luck,0.003,,p: MBA vs hs luck-condition Gini level
pool_luck_hiinc,0.281,,"DL pooled luck-condition Gini, high-income"
pool_perfgini_hiinc,0.540,,"DL pooled performance-condition Gini, high-income"
cohn_perf,0.253,,Cohn top-5% performance effect
cohn_perfgini,0.524,,Cohn top-5% performance-condition Gini
cohn_business_luck,0.350,,Cohn business-owners luck-condition Gini
almas_us_eff,0.011,,Almas representative-US efficiency-cost effect
bench_luck_mean,0.30,,"Unweighted mean Gini, luck, 4 representative samples"
bench_luck_median,0.05,,"Unweighted median Gini, luck, 4 representative samples"
bench_perf_mean,0.59,,"Unweighted mean Gini, performance, 4 representative samples"
bench_perf_median,0.54,,"Unweighted median Gini, performance, 4 representative samples"
usnor_luck_gap,0.196,,"US minus Norway mean inequality, luck condition"
mba_vs_usnor_pct,68,,MBA-genpop luck gap as pct of US-Norway gap
egal_share_btw,9.9,,Pct egalitarian between
lib_share_btw,23.2,,Pct libertarian between
merit_share_btw,22.6,,Pct meritocrat between
other_share_btw,44.3,,Pct other between
egal_share_wth,6.5,,Pct egalitarian within
lib_share_wth,15.3,,Pct libertarian within
merit_share_wth,28.7,,Pct meritocrat within
other_share_wth,49.4,,Pct other within
moderate_redist_pct,18.8,,Moderate mean share redistributed pct
merit_redist_pct,31.7,,Meritocrat mean share redistributed pct
moderate_gini,0.624,,Moderate mean Gini
merit_gini,0.367,,Meritocrat mean Gini
moderate_share_all,19.5,,Moderate share of all spectators pct
moderate_share_of_other,39.5,,Moderate share of unclassified pct
moderate_pattern1_pct,11.1,,Pct redistribute 1-2 luck 0 perf
moderate_pattern2_pct,8.4,,Pct redistribute 2 luck 1 perf
joint_mba_pct_00,15.3,,"Pct redistributing 0 in both conditions (joint_mba cell 0,0)"
joint_mba_pct_33,6.5,,"Pct equalizing (3) in both conditions (joint_mba cell 3,3)"
other_residual_pct,60,,Pct of unclassified with no dominant pattern
attn_fail_pct,6.7,,Pct failing attention check
attn_other_beta,-0.005,,Attn-fail on other-ideal coefficient
attn_other_p,0.88,,Attn-fail on other-ideal p-value
decomp_narr_pct,20,,Pct of redistribution gap accounted for by narratives
pool_other_genpop,19,,"DL pooled other-ideal share, 4 representative samples"
chi2_btw_wth,0.27,,p MBA between vs within
chi2_mba_sce,0.005,,p MBA vs SCE
chi2_mba_almas,0.002,,p MBA vs Almas
chi2_mba_cohn,0.0000,,p MBA vs Cohn
chi2_mba_harrs,0.0000,,p MBA vs Harrs
chi2_mba_sce_hi,0.060,,p MBA vs SCE hi
chi2_mba_almas_hi,0.081,,p MBA vs Almas hi
chi2_mba_cohn_hi,0.0000,,p MBA vs Cohn hi
chi2_mba_harrs_hi,0.004,,p MBA vs Harrs hi
oeNcoded,116,,N coded open-ended responses
oeNmerit,36,,N meritocratic coded
oeNmod,26,,N moderate coded
oeNlib,15,,N libertarian coded
oeNegal,8,,N egalitarian coded
oeNother,31,,N other coded
oeNclassified,85,,N classified non-other coded
oeMlEduN,37,,Multi-label N C01
oeMlEduP,31.9,,Multi-label pct C01
oeMlOppN,28,,Multi-label N C02
oeMlOppP,24.1,,Multi-label pct C02
oeMlDisN,19,,Multi-label N C03
oeMlDisP,16.4,,Multi-label pct C03
oeMlPolN,20,,Multi-label N C04
oeMlPolP,17.2,,Multi-label pct C04
oeMlCorN,13,,Multi-label N C05
oeMlCorP,11.2,,Multi-label pct C05
oeMlHisN,13,,Multi-label N C06
oeMlHisP,11.2,,Multi-label pct C06
oeMlBehN,12,,Multi-label N C07
oeMlBehP,10.3,,Multi-label pct C07
oeMlNonN,3,,Multi-label N C08
oeMlNonP,2.6,,Multi-label pct C08
oeMlMultiN,25,,N coded with >=2 labels
oeMlMultiP,21.6,,Pct coded with >=2 labels
oeMlLibEduP,20.0,,Multi-label pct Libertarian x C01
oeMlLibOppP,6.7,,Multi-label pct Libertarian x C02
oeMlLibDisP,20.0,,Multi-label pct Libertarian x C03
oeMlLibPolP,20.0,,Multi-label pct Libertarian x C04
oeMlLibCorP,20.0,,Multi-label pct Libertarian x C05
oeMlLibHisP,0.0,,Multi-label pct Libertarian x C06
oeMlLibBehP,26.7,,Multi-label pct Libertarian x C07
oeMlLibNonP,6.7,,Multi-label pct Libertarian x C08
oeMlModEduP,30.8,,Multi-label pct Moderate x C01
oeMlModOppP,26.9,,Multi-label pct Moderate x C02
oeMlModDisP,23.1,,Multi-label pct Moderate x C03
oeMlModPolP,19.2,,Multi-label pct Moderate x C04
oeMlModCorP,11.5,,Multi-label pct Moderate x C05
oeMlModHisP,3.8,,Multi-label pct Moderate x C06
oeMlModBehP,11.5,,Multi-label pct Moderate x C07
oeMlModNonP,0.0,,Multi-label pct Moderate x C08
oeMlMeritEduP,38.9,,Multi-label pct Meritocratic x C01
oeMlMeritOppP,19.4,,Multi-label pct Meritocratic x C02
oeMlMeritDisP,11.1,,Multi-label pct Meritocratic x C03
oeMlMeritPolP,19.4,,Multi-label pct Meritocratic x C04
oeMlMeritCorP,11.1,,Multi-label pct Meritocratic x C05
oeMlMeritHisP,25.0,,Multi-label pct Meritocratic x C06
oeMlMeritBehP,8.3,,Multi-label pct Meritocratic x C07
oeMlMeritNonP,0.0,,Multi-label pct Meritocratic x C08
oeMlEgalEduP,37.5,,Multi-label pct Egalitarian x C01
oeMlEgalOppP,62.5,,Multi-label pct Egalitarian x C02
oeMlEgalDisP,25.0,,Multi-label pct Egalitarian x C03
oeMlEgalPolP,0.0,,Multi-label pct Egalitarian x C04
oeMlEgalCorP,0.0,,Multi-label pct Egalitarian x C05
oeMlEgalHisP,12.5,,Multi-label pct Egalitarian x C06
oeMlEgalBehP,0.0,,Multi-label pct Egalitarian x C07
oeMlEgalNonP,12.5,,Multi-label pct Egalitarian x C08
oeLibBarrierP,27,,Pct citing >=1 barrier Libertarian
oeModBarrierP,50,,Pct citing >=1 barrier Moderate
oeMeritBarrierP,28,,Pct citing >=1 barrier Meritocratic
oeEgalBarrierP,75,,Pct citing >=1 barrier Egalitarian
oeAgreeMiniNanoRate,92.2,,Agreement rate gpt-mini vs gpt-nano
oeAgreeMiniNanoN,107,,Agreement N gpt-mini vs gpt-nano
oeAgreeMiniSonnetRate,87.9,,Agreement rate gpt-mini vs sonnet
oeAgreeMiniSonnetN,102,,Agreement N gpt-mini vs sonnet
oeAgreeMiniHaikuRate,85.3,,Agreement rate gpt-mini vs haiku
oeAgreeMiniHaikuN,99,,Agreement N gpt-mini vs haiku
oeAgreeNanoSonnetRate,89.7,,Agreement rate gpt-nano vs sonnet
oeAgreeNanoSonnetN,104,,Agreement N gpt-nano vs sonnet
oeAgreeNanoHaikuRate,90.5,,Agreement rate gpt-nano vs haiku
oeAgreeNanoHaikuN,105,,Agreement N gpt-nano vs haiku
oeAgreeSonnetHaikuRate,92.2,,Agreement rate sonnet vs haiku
oeAgreeSonnetHaikuN,107,,Agreement N sonnet vs haiku
oeAgreeAllRate,82.8,,Four-way agreement rate
oeAgreeAllN,96,,Four-way agreement N
oeNdisagree,20,,N inter-model disagreements
oeNsplitThreeOne,11,,N 3-vs-1 disagreement splits
oeNsplitTwoTwo,9,,N 2-vs-2 disagreement splits
worker_age_mean,40,,Worker mean age
worker_age_sd,13.8,,Worker SD age
worker_pct_male,49,,Worker pct male
worker_pct_born_usa,94,,Worker pct US-born
worker_pct_white,70,,Worker pct White
worker_pct_black,12,,Worker pct Black
worker_pct_asian,9,,Worker pct Asian
worker_enc_mean,19.67,,Worker mean encryptions completed
worker_enc_sd,5.91,,Worker SD encryptions completed
worker_enc_mean_1dp,19.7,,Worker mean encryptions completed (1dp)
worker_enc_sd_1dp,5.9,,Worker SD encryptions completed (1dp)
worker_att_mean,20.92,,Worker mean encryption attempts
dyson_age_mean,19.5,,Dyson mean age
dyson_pct_male,54.3,,Dyson pct male
dyson_pct_born_usa,74.7,,Dyson pct US-born
dyson_pct_comfort,71.8,,Dyson pct comfortable upbringing
dyson_pct_manager,80.7,,Dyson pct aspire managerial
dyson_pct_private,72.7,,Dyson pct private sector
dyson_pct_mba,51.6,,Dyson pct plan MBA
dyson_pct_voted,20.4,,Dyson pct voted
dyson_gini_b0,0.416,,Dyson luck Gini (beta0)
dyson_gini_b1,0.311,,Dyson performance effect (beta1)
dyson_gini_b2,0.198,,Dyson efficiency effect (beta2)
dyson_gini_b1_pct,74.8,,Dyson perf effect pct of luck Gini
dyson_gini_b2_pct,47.6,,Dyson eff effect pct of luck Gini
dyson_gini_perfgini,0.727,,"Dyson performance-condition Gini (3dp, R3.5)"
dyson_egal_btw,1.9,,Dyson pct egalitarian between
dyson_lib_btw,23.9,,Dyson pct libertarian between
dyson_merit_btw,37.6,,Dyson pct meritocrat between
dyson_other_btw,36.6,,Dyson pct other between
dyson_moderate_share,24.5,,Dyson pct moderate
fisman_rho_mba,0.500,,"Median implied CES rho, MBA"
fisman_rho_dyson,0.500,,"Median implied CES rho, Dyson"
fisman_rho_yls_all,0.699,,"Median rho, Fisman YLS all subjects"
fisman_n_yls_all,208,,"N, Fisman YLS all subjects"
fisman_rho_yls_fair,0.533,,"Median rho, Fisman YLS fair-minded"
fisman_n_yls_fair,30,,"N, Fisman YLS fair-minded"
fisman_rho_genpop_fair,0.276,,"Median rho, Fisman general-pop fair-minded"
fisman_p_yls_all,0.432,,"Rank-sum p, MBA implied rho vs Fisman YLS all"
rush_cutoff_min,2.20,,"Rush-time cutoff, p5 minutes"
eff_excl_n,46,,N excl eff-cost misunderstanders
eff_excl_pct,17,,Pct excl eff-cost misunderstanders
rob_oth_perf,0.296,,Perf effect excl other (col3)
rob_oth_eff,0.199,,Eff effect excl other (col3)
rob_eff_perf,0.234,,Perf effect excl eff-misund (col1)
rob_eff_eff,0.243,,Eff effect excl eff-misund (col1)
rob_eff_cons,0.447,,Constant excl eff-misund (col1)
rob_coh23_perf,0.235,,Perf 2023 (col1)
rob_coh23_eff,0.182,,Eff 2023 (col1)
rob_coh23_cons,0.419,,Const 2023 (col1)
rob_coh24_perf,0.245,,Perf 2024 (col1)
rob_coh24_eff,0.195,,Eff 2024 (col1)
rob_coh24_cons,0.421,,Const 2024 (col1)
rob_round_perf,0.240,,Perf round FE (col1)
rob_round_eff,0.188,,Eff round FE (col1)
\end{filecontents*}

	
	\title{Are Elites Meritocratic and Efficiency-Seeking? Evidence from MBA Students\thanks{Preuss: Cornell University, email: \href{mailto:mp2222@cornell.edu}{mp2222@cornell.edu}; Reyes (corresponding author): Middlebury College and IZA, email: \href{mailto:greyes@middlebury.edu}{greyes@middlebury.edu}; Somerville: University of California, Santa Barbara, email: \href{mailto:jasonsomerville@ucsb.edu}{jasonsomerville@ucsb.edu}; Wu: University of British Columbia, email: \href{mailto:joy.wu@sauder.ubc.ca}{joy.wu@sauder.ubc.ca}. For helpful comments and suggestions, we thank Peter Andre, Alexander Cappelen, Peter Matthews, Ricardo P\'erez-Truglia, Erik S\o{}rensen, Bertil Tungodden, and participants in FAIR NHH and the Cornell Behavioral Economics Research Group. The experiment reported in this paper was registered in the American Economic Association's registry for randomized controlled trials under ID AEARCTR-0015448. The experiment was reviewed and granted an exemption by the Institutional Review Board at Cornell University (Protocol \#0147786).}}

	\author{ Marcel Preuss
		\and Germán Reyes
		\and Jason Somerville
		\and Joy Wu
	}
	
	\renewcommand{\today}{\ifcase \month \or January\or February\or March\or %
		April\or May \or June\or July\or August\or September\or October\or November\or %
		December\fi \ \number \year} 
	
	\date{\today}

	\maketitle
		
	\begin{abstract} 			
			\noindent Elites disproportionately influence policymaking, yet little is known about their fairness and efficiency preferences---key determinants of support for redistributive policies. We investigate these preferences in an incentivized lab experiment with future elites: Ivy League MBA students. We find that MBA students implement substantially more unequal earnings distributions than the average American, regardless of whether inequality stems from luck or merit. Their redistributive choices are also far more responsive to efficiency costs than the near-zero response found in representative U.S.\ samples. These patterns partly reflect distinct fairness ideals: a large share of MBA students falls outside standard classifications, instead displaying ``weak meritocratic'' tendencies that tolerate inequality even when it stems from luck. These findings identify a channel through which elite preferences may sustain U.S.\ inequality.
	\end{abstract}
	
	
	
	\clearpage

	\section{Introduction} \label{sec:intro}
	
	Support for redistributive policies depends on how individuals weigh fairness considerations against efficiency costs \citep[e.g.,][]{alesina.angeletos2005a, benabou.tirole2006a}. Studies of representative U.S.\ samples find that individuals prefer to redistribute income when inequality stems from luck rather than effort, but show little sensitivity to efficiency costs \citep{almas.etal2020, almas2025fairness, cohn.etal2023}.
	
	Policy outcomes, however, are not determined by the preferences of the general public alone. Theories of ``elite control'' in political science argue that elites---individuals with substantial economic, political, or social capital---largely determine which policies are adopted.\footnote{See, among others, \cite{mills1956power}, \cite{burch198081elites}, \cite{ferguson1995golden}, \cite{winters2009oligarchy}, \cite{winters2011oligarchy}, \cite{beard2012economic}, and \cite{domhoff2013who, domhoff2017power}.} Consistent with these theories, empirical evidence shows that elites have a disproportionate influence on policymaking, while the median voter has little or no influence. For example, \cite{gilens.page2014} show that policy adoption in the U.S.\ is strongly related to support from business groups and the richest ten percent, and is uncorrelated with the preferences of the average American.\footnote{For further evidence in the U.S., see \cite{hacker2010}, \cite{gilens2012}, \cite{rigby2013political}, \cite{gilens.page2014}, \cite{bartels2016unequal}, \cite{page2017}, \cite{page2018billionaires}, and \cite{hertel2019state}. Evidence on the influence of elites on policymaking from other countries is mixed. Single-country studies from Germany, Norway, and the Netherlands show that elites exert disproportionate influence on policymaking \citep{elsasser2021not, schakel2021unequal, mathisen2023}. However, in cross-country regressions, the preferences of low-income individuals are more predictive of redistribution than those of high-income individuals \citep{marechal2025whose}.} Understanding U.S.\ inequality therefore requires measuring elites' fairness and efficiency preferences.
	
	Yet little work has estimated these preferences, partly because elites are difficult to reach. We address this gap by eliciting incentivized redistribution choices from two cohorts of Cornell MBA students---\textit{future} elites. Students from elite universities are disproportionately likely to reach the top one percent \citep{chetty.etal2020}, and graduates from these programs are overrepresented in U.S.\ leadership positions \citep{wai2024most, chetty2026diversifying}. Alumni of the Cornell MBA program include CEOs of Fortune 500 companies, members of Congress, board members, and founders of major companies \citep{johnson2026}.
	
	To estimate MBA students' redistributive preferences, we follow the impartial-spectator paradigm \citep{cappelen.etal2013}. This experimental design involves ``workers'' and ``impartial spectators'' across three stages. In the production stage, workers complete a real-effort task. In the earnings stage, we randomly pair workers and determine their earnings based on either task performance or chance. Thus, earnings inequality is either due to merit or luck. In the redistribution stage, MBA students acting as impartial spectators choose the final earnings allocations for three worker pairs. Each worker pair differs in the source of inequality (merit vs. luck) and the cost of redistributing earnings (no cost vs. costly redistribution). Our main analysis uses spectators' first redistributive choice, the standard between-subject design in the literature. We complement this with a within-subject analysis that uses all three decisions per spectator.
	
	We report two main findings. First, MBA students implement more unequal earnings distributions than those chosen by the average American, regardless of whether worker earnings are due to luck or merit. MBA students implement a Gini coefficient of $\getval{cmp_mba_b0_2dp}$ when worker earnings reflect luck and $\getval{cmp_mba_perfgini_2dp}$ when they reflect task performance. These levels exceed the corresponding pooled estimates of $\getval{pool_luck_genpop_2dp}$ and $\getval{pool_perfgini_genpop_2dp}$ across four representative U.S.\ samples \citep{almas.etal2020, cohn.etal2023, preuss2025inequality, harrs2025fairness}.
	
	Second, MBA students' redistributive choices respond to the efficiency cost of redistribution. When worker earnings are randomly assigned, introducing an efficiency cost increases the implemented Gini coefficient by $\getval{gini_b2_fd}$ points. This response is much larger than in representative U.S.\ samples, where estimated responses to efficiency costs are typically statistically
	indistinguishable from zero \citep{almas.etal2020, almas2025fairness}. We replicate these findings with Cornell undergraduate business students---another sample of future elites. 

	To understand why future elites implement more unequal distributions than the average American, we classify them according to established fairness ideals in the literature. Following the standard methodology of \cite{almas.etal2020}, we estimate that $\getval{egal_share_btw}$ percent of MBA students are egalitarians, $\getval{lib_share_btw}$ percent are libertarians, $\getval{merit_share_btw}$ percent are meritocrats, and $\getval{other_share_btw}$ percent do not fit any of these three fairness ideals. Relative to representative U.S.\ samples, the fraction of egalitarians is similar, the fractions of meritocrats and libertarians are lower, and the unclassified fraction is much larger \citep{almas.etal2020, almas2025fairness, cohn.etal2023, preuss2025inequality, harrs2025fairness}. We use our repeated-measures design and find that many of these unclassified spectators display weak meritocratic tendencies---they redistribute more when inequality stems from luck than from performance, but still allow luck-based winners to retain an earnings premium. We refer to these unclassified weak meritocrats as ``moderates.'' Open-ended responses on the drivers of U.S.\ inequality show that moderates attribute inequality to multiple forces---education, unequal opportunities, and discrimination---a profile distinct from any standard fairness ideal.
	
	Our findings suggest that U.S.\ redistributive policies remain more limited than the average citizen would prefer because elites, exerting disproportionate influence on policy, tolerate more inequality. This mismatch fits a broader puzzle: the U.S.\ remains a highly unequal country despite widespread support for a more egalitarian distribution \citep{norton.ariely2011, norton.ariely2013}. This preference gap is also consistent with recent evidence that the welfare weights implied by actual U.S.\ tax and transfer policies are less progressive than those of the general population, with high-income individuals receiving disproportionate weight \citep{capozza2024should}.
	
	We contribute to the literature on the relationship between fairness views and income redistribution. Empirical evidence shows that the source of inequality---effort versus luck---shapes individuals' attitudes toward redistribution.\footnote{Empirical work using lab experiments includes \cite{cappelen.etal2007}, \cite{cappelen.etal2010}, \cite{cappelen.etal2013}, \cite{cappelen.etal2022}, \cite{almas.etal2010}, \cite{almas.etal2011}, \cite{durante.etal2014}, \cite{mollerstrom.etal2015}, \cite{andre2025}, \cite{dong2025they}, \cite{bhattacharyaandmollerstrom2025}, \cite{preuss2025inequality}, \cite{harrs2025fairness}, and \cite{yusof2025market}.} Recent work estimates how fairness ideals are distributed in representative population samples \citep{almas.etal2020, almas2025fairness, muller.renes2021, cohn.etal2023, harrs2025fairness}. We estimate fairness ideals in a sample of \textit{future} elites who will likely influence policymaking and business decisions that directly affect inequality, such as worker compensation schemes.

	Two related papers also study elites' distributional preferences: \citet{fisman.etal2015a} and \citet{cohn.etal2023}. \citet{fisman.etal2015a} study Yale Law students' distributional preferences using modified dictator games in which subjects divide an endowment between themselves and another participant. They find that elites place more weight on their own payoff and prioritize efficiency over \textit{altruism}---i.e., reducing their own payoff to help others. By contrast, we find that elites prioritize efficiency over concerns for \textit{equality}. Moreover, when subjects have a personal stake in the allocation as in \citet{fisman.etal2015a}, separating self-interest from fairness preferences requires strong functional-form assumptions.\footnote{Individuals use uncertainty or inefficiencies as ``moral wiggle room'' to excuse selfishness \citep{dana2007exploiting, exley2016excusing, exley2020using}. The functional form used to estimate efficiency concerns neglects this possible confound, which may inflate efficiency estimates for more selfish subjects.} Further, \citet{nax2021elites} report a failed replication and argue that the observed elite--non-elite differences may reflect protocol variations. Our impartial-spectator design addresses both concerns: by removing subjects' stake in the outcome, we measure fairness and efficiency preferences directly without structural assumptions, and by using a consistent protocol across populations, we ensure comparability with existing studies. \citet{cohn.etal2023} find that high-income or high-wealth Americans---households in the top 5 percent by income or wealth, identified using self-reported survey data---are less inequality-averse than the general population. Beyond studying a different population, our paper complements and is distinct from their work in two main ways. First, we examine how efficiency concerns shape redistribution---a parameter that prior representative-sample studies estimate to be near zero, a puzzling null result. Second, our in-class recruitment mitigates the self-selection bias of survey-based samples, in which individuals with particular social preferences are more likely to participate \citep{levitt.list2007}.
	
	Finally, we contribute to the literature on the determinants of labor market inequality. While firm wage-setting policies drive substantial earnings inequality \citep[e.g.,][]{card.etal2018}, why firms pay
	different wages to observationally equivalent workers remains poorly understood.\footnote{Recent work studying firm wage-setting includes \cite{hall2012evidence}, \cite{caldwell2019outside}, \cite{lachowska2022wage}, \cite{cullen2025}, \cite{derenoncourt2025}, \cite{dube2025monopsony}, \cite{reyes2025}, \cite{hazell2026national}, and \cite{hjort2026across}.} One potential explanation centers on managers' roles in designing compensation schemes \citep{frank2015performance, acemoglu.etal2025, he.lemaire2025}. If managers' fairness preferences shape wage-setting decisions, then heterogeneity in these preferences could generate firm-level wage dispersion. We address this hypothesis by having future managers allocate real earnings between observationally equivalent workers, varying only the stated source of worker earnings across decisions. We document substantial heterogeneity in implemented earnings inequality across participants, suggesting that variation in managers' fairness preferences may contribute to wage differentials among otherwise similar workers.

	\section{Experimental Design} \label{sec:experiment}			
	
	We follow the standard impartial-spectator paradigm \citep{cappelen.etal2013}. The experiment has two types of participants---workers and impartial spectators---across three stages: a production stage, an earnings stage, and a redistribution stage (Appendix Figure~\ref{fig:flow} shows the flow of the experiment).
	
	\subsection{Stages of the Experiment}\label{sec:stages}
	
	In the production stage, workers have five minutes to complete a real-effort encryption task (Appendix Figure~\ref{fig:encryptiontask} shows an example). We inform workers that their earnings depend on their relative performance and a third party's decision, but we do not describe the redistribution mechanism.
	
	In the earnings stage, workers are randomly paired and compete in a winner-take-all format. We determine the winner of each pair either by the number of encryptions completed (``performance'') or by a coin flip (``luck''). Winners receive an initial allocation of $\$6$, while losers receive $\$0$. Workers never learn the outcome of their competition---they observe only their final payment after redistribution.
	
	In the redistribution stage, spectators (MBA students) choose final earnings allocations for three worker pairs. Each pair differs in how the winner is determined and whether redistribution is costly. Spectators know which worker won each competition but not how many encryptions workers completed. Spectators can redistribute any amount ranging from \$0 to \$6 in \$0.50 increments. We present each decision as an adjustment schedule. We incentivize spectators by randomly selecting one of their decisions to implement.\footnote{To mitigate the impact of anchoring effects on redistribution decisions, we tell spectators that workers were not informed about whether they won or lost their match, but would only be told their final earnings.}
	
	After the redistribution stage, spectators complete an exit survey covering demographics, socioeconomic background, career aspirations, social views, and an attention check. The second MBA cohort also answers an open-ended question on the main drivers of income inequality in the U.S.

	\subsection{Treatment Conditions}\label{sec:environments}
	
	We study three conditions, presented to spectators in a randomized order (see Appendix Figure~\ref{fig:screenshots} for screenshots of each treatment). In the \textit{luck} treatment, the winner is determined by a coin flip. In the \textit{performance} treatment, the winner is the worker who completed more encryptions. In the \textit{efficiency cost} treatment, the winner is determined by a coin flip, and redistribution incurs an ``adjustment cost'' that reduces total participant earnings. For every \$1.00 reduction in the winner's earnings, the loser's earnings increase by only \$0.50, resulting in a net loss of total income.\footnote{One potential concern is that spectators might try to minimize cognitive effort by selecting default options, which could bias our results. Our experimental design addresses this in two ways: no option was pre-selected, requiring spectators to actively choose, and we presented the options horizontally rather than vertically to prevent defaulting to ``top'' choices. No spectator clicked through all redistributive decisions without making at least one active choice, suggesting effort-minimizing concerns are unlikely to bias our results.}
	
	\section{Data and Summary Statistics} \label{sec:data}

	\subsection{Recruitment of Workers and Spectators} 

	We recruited $\getval{n_workers_paired}$ U.S.-based individuals on Prolific to work on the encryption task.\footnote{These were paired into \getval{n_worker_pairs} worker pairs, which were matched to redistributive decisions from our \getval{n_mba} MBA students and \getval{n_dyson_recruited} undergraduate business students.} We paid workers a \$1.50 participation fee for completing the task, plus a bonus of up to \$6 based on a randomly chosen spectator's decision. The average bonus was \$\getval{worker_bonus_mean}, resulting in average total compensation of \$\getval{worker_comp_mean} per worker. (See Appendix~\ref{app:worker} for additional details on the worker sample.)
	
	We collected MBA students' redistributive decisions from two consecutive cohorts (2023 and 2024) during mandatory classes. This approach mitigates the self-selection concerns that can arise in experiments \citep{levitt.list2007}. Of the \getval{n_mba_started} MBA students who started the experiment, we excluded \getval{n_mba_excluded} who quit before reaching the first redistribution screen, leaving an analysis sample of \getval{n_mba} students.\footnote{The experiment was conducted at the end of class sessions with voluntary participation clearly stated. MBA students often have commitments immediately following class, including case interview practice sessions, recruiting calls, and networking meetings, which may have led some students to leave before completing the experiment.}

	\subsection{Summary Statistics and Balance} 

	Appendix Table~\ref{tab:summ_trt} presents summary statistics for the spectators, both overall (column 1) and by the first treatment shown (columns 2--4). Most MBA students are aged 24--31 (\getval{mba_pct_age_2431} percent), male (\getval{mba_pct_male} percent), and U.S.-born (\getval{mba_pct_born_usa} percent). They strongly prefer private-sector careers (\getval{mba_pct_private} percent) and managerial roles (\getval{mba_pct_manager} percent), with most planning to work in the U.S.\ after graduation (\getval{mba_pct_stay_usa} percent). The majority voted in recent elections (\getval{mba_pct_voted} percent) and believe that hard work leads to a better life (\getval{mba_pct_hardwork} percent). Observable characteristics are balanced across the first treatment shown: F-tests of joint equality fail to reject for any condition.
	
	To benchmark our results, we compare MBA students' redistributive choices with those in four studies that estimate the distribution of fairness ideals in the U.S.\ general population: \cite{almas.etal2020}, \cite{cohn.etal2023}, \cite{harrs2025fairness}, and the Survey of Consumer Expectations (SCE) collected by \cite{preuss2025inequality}. For studies that report the share of earnings redistributed, we use that measure to construct the corresponding Gini coefficient.\footnote{While these comparison studies were conducted at different times, research shows that distributional preferences remain stable over time \citep{fisman2023distributional, harrs2026stable}, making these cross-study comparisons informative.} We also report a pooled estimate across all representative samples using a DerSimonian--Laird random-effects meta-analysis \citep{dersimonian1986meta}. Appendix~\ref{app:benchmark} compares the experimental design, recruitment, and sample characteristics across
	benchmark studies.\footnote{Another relevant comparison is \citet{fisman.etal2015a}, who also study elites' distributional and efficiency preferences. Because their experimental design, data, and identification strategy differ substantially from ours---they use a dictator game with structural estimation of a CES utility function---we do not include them in the main benchmarking exercise. In Appendix~\ref{app:fisman_comparison}, we compare the implied efficiency parameters in our paper using their methodology.}
	
	\section{Origin of Income Differences and Implemented Inequality} \label{sec:results}
	
	\subsection{Effects of Merit and Efficiency on Redistribution}
	
	We estimate linear models of the form:
	\begin{align}\label{eq:gini}
		\text{Gini}_{ip} = \beta_0 + \beta_1 \text{Performance}_{p} + \beta_2 \text{EfficiencyCost}_{p} + \varepsilon_{ip},
	\end{align}
	where $\text{Gini}_{ip}$ is the Gini coefficient of the final earnings allocation in worker pair $p$ implemented by spectator $i$, $\text{Performance}_{p}$ and $\text{EfficiencyCost}_{p}$ are indicators that equal one if pair $p$ competed in the performance or efficiency cost treatments and zero if they competed in the luck treatment. $\beta_0$ measures the average Gini when worker earnings are randomly assigned and there is no redistribution cost. $\beta_1$ measures the causal impact of assigning worker earnings based on performance on implemented inequality. $\beta_2$ measures the effect of introducing an efficiency cost on the Gini.\footnote{While $\beta_2$ represents the effect of switching $\text{EfficiencyCost}_{p}$ from 0 to 1, we can derive the implied arc elasticity with respect to the efficiency cost $c$. With $c = 0$ in the no-cost condition and $c = 0.5$ in the efficiency cost condition, the arc elasticity is approximately $\epsilon \approx (\beta_2/\bar{G}) / (\Delta c/\bar{c})$, where $\bar{G}$ is the mean Gini coefficient, $\Delta c = 0.5$, and $\bar{c} = 0.25$ is the midpoint of the two cost levels.}
	
	Table~\ref{tab:gini} presents regression estimates of equation \eqref{eq:gini}. Columns 1 and 2 use all redistributive decisions from spectators, with column 1 including no controls (thus, identification is based on both between- and within-subject variation) and column 2 including spectator fixed effects (identification based on within-subject variation). Columns 3 and 4 use only spectators' first decisions (identification based on between-subject variation), though the order in which we presented the treatments has no statistically significant effect on redistributive choices (Appendix Table~\ref{tab:gini_order}). We cluster standard errors at the spectator level throughout. Appendix~\ref{app:descriptive} shows descriptive evidence on implemented inequality and redistribution choices across experimental conditions.

	Merit-based earnings inequality and efficiency costs increase implemented inequality. When worker earnings are randomly assigned, spectators implement a Gini coefficient of $\hat{\beta}_0 = \getval{gini_b0}$ (columns 1 and 2). Assigning worker earnings based on their performance increases the implemented Gini coefficient by $\hat{\beta}_1 = \getval{gini_b1}$ Gini points ($p<0.01$), or $\getval{gini_b1_pct}$ percent of the luck condition Gini. Similarly, introducing a redistribution cost increases the Gini coefficient by $\hat{\beta}_2 = \getval{gini_b2}$ Gini points ($p<0.01$), or $\getval{gini_b2_pct}$ percent of the luck condition Gini. This corresponds to an implied elasticity of approximately $\getval{gini_elast}$: a one percent increase in the cost is associated with roughly a \getval{gini_elast} percent increase in the Gini.
	
	The coefficients are similar when using only the first decision of each spectator, albeit with a slightly smaller impact of the performance condition. Column 3 shows that assigning worker earnings based on performance (relative to luck) increases the Gini coefficient by $\hat{\beta}_1 = \getval{gini_b1_fd}$ Gini points ($p<0.01$), or $\getval{gini_b1_fd_pct}$ percent of the luck condition Gini. Still, given the standard errors, we cannot reject equality of coefficients between columns 1 and 3 at conventional levels.
	
	The results are robust to several sample restrictions, specification checks, and alternative dependent variables. Appendix~\ref{sec:robustness} shows robustness to (i) excluding spectators who failed the attention check (Appendix Table~\ref{tab:rob_att}); (ii) excluding spectators who rushed through the experiment (Appendix Table~\ref{tab:rob_tim}); (iii) excluding spectators who allocated strictly more earnings to the loser than to the winner (Appendix Table~\ref{tab:rob_los}); (iv) excluding spectators whom we classify as having non-standard fairness ideals (Appendix Table~\ref{tab:rob_oth}); (v) excluding spectators who redistributed more when there was an efficiency cost (Appendix Table~\ref{tab:rob_eff}); (vi) estimating the regression separately for each cohort (Appendix Table~\ref{tab:rob_coh}); (vii) including round fixed effects (Appendix Table~\ref{tab:rob_round}); and (viii) using the net-of-efficiency-cost share of earnings redistributed as the outcome (Appendix Table~\ref{tab:rob_red}).
	
	\subsection{Are Elite Redistributive Choices Different from the Average American's?}
	
	To compare MBA students with U.S.\ benchmark samples, Figure~\ref{fig:gini_comparison} plots mean Gini coefficients by treatment condition and Table~\ref{tab:comparison} presents side-by-side regression estimates of equation \eqref{eq:gini}, with Panel~A comparing MBA students to the general population and Panel~B to higher-income subsamples. Column 1 reproduces the first-decision specification from column~4 of Table~\ref{tab:gini}, which is the design most comparable to the benchmark studies.

	Future elites are more tolerant of inequality when earnings differences stem from luck. In the luck condition, MBA students implement a Gini of $\getval{cmp_mba_b0}$ (Table~\ref{tab:comparison}, Panel A, column 1), compared with a pooled estimate of $\getval{pool_luck_genpop}$ across the four representative U.S.\ samples (column 6). The MBA Gini significantly exceeds those of \citet{cohn.etal2023} ($p<0.001$) and \citet{harrs2025fairness} ($p=\getval{p_mba_vs_hs_luck}$), and is statistically indistinguishable from the higher-inequality \citet{almas.etal2020} and \citet{preuss2025inequality} samples ($p=\getval{p_mba_vs_almas_luck}$ and $p=\getval{p_mba_vs_sce_luck}$, respectively). The MBA--pooled gap amounts to about $\getval{mba_vs_usnor_pct}$ percent of the U.S.--Norway implemented-inequality gap in \citet{almas.etal2020}. The conclusion extends to higher-income benchmark samples (Panel B): the pooled luck-condition Gini is $\getval{pool_luck_hiinc}$. The gap persists even against \citet{cohn.etal2023}'s top-5\%-by-income-or-wealth sample, which implements a luck-condition Gini of $\getval{cohn_top5_luck}$---about $\getval{mba_vs_cohn_pct}$ percent smaller than the MBA estimate.\footnote{\label{fn:cohn_business}\cite{cohn.etal2023} find that business owners are the most tolerant of inequality. Inequality tolerance in this subgroup is also the closest to our sample, although a gap remains in the luck treatment where business owners still implement less inequality than MBA students (a Gini of $\getval{cohn_business_luck}$ vs.\ $\getval{mba_luck_gini_fd}$).}
	
	Future elites are less responsive to the source of inequality than the average American. The performance condition increases the implemented Gini by $\getval{cmp_mba_perf}$ points among MBA students, compared with a pooled estimate of $\getval{pool_perf_genpop}$ across the representative samples (Table~\ref{tab:comparison}, column 6). Yet MBA students start from a much higher baseline in the luck condition, so they still implement more inequality in absolute terms: their performance-condition Gini is $\getval{cmp_mba_perfgini}$, compared with a pooled estimate of $\getval{pool_perfgini_genpop}$ across representative samples. The pattern is similar for higher-income subsamples: the pooled performance-condition Gini is $\getval{pool_perfgini_hiinc}$ (Panel B), still below the MBA estimate. The same pattern holds against \citet{cohn.etal2023}'s top-5\%-by-income-or-wealth sample: although their performance effect ($\getval{cohn_perf}$ points) exceeds the MBA estimate, their lower luck-condition baseline ($\getval{cohn_top5_luck}$) leaves their performance-condition Gini at $\getval{cohn_perfgini}$---still below the MBA level.\footnote{Some political economy models suggest that the median voter's preferences, rather than the mean, determine policy outcomes \citep[e.g.,][]{downs1957economic}. To assess whether our comparisons are sensitive to this distinction, we compute the median and mean implemented Gini in each of the representative samples and take a simple unweighted average across studies (distinct from the random-effects pooled estimate of $\getval{pool_luck_genpop}$ reported above). In the performance condition, the mean and median Gini are similar ($\getval{bench_perf_mean}$ vs.\ $\getval{bench_perf_median}$). In the luck condition, however, the median Gini ($\getval{bench_luck_median}$) is substantially lower than the mean ($\getval{bench_luck_mean}$). Therefore, comparing future elites to the median citizen yields an even larger preference gap than comparing to the mean citizen.}

	The largest gap between MBA students and the general population is in their responsiveness to efficiency costs. The efficiency-cost condition increases the Gini by $\getval{cmp_mba_eff}$ points among MBA students, compared with a statistically insignificant $\getval{almas_us_eff}$ points in the representative U.S.\ sample of \citet{almas.etal2020}---the only other study with an equivalent efficiency-cost treatment (same 50 percent cost of redistribution). The difference between these two efficiency responses is statistically significant ($p=\getval{p_mba_vs_almas_eff}$). In a re-analysis of \citet{almas.etal2020}'s data, we find that respondents with household income above \$100{,}000 also respond significantly to efficiency costs ($\getval{almas_hi_eff}$ Gini points, $p<0.05$), though MBA students' response remains larger.\footnote{Recent work by \citet{almas2025fairness}, for which microdata are not available, estimates that the efficiency-cost condition increases the Gini by about $0.08$ points in their U.S.\ sample (see their Figure 5, panel b), a coefficient that is also statistically indistinguishable from zero. The MBA efficiency response exceeds that of all countries in their worldwide sample, confirming that future elites are substantially more responsive to efficiency costs than representative populations.}

	\subsection{Replication with Elite Undergraduate Business Students} \label{sec:replication}
	
	To assess the robustness of these results, we replicate our analysis with \textit{undergraduate} Cornell business students from the Dyson School of Applied Economics and Management (henceforth ``Dyson students''; see Appendix~\ref{app:dyson} for details). Dyson students are another future-elite sample: Ivy League universities show high mobility to the top one percent \citep{chetty.etal2020} and disproportionate representation in leadership \citep{chetty2026diversifying}.	
	
	Dyson students implement earnings distributions as unequal as those of MBA students: a Gini coefficient of $\getval{dyson_gini_b0}$ when worker earnings are determined by luck and $\getval{dyson_gini_perfgini}$ when determined by performance---substantially higher than the Ginis in representative U.S.\ samples. Dyson students' sensitivity to efficiency costs also mirrors that of MBA students: redistribution costs increase the Gini by $\getval{dyson_gini_b2}$ points, nearly identical to the MBA estimate and substantially larger than the near-zero response in representative samples. This consistency across undergraduate and graduate business populations supports our conclusion that future elites tolerate more inequality and respond more strongly to efficiency costs than the general population.
	
	\section{The Fairness Ideals of the Elite} \label{sec:ideals}
	
	\subsection{Measuring Fairness Ideals}
	
	To understand why future elites implement more inequality than the general population, we classify MBA students by their fairness ideal. We develop a statistical framework that models fairness ideals as mappings from initial earnings distributions to final allocations (see Appendix~\ref{app:framework}). Following the literature, we focus on three fairness ideals: egalitarian, libertarian, and meritocratic. Egalitarians equalize workers' earnings regardless of how the earnings were generated. Libertarians leave the initial earnings unchanged regardless of how earnings differences arose. Meritocrats condition their redistributive decisions on the source of inequality: they equalize earnings when inequality is entirely luck-based (i.e., there is no merit to the earnings allocation), but redistribute less when inequality reflects performance differences. These fairness ideals predict individuals' social preferences for redistribution \citep{harrs2025fairness}.
	
	We use two complementary designs to measure fairness ideals. The first approach relies on between-subject comparisons, using only each spectator's first redistributive decision and excluding those who saw the efficiency cost environment first. This is the standard design used in the literature \citep[e.g.,][]{almas.etal2020, almas2025fairness, cohn.etal2023} and lets us benchmark our results against prior work. The second approach uses within-subject variation from multiple decisions per spectator across the luck and performance treatments. This approach, also used by \citet{harrs2025fairness}, lets us identify each spectator's fairness ideal, not just aggregate shares. We extend the standard identification assumptions to accommodate these multiple observations per spectator (see Appendix~\ref{app:framework}).
	
	The first-choice design provides clean comparisons across individuals and avoids biases arising from sequential decisions, but is more vulnerable to measurement error and choice noise. The repeated-measures design reduces measurement error by incorporating more information about each spectator's behavior, but requires assuming that initial choices do not influence subsequent decisions.\footnote{Recent evidence shows that individual-level fairness type classifications are highly stable over time \citep{harrs2026stable}, further supporting the use of within-subject variation to identify fairness ideals.} We present results from both approaches to establish robustness and to ease comparison with existing research.
	
	\subsection{Estimates of Fairness Ideals}
	
	Based on the spectators' first choices, we estimate that $\getval{egal_share_btw}$ percent of MBA students are egalitarians, $\getval{lib_share_btw}$ percent are libertarians, $\getval{merit_share_btw}$ percent are meritocrats, and $\getval{other_share_btw}$ percent follow other fairness ideals (Table~\ref{tab:ideals}, Panel A; Figure~\ref{fig:ideals} plots these estimates alongside the pooled random-effects benchmark). These estimates are similar to those from the repeated-measures design, which classifies $\getval{egal_share_wth}$ percent as egalitarians, $\getval{lib_share_wth}$ percent as libertarians, $\getval{merit_share_wth}$ percent as meritocrats, and $\getval{other_share_wth}$ percent as following other fairness ideals. Differences in proportions across the two designs are not statistically significant at conventional levels (Appendix Figure~\ref{fig:ideals_btw_wth}). A chi-squared Wald test fails to reject equality of distributions across the two designs ($p = \getval{chi2_btw_wth}$), suggesting that our fairness-ideal estimates are robust to the elicitation method.\footnote{Some cross-design differences are sizable despite being statistically insignificant, particularly for libertarians (\getval{lib_share_btw} vs.\ \getval{lib_share_wth} percent).}
	
	The distribution of fairness ideals among MBA students differs markedly from that of the general population (Table~\ref{tab:ideals}, Panel B). Relative to the pooled benchmark across all four representative U.S.\ samples, the shares of libertarians and meritocrats are smaller among MBA students. The largest difference is in the ``other'' category: $\getval{other_share_btw}$ percent of future elites are unclassified, compared with a pooled estimate of $\getval{pool_other_genpop}$ percent across the four benchmark samples.\footnote{The high proportion of individuals in the ``other'' category is not an artifact of our definitions of fairness ideals---we employ the same classification approach as \cite{almas.etal2020} in our between-subject analysis, yet still find a substantially larger share of unclassified individuals than in prior studies.} A chi-squared Wald test rejects the equality of distributions between the MBA sample and each of the four representative samples individually: $p = \getval{chi2_mba_sce}$ for the SCE, $p = \getval{chi2_mba_almas}$ for \citet{almas.etal2020}, $p < 0.001$ for \citet{cohn.etal2023}, and $p < 0.001$ for \citet{harrs2025fairness}.
	
	MBA students' fairness preferences differ systematically from those of higher-income Americans (Table~\ref{tab:ideals}, Panel C). MBA students are less likely to be meritocrats, and a larger proportion falls outside the three conventional fairness ideal classifications. This contrasts with prior work showing that higher-income Americans are less frequently egalitarian and more frequently meritocratic than the average American, though findings vary across studies. A chi-squared Wald test rejects equal distributions between the MBA sample and the higher-income samples in \citet{cohn.etal2023} ($p < 0.001$) and \citet{harrs2025fairness} ($p = \getval{chi2_mba_harrs_hi}$), and yields marginal rejections for the SCE ($p = \getval{chi2_mba_sce_hi}$) and \citet{almas.etal2020} ($p = \getval{chi2_mba_almas_hi}$).
	
	\subsection{Understanding the Fairness Views of Future Elites} \label{sub:views_elites}
	
	Who are the spectators who do not conform to standard fairness ideals? To answer this, we first assess whether unclassified fairness views reflect deliberate choices. Then, we identify systematic patterns in redistribution decisions using the repeated-measures design.
	
	\subsubsection{Are Non-Standard Fairness Views Deliberate or Random?}
	
	Spectators classified as ``other'' make deliberate choices rather than randomly clicking through the survey. Four pieces of evidence support this. First, unclassified spectators spend just as much time making their redistribution decisions as classified spectators (Appendix Table~\ref{tab:other_time}). Second, they pass the embedded attention check at the same rate as other spectators.\footnote{In our data, $\getval{attn_fail_pct}$ percent of all spectators fail the attention check. A bivariate regression of an indicator for failing the attention check on the ``other ideal'' dummy yields a statistically insignificant $\hat\beta = \getval{attn_other_beta}$ ($p = \getval{attn_other_p}$).} Third, the joint distribution of their redistribution choices across the luck and performance conditions shows systematic patterns rather than the uniform distribution we would expect from random clicking (discussed below). Fourth, our replication study with another future-elite sample---the Dyson students---finds similar rates of unclassified spectators (Section~\ref{sec:replication}).
	
	\subsubsection{Redistributive Patterns Leading to Unclassified Elites.}
	
	To identify the patterns behind unclassified fairness ideals, we examine the joint distribution of redistribution choices in the performance and luck conditions (Appendix Table~\ref{tab:joint}). The most common unclassified pattern is to redistribute more in the luck condition than in the performance condition, but without fully equalizing earnings under luck. Specifically, $\getval{moderate_pattern1_pct}$ percent of spectators redistribute \$1 or \$2 in the luck condition while redistributing \$0 in the performance condition, and $\getval{moderate_pattern2_pct}$ percent redistribute \$2 in the luck condition and \$1 in the performance condition. These spectators behave as ``weak meritocrats'': they reward effort differences but still let luck-based winners retain a premium by not fully equalizing earnings in the luck condition. Together, these choices account for $\getval{moderate_share_all}$ percent of all spectators (or $\getval{moderate_share_of_other}$ percent of unclassified spectators). We refer to these spectators as ``moderates.''\footnote{The remaining \getval{other_residual_pct} percent of unclassified spectators show no single dominant pattern---the largest subgroup redistributes the same intermediate amount regardless of the source of inequality, and the rest are scattered across the joint distribution (Appendix Table~\ref{tab:joint})---so they fit none of the standard fairness ideals and are best understood as a heterogeneous residual rather than a coherent type. The only characteristic that predicts membership is a marginally lower propensity to have voted in recent elections (Appendix Table~\ref{tab:ideals_correlates_bonf}; $p < 0.10$).}
	
	Moderates implement more unequal earnings distributions than meritocrats. On average, moderates redistribute $\getval{moderate_redist_pct}$ percent of earnings and implement a Gini coefficient of $\getval{moderate_gini}$, whereas meritocrats redistribute $\getval{merit_redist_pct}$ percent of earnings and implement a Gini coefficient of $\getval{merit_gini}$. These differences are statistically significant at $p < 0.01$. The key distinction between moderates and meritocrats is that moderates preserve some inequality even when it results purely from chance, contradicting the meritocratic principle that only merit-based differences should be rewarded.\footnote{Several individual characteristics correlate significantly with fairness type (Appendix Table~\ref{tab:ideals_correlates_bonf}). Men are more likely than women to be libertarians ($p < 0.01$), respondents who believe that working long hours is necessary are more likely to be moderates ($p < 0.05$), and politically active students---measured by voting in recent elections---are more likely to be meritocrats ($p < 0.05$).}

	\subsubsection{Inequality Narratives.}

	To further characterize moderates relative to other fairness types, we analyze the second MBA cohort's open-ended responses to the question: ``What do you believe is the main driver of income inequality in the United States?'' We use large language models (LLMs) to classify the responses into eight categories developed inductively from the data: education, unequal opportunities, discrimination, government policy, corporate/elite power, historical legacy, behavioral/cultural explanations, and other. For each response, the LLMs produced a single-best label and multi-label coding. For example, responses mentioning ``school,'' ``college,'' or ``studying'' are classified under education, while mentions of race, gender, or bias are classified under discrimination. Appendix~\ref{app:quotes} details the methodology, robustness checks, and representative quotes.

	The classification shows systematic differences across fairness types (Appendix Table~\ref{tab:oe_crosstab} and Figure~\ref{fig:ineq_drivers}). Meritocrats emphasize education (\getval{oeMlMeritEduP} percent), libertarians emphasize behavioral and cultural explanations (\getval{oeMlLibBehP} percent), and egalitarians emphasize barriers---unequal opportunities (\getval{oeMlEgalOppP} percent) and discrimination (\getval{oeMlEgalDisP} percent). Moderates share meritocrats' emphasis on education but cite barriers---discrimination (\getval{oeMlModDisP} vs.\ \getval{oeMlMeritDisP} percent) and unequal opportunities (\getval{oeMlModOppP} vs.\ \getval{oeMlMeritOppP} percent)---more often. Unlike meritocrats, who frequently cite historical and systemic forces (\getval{oeMlMeritHisP} percent), moderates almost never do (\getval{oeMlModHisP} percent), instead attributing inequality to multiple proximate causes. Moderates differ from libertarians in their low emphasis on behavioral explanations (\getval{oeMlModBehP} vs.\ \getval{oeMlLibBehP} percent) and high emphasis on barriers. Taken together, these patterns indicate that moderates hold a distinct view of why inequality exists, one not captured by standard fairness type classifications.\footnote{As an exploratory accounting exercise, we ask how much of the moderate--meritocrat gap in redistribution can be explained by their differing narratives. We regress the share of earnings redistributed on indicators for each narrative category, pooling moderates and meritocrats, and multiply each coefficient by the moderate--meritocrat difference in how often each category is cited. Summing across categories, narratives explain about \getval{decomp_narr_pct} percent of the gap, driven mainly by meritocrats citing historical and systemic causes more often, which is associated with more redistribution.}
	
	\section{Conclusion}
	
	We examine which types of inequality future elites consider worthy of redistribution and how much weight they give to efficiency. We find that MBA students implement substantially more unequal earnings distributions than the average American and are far more responsive to efficiency costs.

	Our results speak to a puzzle in the political economy of inequality: while most Americans express strong preferences for more equal income distributions \citep{norton.ariely2011, norton.ariely2013}, the U.S.\ remains a highly unequal nation \citep{chancel2022world, saez2022top}. Our results suggest one explanation: future elites, who are likely to enter positions with disproportionate influence over policy and organizational decisions, tolerate substantially more unequal earnings distributions than the average American, and may therefore be more reluctant to adopt redistributive policies. This aligns with recent evidence that the general population's welfare weights are substantially more progressive than those implied by actual tax and transfer policies \citep{capozza2024should}. The prevalence of moderates among future elites further suggests that elites do not just tolerate more inequality---they think about its causes differently, in ways existing fairness ideal classifications miss.
	
	Although we document differences in redistributive preferences between future elites and the general population, we cannot pin down why they differ. Observable characteristics such as education, socioeconomic background, or career aspirations may explain these differences, but the direction of causality remains unclear. Individuals with efficiency-seeking preferences may self-select into elite business programs, or these programs may shape their preferences through training and socialization \citep{bauman2011selection, sundemo2025business}. While these causal mechanisms warrant future research, our core finding is unchanged: a sizable preference gap exists between the general population and those who will likely shape policy.
	
	Whether these preferences persist as MBA students advance into positions of power is an open question. Three pieces of evidence suggest these preferences may endure. First, MBA and undergraduate business students yield similar results, indicating that future elites'
	redistributive preferences are stable across educational stages. Second, distributional preferences tend to be stable over time \citep{fisman2023distributional, harrs2026stable}. Third, \cite{almas.etal2020} find that age has no economically significant effect on redistributive preferences, suggesting that preference gaps between future elites and the general population are unlikely to narrow with age alone.

	Another promising direction is exploring preference heterogeneity across elite groups. Demographic and ideological differences exist across Silicon Valley, Wall Street, and Capitol Hill elites \citep{broockman.etal2019, buhlmann2025varieties}. Business leaders may have different preferences from political, academic, or cultural elites. Measuring these preferences across elite groups and tracing how they translate into policy influence are natural extensions.
	
	\clearpage
	\section*{Figures and Tables}

	\begin{figure}[H]
		\caption{Implemented Inequality Across Treatments: Elites vs.\ Benchmark Studies}\label{fig:gini_comparison}
		\centering
		\includegraphics[width=.75\linewidth]{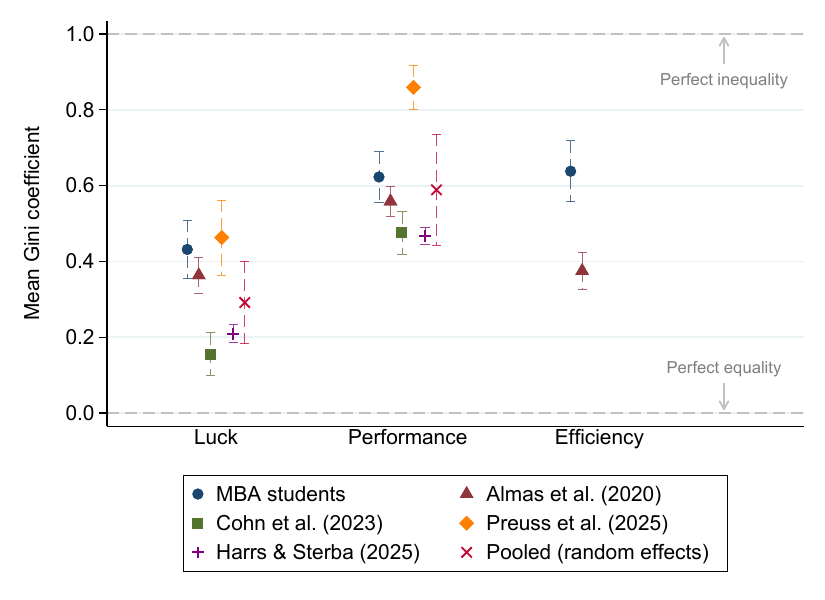}
		
		\hfill
		{\footnotesize
			\singlespacing \justify
			
			\textit{Notes:} This figure plots mean Gini coefficients with 95 percent confidence intervals for each treatment condition across studies. The Gini coefficient measures implemented inequality in the final earnings allocation, ranging from zero (perfect equality) to one (maximum inequality). For \cite{cohn.etal2023}, we construct the Gini as $1 - 2 \times (\text{share redistributed})$; observations are weighted to be representative of the U.S.\ population. For \cite{harrs2025fairness}, we construct the Gini as $|1 - \text{transfer}/2|$ using each respondent's first decision only. The efficiency treatment is available only in our experiment and in \cite{almas.etal2020}. The pooled estimate is a DerSimonian--Laird random-effects meta-analysis across the four benchmark studies \citep{dersimonian1986meta}. \par
			
		}
		
	\end{figure}

	\clearpage
	\begin{figure}[H]
		\caption{The Distribution of Fairness Ideals Among MBA Students and in the U.S.}\label{fig:ideals}
		\centering
		\includegraphics[width=.75\linewidth]{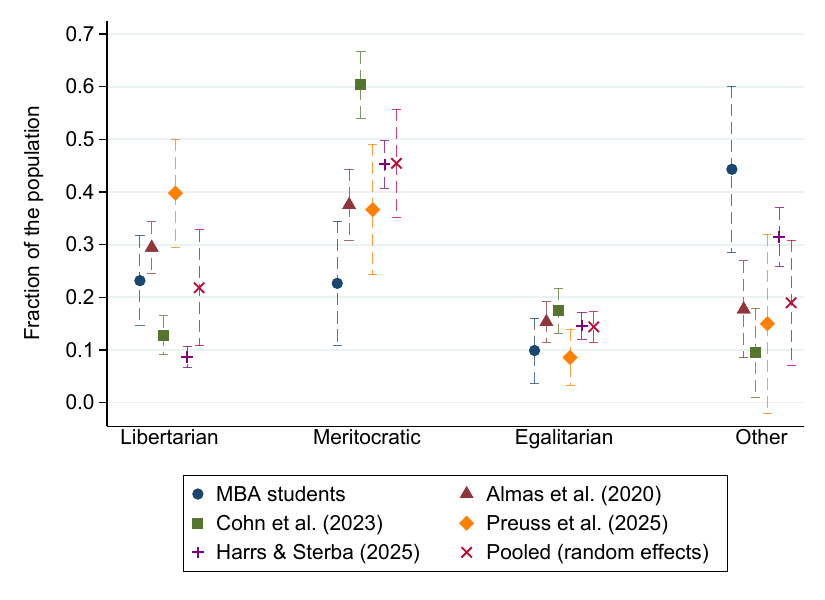}
		
		\hfill						
		{\footnotesize
			\singlespacing \justify
			
			\textit{Notes:} This figure shows estimated fractions of egalitarians, libertarians, meritocrats, and unclassified (``other'') spectators across studies, based on a between-subject design using one redistributive choice per spectator. Egalitarians are spectators who equalize earnings when the winner is determined by performance. Libertarians are spectators who do not redistribute when earnings are randomly assigned. Meritocrats are identified as the difference between the share of spectators allocating more to the winner in the performance condition and the share allocating more to the winner in the luck condition (following \citealp{almas.etal2020}). Remaining spectators are classified as having ``other'' ideals. Vertical dashed lines represent 95 percent confidence intervals from robust standard errors clustered at the spectator level. \par
			
			We estimate the distribution of fairness ideals in the \cite{preuss2025inequality} data using only spectators' first decision in the ``lucky outcomes'' condition, in which the winner is determined by a coin flip with some probability and by performance otherwise. We keep only decisions in which the coin flip determined the outcome with probability one (equivalent to our ``luck'' condition) or performance did so with probability one (equivalent to our ``performance'' condition). For \cite{harrs2025fairness}, we use their U.S.\ Prolific sample, in which spectators make redistribution decisions in both luck and merit conditions, presented in randomized order. For the between-subject estimates shown here, we use only the condition shown first. The winner receives \$4 and the loser \$0; spectators can transfer \$0--\$4. We classify fairness ideals using the same definitions as in the other studies.
			
			We also report a pooled estimate across all representative samples using a DerSimonian--Laird random-effects meta-analysis \citep{dersimonian1986meta}, which accounts for within-study sampling error and between-study heterogeneity by weighting each estimate by $1/(\text{se}_i^2 + \hat{\tau}^2)$, where $\hat{\tau}^2$ is the estimated between-study variance.
			
		}

	\end{figure}

	\clearpage
	\begin{table}[H]{\footnotesize
			\begin{center}
				\caption{Implemented Gini Coefficient Across Environments} \label{tab:gini}
				\newcommand\w{1.65}
				\begin{tabular}{l@{}lR{\w cm}@{}L{0.5cm}R{\w cm}@{}L{0.5cm}R{\w cm}@{}L{0.5cm}R{\w cm}@{}L{0.5cm}}
					\midrule
					&& \multicolumn{8}{c}{Dependent variable: Gini coefficient} \\ \cmidrule{3-10}
					& \multicolumn{4}{c}{All redistributive decisions} && \multicolumn{4}{c}{First decision only} \\ 	\cmidrule{3-5} \cmidrule{7-10}				
					&& (1) && (2) && (3) && (4)  \\
					\midrule
					Performance condition&&0.240&\sym{***}&0.242&\sym{***}&0.191&\sym{***}&0.184&\sym{***}\\
&&(0.029&)&(0.035&)&(0.052&)&(0.054&)\\
Efficiency condition&&0.188&\sym{***}&0.187&\sym{***}&0.206&\sym{***}&0.211&\sym{***}\\
&&(0.024&)&(0.029&)&(0.057&)&(0.060&)\\
Constant (Mean Gini Luck)&&0.420&\sym{***}&0.420&\sym{***}&0.432&\sym{***}&0.425&\sym{***}\\
&&(0.024&)&(0.018&)&(0.040&)&(0.071&)\\
  \midrule
					Additional controls? 				&& No && Yes && No  && Yes \\	
					\(N\) (redistributive decisions)&&793&&793&&271&&271&\\

					\(N\) (individuals)&&271&&271&&271&&271&\\
 \midrule
					
				\end{tabular}
			\end{center}
			\begin{singlespace}  \vspace{-.5cm}
				\noindent \justify \textit{Notes:} This table displays estimates of $\beta_0$, $\beta_1$, and $\beta_2$ from equation \eqref{eq:gini}. The omitted treatment is the luck condition. The outcome is the Gini coefficient of the final earnings allocation in worker pair $p$ implemented by spectator $i$, $\text{Gini}_{ip}$. The Gini coefficient takes a value between zero and one, with zero denoting perfect equality (both workers have the same post-redistribution earnings) and one denoting perfect inequality (one worker holds all earnings). We calculate the Gini coefficient as: \par
				\begin{align*}
					\text{Gini}_{ip} = \frac{|\text{Income Worker 1} - \text{Income Worker 2}|}{\text{Income Worker 1} + \text{Income Worker 2}}.
				\end{align*}	
				
				The specifications in columns 2 and 4 include additional controls. In column 2, we include spectator fixed effects. In column 4, we control for gender, age bins, parental financial situation while growing up, and a dummy for being born in the U.S. Heteroskedasticity-robust standard errors clustered at the spectator level in parentheses. {*} $p<0.10$, {*}{*} $p<0.05$, {*}{*}{*} $p<0.01$.
			\end{singlespace} 	
		}
	\end{table}

	\begin{table}[H]{\footnotesize
		\begin{center}
			\caption{Implemented Gini Coefficient: Elites vs.\ Benchmark Studies} \label{tab:comparison}
			\newcommand\w{1.55}
			\begin{tabular}{l@{}lR{\w cm}@{}L{0.45cm}R{\w cm}@{}L{0.45cm}R{\w cm}@{}L{0.45cm}R{\w cm}@{}L{0.45cm}R{\w cm}@{}L{0.45cm}R{\w cm}@{}L{0.45cm}}
				\midrule
				&& \multicolumn{12}{c}{Dependent variable: Gini coefficient} \\ \cmidrule{3-14}
				&& && \multicolumn{10}{c}{Representative U.S. samples} \\ \cmidrule{5-14}
				&& MBA students && \cite{almas.etal2020} && \cite{cohn.etal2023} && \cite{preuss2025inequality} && \cite{harrs2025fairness} && Pooled \\
				&& (1) && (2) && (3) && (4) && (5) && (6) \\
				\midrule
				\multicolumn{14}{l}{\hspace{-1em} \textbf{Panel A. General population samples}} \\\addlinespace
				Performance&&0.184&\sym{***}&0.195&\sym{***}&0.320&\sym{***}&0.396&\sym{***}&0.258&\sym{***}&0.279&\sym{***}\\
&&(0.054&)&(0.032&)&(0.040&)&(0.059&)&(0.017&)&(0.033&)\\
Efficiency&&0.211&\sym{***}&0.011&&&&&&&&0.011&\\
&&(0.060&)&(0.035&)&&&&&&&(0.035&)\\
Constant&&0.425&\sym{***}&0.363&\sym{***}&0.156&\sym{***}&0.463&\sym{***}&0.209&\sym{***}&0.292&\sym{***}\\
&&(0.071&)&(0.024&)&(0.029&)&(0.050&)&(0.012&)&(0.055&)\\
  \midrule
				\(N\)&&271&&1,000&&614&&197&&1,476&&3,287&\\
 \midrule
				\multicolumn{14}{l}{\hspace{-1em} \textbf{Panel B. Higher-income subsamples}} \\\addlinespace
				Performance&&0.184&\sym{***}&0.271&\sym{***}&0.253&\sym{***}&0.467&\sym{***}&0.244&\sym{***}&0.259&\sym{***}\\
&&(0.054&)&(0.065&)&(0.045&)&(0.124&)&(0.033&)&(0.025&)\\
Efficiency&&0.211&\sym{***}&0.174&\sym{**}&&&&&&&0.174&\sym{**}\\
&&(0.060&)&(0.077&)&&&&&&&(0.077&)\\
Constant&&0.425&\sym{***}&0.316&\sym{***}&0.271&\sym{***}&0.456&\sym{***}&0.268&\sym{***}&0.281&\sym{***}\\
&&(0.071&)&(0.048&)&(0.033&)&(0.118&)&(0.024&)&(0.019&)\\
  \midrule
				\(N\)&&271&&213&&316&&49&&398&&976&\\
 \midrule
				
			\end{tabular}
		\end{center}
		\begin{singlespace}  \vspace{-.5cm}
			\noindent \justify \textit{Notes:} This table displays estimates of equation \eqref{eq:gini} across studies. The dependent variable is the Gini coefficient of the final earnings allocation. The omitted treatment is the luck condition, so the constant estimates the average Gini when earnings are randomly assigned. ``Performance condition'' and ``Efficiency condition'' display the treatment effects of switching from the luck condition to the performance and efficiency conditions, respectively. Column 1 reports estimates from our MBA sample using spectators' first decisions and controlling for gender, age bins, parental financial situation while growing up, and a dummy for being born in the U.S. Columns 2--5 report estimates from representative U.S.\ samples in the indicated studies, without additional controls. Column 6 reports pooled estimates from a DerSimonian--Laird random-effects meta-analysis across the four benchmark studies \citep{dersimonian1986meta}. The efficiency-cost condition is available in our experiment and in \cite{almas.etal2020}, but not in the other studies; the pooled efficiency estimate therefore reflects a single study. For \cite{preuss2025inequality}, we compute the Gini directly from each spectator's allocation between the two workers, using the first decision per respondent. For \cite{cohn.etal2023}, we construct the Gini coefficient as $1 - 2 \times (\text{share redistributed})$; observations are weighted to be representative of the U.S.\ population. For \cite{harrs2025fairness}, we construct the Gini as $|1 - \text{transfer}/2|$, where the transfer ranges from \$0 to \$4; we use each respondent's first decision only. Panel A uses the full sample from each study. Panel B restricts to higher-income subsamples: individuals with household income $\ge$\$100,000 in \cite{almas.etal2020}, the SCE, and \cite{harrs2025fairness}, and the top 5\% by income or wealth in \cite{cohn.etal2023}. The MBA column is identical across panels. Heteroskedasticity-robust standard errors in parentheses; pooled standard errors account for between-study heterogeneity. {*} $p<0.10$, {*}{*} $p<0.05$, {*}{*}{*} $p<0.01$. \par
		\end{singlespace}
	}
\end{table}

	\begin{table}[H]\caption{Distribution of Fairness Ideals: Elites vs. Average Citizens} \label{tab:ideals}
		{\footnotesize
			\begin{center} 
				\protect
				\begin{tabular}{lcccc}
					\midrule 
					& \multicolumn{4}{c}{Percentage of...} \\ \cmidrule{2-5}
					& Libertarians & Meritocrats & Egalitarians & Other ideals  \\
					& (1) & (2) & (3) & (4) \\\hline \addlinespace
					
					\multicolumn{5}{l}{\hspace{-1em} \textbf{Panel A. MBA students}} \\\addlinespace
					
					MBA students (spectators' first choices) &23.16 &22.65 &9.89 &44.30 \\
 &(4.35) &(5.99) &(3.15) &(8.05) \\
 
					MBA students (repeated-measures design) &15.33 &28.74 &6.51 &49.43 \\
 &(1.58) &(1.98) &(1.08) &(2.76) \\
 					
					\midrule
					\multicolumn{5}{l}{\hspace{-1em} \textbf{Panel B. Average American}} \\\addlinespace

					SCE -- Average American &39.77 &36.67 &8.57 &14.99 \\
 &(5.25) &(6.31) &(2.75) &(8.66) \\

					\cite{almas.etal2020} -- Average American &29.43 &37.54 &15.32 &17.72 \\
 &(2.50) &(3.43) &(1.98) &(4.69) \\

					\cite{cohn.etal2023} -- Average American &12.80 &60.33 &17.45 &9.42 \\
 &(1.89) &(3.22) &(2.19) &(4.33) \\

					\cite{harrs2025fairness} -- Average American &8.68 &45.24 &14.61 &31.46 \\
 &(1.04) &(2.30) &(1.30) &(2.84) \\

					%

					\midrule
					\multicolumn{5}{l}{\hspace{-1em} \textbf{Panel C. Higher-income samples}} \\\addlinespace
					SCE -- USA income over 100k &44.44 &40.32 &3.23 &12.01 \\
 &(12.05) &(13.27) &(3.23) &(18.22) \\

					\cite{almas.etal2020} -- USA income over 100k &22.67 &41.92 &14.08 &21.33 \\
 &(4.87) &(7.11) &(4.16) &(9.57) \\

					\cite{cohn.etal2023} -- USA top 5 percent &25.15 &59.14 &8.97 &6.75 \\
 &(3.33) &(4.37) &(2.38) &(5.99) \\

					\cite{harrs2025fairness} -- USA income over 100k &8.96 &41.44 &7.61 &41.99 \\
 &(2.02) &(4.42) &(1.89) &(5.21) \\

					%
					%

					\midrule
					
				\end{tabular}
			\end{center}
			\begin{singlespace}  \vspace{-.5cm}
				\justify \textit{Notes:} This table shows estimates of the fraction of egalitarians, libertarians, meritocrats, and unclassified (``other'') spectators in various studies. Estimates in Panels B and C are based on a between-subject design that uses one redistributive choice per spectator. See Appendix~\ref{app:framework} for the definitions and estimation of each fairness ideal. \par
				
				We estimate the distribution of fairness ideals in the Survey of Consumer Expectations (SCE), collected by \cite{preuss2025inequality}, using only data from the first decision of spectators in the ``lucky outcomes'' condition, in which the winner of a worker pair is determined by a coin flip with some probability and by performance with the remaining probability. We keep only decisions in which the coin flip determined the outcome with probability one (equivalent to our ``luck'' condition) or performance determined the outcome with probability one (equivalent to our ``performance'' condition).

				For \cite{harrs2025fairness}, we use their U.S.\ Prolific sample, in which spectators redistribute earnings in both a luck and a merit condition. Each respondent sees both conditions in randomized order; we use only the condition shown first. The winner receives \$4 and the loser \$0; spectators can transfer \$0--\$4 in \$0.10 increments. Income is reported in seven categories; we define ``high-income'' as categories 6 (\$100--150k) and 7 (\$150k+).

				Panel C shows estimates for higher-income samples. In \cite{almas.etal2020}, the SCE data, and \cite{harrs2025fairness}, these are individuals with annual household income above \$100,000. In \cite{cohn.etal2023}, these are households in the top 5 percent by income or wealth (i.e., annual household income above \$250,000 or gross liquid assets of at least \$1 million).
				
				
				
				
			\end{singlespace}
		}
	\end{table}	
	
	\clearpage
	\begin{singlespace}
		\bibliographystyle{apalike}
		\bibliography{add_elites}

\begin{thebibliography}{}

\bibitem[Acemoglu et~al., 2025]{acemoglu.etal2025}
Acemoglu, D., He, A., and {le Maire}, D. (2025).
\newblock Eclipse of rent-sharing: The effects of managers' business education
  on wages and the labor share in the {US} and {Denmark}.
\newblock {\em American Economic Review (conditionally accepted)}.

\bibitem[Alesina and Angeletos, 2005]{alesina.angeletos2005a}
Alesina, A. and Angeletos, G.-M. (2005).
\newblock Fairness and redistribution.
\newblock {\em American Economic Review}, 95(4):960--980.

\bibitem[Alm{\aa}s et~al., 2011]{almas.etal2011}
Alm{\aa}s, I., Cappelen, A.~W., Lind, J.~T., S{\o}rensen, E.~{\O}., and
  Tungodden, B. (2011).
\newblock Measuring unfair (in)equality.
\newblock {\em Journal of Public Economics}, 95(7-8):488--499.

\bibitem[Alm{\aa}s et~al., 2010]{almas.etal2010}
Alm{\aa}s, I., Cappelen, A.~W., Sorensen, E.~O., and Tungodden, B. (2010).
\newblock Fairness and the development of inequality acceptance.
\newblock {\em Science}, 328(5982):1176--1178.

\bibitem[Alm{\aa}s et~al., 2025]{almas2025fairness}
Alm{\aa}s, I., Cappelen, A.~W., S{\o}rensen, E.~{\O}., and Tungodden, B.
  (2025).
\newblock Fairness across the world.
\newblock Working Paper 06/2025, NHH Norwegian School of Economics, Department
  of Economics.

\bibitem[Alm{\aa}s et~al., 2020]{almas.etal2020}
Alm{\aa}s, I., Cappelen, A.~W., and Tungodden, B. (2020).
\newblock Cutthroat capitalism versus cuddly socialism: Are {Americans} more
  meritocratic and efficiency-seeking than {Scandinavians}?
\newblock {\em Journal of Political Economy}, 128(5):1753--1788.

\bibitem[Andre, 2025]{andre2025}
Andre, P. (2025).
\newblock Shallow meritocracy.
\newblock {\em Review of Economic Studies}, 92(2):772--807.

\bibitem[Bartels, 2016]{bartels2016unequal}
Bartels, L.~M. (2016).
\newblock {\em Unequal Democracy: The Political Economy of the New Gilded Age}.
\newblock Princeton University Press, Princeton, NJ, 2 edition.

\bibitem[Bauman and Rose, 2011]{bauman2011selection}
Bauman, Y. and Rose, E. (2011).
\newblock Selection or indoctrination: Why do economics students donate less
  than the rest?
\newblock {\em Journal of Economic Behavior \& Organization}, 79(3):318--327.

\bibitem[Beard, 2012]{beard2012economic}
Beard, C.~A. (2012).
\newblock {\em An Economic Interpretation of the Constitution of the United
  States}.
\newblock Simon and Schuster, New York.

\bibitem[Bellavance et~al., 2009]{bellavance.etal2009}
Bellavance, F., Dionne, G., and Lebeau, M. (2009).
\newblock The value of a statistical life: A meta-analysis with a mixed effects
  regression model.
\newblock {\em Journal of Health Economics}, 28(2):444--464.

\bibitem[B{\'e}nabou and Tirole, 2006]{benabou.tirole2006a}
B{\'e}nabou, R. and Tirole, J. (2006).
\newblock Belief in a just world and redistributive politics.
\newblock {\em The Quarterly Journal of Economics}, 121(2):699--746.

\bibitem[Bhattacharya and Mollerstrom, 2025]{bhattacharyaandmollerstrom2025}
Bhattacharya, P. and Mollerstrom, J. (2025).
\newblock Lucky to work.
\newblock {\em Review of Economics and Statistics (forthcoming)}.

\bibitem[Broockman et~al., 2019]{broockman.etal2019}
Broockman, D.~E., Ferenstein, G., and Malhotra, N. (2019).
\newblock Predispositions and the political behavior of {American} economic
  elites: Evidence from technology entrepreneurs.
\newblock {\em American Journal of Political Science}, 63(1):212--233.

\bibitem[B{\"u}hlmann et~al., 2025]{buhlmann2025varieties}
B{\"u}hlmann, F., Christesen, C.~A., Cousin, B., Denord, F., Ellersgaard,
  C.~H., Lagneau-Ymonet, P., Larsen, A.~G., Savage, M., Thine, S., Young, K.,
  et~al. (2025).
\newblock Varieties of economic elites? preliminary results from the {World
  Elite Database (WED)}.
\newblock {\em The British Journal of Sociology}, 76(3):663--673.

\bibitem[Burch, 1980]{burch198081elites}
Burch, P.~H. (1980).
\newblock {\em Elites in American History}.
\newblock Holmes and Meier, New York.

\bibitem[Caldwell and Harmon, 2019]{caldwell2019outside}
Caldwell, S. and Harmon, N. (2019).
\newblock Outside options, bargaining and wages: Evidence from coworker
  networks.
\newblock {\em AEA Papers and Proceedings}, 109:203--207.

\bibitem[Capozza and Srinivasan, 2024]{capozza2024should}
Capozza, F. and Srinivasan, K. (2024).
\newblock Who should get money? estimating welfare weights in the {US}.
\newblock URPP Equality of Opportunity Discussion Paper Series~50, University
  of Zurich.

\bibitem[Cappelen et~al., 2007]{cappelen.etal2007}
Cappelen, A.~W., Hole, A.~D., S{\o}rensen, E.~{\O}., and Tungodden, B. (2007).
\newblock The pluralism of fairness ideals: An experimental approach.
\newblock {\em American Economic Review}, 97(3):818--827.

\bibitem[Cappelen et~al., 2013]{cappelen.etal2013}
Cappelen, A.~W., Konow, J., S{\o}rensen, E.~{\O}., and Tungodden, B. (2013).
\newblock Just luck: An experimental study of risk-taking and fairness.
\newblock {\em American Economic Review}, 103(4):1398--1413.

\bibitem[Cappelen et~al., 2022]{cappelen.etal2022}
Cappelen, A.~W., Mollerstrom, J., Reme, B.-A., and Tungodden, B. (2022).
\newblock A meritocratic origin of egalitarian behaviour.
\newblock {\em The Economic Journal}, 132(646):2101--2117.

\bibitem[Cappelen et~al., 2010]{cappelen.etal2010}
Cappelen, A.~W., S{\o}rensen, E.~{\O}., and Tungodden, B. (2010).
\newblock Responsibility for what? fairness and individual responsibility.
\newblock {\em European Economic Review}, 54(3):429--441.

\bibitem[Card et~al., 2018]{card.etal2018}
Card, D., Cardoso, A.~R., Heining, J., and Kline, P. (2018).
\newblock Firms and labor market inequality: Evidence and some theory.
\newblock {\em Journal of Labor Economics}, 36(S1):S13--S70.

\bibitem[Chancel et~al., 2022]{chancel2022world}
Chancel, L., Piketty, T., Saez, E., and Zucman, G. (2022).
\newblock {\em World Inequality Report 2022}.
\newblock Harvard University Press.

\bibitem[Chetty et~al., 2026]{chetty2026diversifying}
Chetty, R., Deming, D.~J., and Friedman, J.~N. (2026).
\newblock Diversifying society's leaders? the determinants and causal effects
  of admission to highly selective private colleges.
\newblock {\em Quarterly Journal of Economics}, 141(1):51--145.

\bibitem[Chetty et~al., 2020]{chetty.etal2020}
Chetty, R., Friedman, J.~N., Saez, E., Turner, N., and Yagan, D. (2020).
\newblock Income segregation and intergenerational mobility across colleges in
  the {United States}.
\newblock {\em The Quarterly Journal of Economics}, 135(3):1567--1633.

\bibitem[Cohn et~al., 2023]{cohn.etal2023}
Cohn, A., Jessen, L.~J., Kla{\v s}nja, M., and Smeets, P. (2023).
\newblock Wealthy {Americans} and redistribution: The role of fairness
  preferences.
\newblock {\em Journal of Public Economics}, 225:104977.

\bibitem[{Cornell University}, 2026]{johnson2026}
{Cornell University} (2026).
\newblock Notable alumni.
\newblock \url{https://www.johnson.cornell.edu/about/75th/alumni/}. Accessed:
  May 30, 2026.

\bibitem[Cullen et~al., 2025]{cullen2025}
Cullen, Z.~B., Li, S., and Perez-Truglia, R. (2025).
\newblock What's my employee worth? the effects of salary benchmarking.
\newblock {\em Review of Economic Studies}.

\bibitem[Dana et~al., 2007]{dana2007exploiting}
Dana, J., Weber, R.~A., and Kuang, J.~X. (2007).
\newblock Exploiting moral wiggle room: {E}xperiments demonstrating an illusory
  preference for fairness.
\newblock {\em Economic Theory}, 33(1):67--80.

\bibitem[Derenoncourt and Weil, 2025]{derenoncourt2025}
Derenoncourt, E. and Weil, D. (2025).
\newblock Voluntary minimum wages: The local labor market effects of national
  retailer policies.
\newblock {\em Quarterly Journal of Economics}, 140(3):1901--1958.

\bibitem[DerSimonian and Laird, 1986]{dersimonian1986meta}
DerSimonian, R. and Laird, N. (1986).
\newblock Meta-analysis in clinical trials.
\newblock {\em Controlled Clinical Trials}, 7(3):177--188.

\bibitem[Domhoff, 2013]{domhoff2013who}
Domhoff, G.~W. (2013).
\newblock {\em Who Rules America? The Triumph of the Corporate Rich}.
\newblock McGraw-Hill, New York, 7 edition.

\bibitem[Domhoff, 2017]{domhoff2017power}
Domhoff, G.~W. (2017).
\newblock {\em The Power Elite and the State: How Policy is Made in America}.
\newblock Routledge.

\bibitem[Dong et~al., 2025]{dong2025they}
Dong, L., Huang, L., and Lien, J.~W. (2025).
\newblock `they never had a chance': Unequal opportunities and fair
  redistributions.
\newblock {\em The Economic Journal}, 135(667):914--942.

\bibitem[Downs, 1957]{downs1957economic}
Downs, A. (1957).
\newblock {\em An Economic Theory of Democracy}.
\newblock Harper and Row, New York.

\bibitem[Dube et~al., 2025]{dube2025monopsony}
Dube, A., Manning, A., and Naidu, S. (2025).
\newblock Monopsony and employer misoptimization explain why wages bunch at
  round numbers.
\newblock {\em American Economic Review}, 115(8):2689--2721.

\bibitem[Durante et~al., 2014]{durante.etal2014}
Durante, R., Putterman, L., and {van der Weele}, J. (2014).
\newblock Preferences for redistribution and perception of fairness: An
  experimental study.
\newblock {\em Journal of the European Economic Association}, 12(4):1059--1086.

\bibitem[Els{\"a}sser et~al., 2021]{elsasser2021not}
Els{\"a}sser, L., Hense, S., and Sch{\"a}fer, A. (2021).
\newblock Not just money: Unequal responsiveness in egalitarian democracies.
\newblock {\em Journal of European Public Policy}, 28(12):1890--1908.

\bibitem[Exley, 2016]{exley2016excusing}
Exley, C.~L. (2016).
\newblock Excusing selfishness in charitable giving: The role of risk.
\newblock {\em The Review of Economic Studies}, 83(2):587--628.

\bibitem[Exley, 2020]{exley2020using}
Exley, C.~L. (2020).
\newblock Using charity performance metrics as an excuse not to give.
\newblock {\em Management Science}, 66(2):553--563.

\bibitem[Ferguson, 1995]{ferguson1995golden}
Ferguson, T. (1995).
\newblock {\em Golden Rule: The Investment Theory of Party Competition and the
  Logic of Money-Driven Political Systems}.
\newblock University of Chicago Press.

\bibitem[Fisman et~al., 2015]{fisman.etal2015a}
Fisman, R., Jakiela, P., Kariv, S., and Markovits, D. (2015).
\newblock The distributional preferences of an elite.
\newblock {\em Science}, 349(6254):aab0096.

\bibitem[Fisman et~al., 2023]{fisman2023distributional}
Fisman, R., Jakiela, P., Kariv, S., and Vannutelli, S. (2023).
\newblock The distributional preferences of {Americans}, 2013--2016.
\newblock {\em Experimental Economics}, 26(4):727--748.

\bibitem[Frank et~al., 2015]{frank2015performance}
Frank, D.~H., Wertenbroch, K., and Maddux, W.~W. (2015).
\newblock Performance pay or redistribution? cultural differences in just-world
  beliefs and preferences for wage inequality.
\newblock {\em Organizational Behavior and Human Decision Processes},
  130:160--170.

\bibitem[Gilens, 2012]{gilens2012}
Gilens, M. (2012).
\newblock {\em Affluence and Influence: Economic Inequality and Political Power
  in America}.
\newblock Princeton University Press.

\bibitem[Gilens and Page, 2014]{gilens.page2014}
Gilens, M. and Page, B.~I. (2014).
\newblock Testing theories of {American} politics: Elites, interest groups, and
  average citizens.
\newblock {\em Perspectives on Politics}, 12(3):564--581.

\bibitem[Hacker and Pierson, 2010]{hacker2010}
Hacker, J.~S. and Pierson, P. (2010).
\newblock {\em Winner-Take-All Politics: How Washington Made the Rich
  Richer---and Turned Its Back on the Middle Class}.
\newblock Simon \& Schuster, New York, NY.

\bibitem[Hall and Krueger, 2012]{hall2012evidence}
Hall, R.~E. and Krueger, A.~B. (2012).
\newblock Evidence on the incidence of wage posting, wage bargaining, and
  on-the-job search.
\newblock {\em American Economic Journal: Macroeconomics}, 4(4):56--67.

\bibitem[Harrs and Sterba, 2025]{harrs2025fairness}
Harrs, S. and Sterba, M.-B. (2025).
\newblock Fairness and support for redistribution: The role of preferences and
  beliefs.
\newblock Working Paper~47, Cluster of Excellence ``The Politics of
  Inequality'', University of Konstanz.

\bibitem[Harrs and Sterba, 2026]{harrs2026stable}
Harrs, S. and Sterba, M.-B. (2026).
\newblock How stable is meritocratic ideology in times of crises?
\newblock Work in progress.

\bibitem[Hazell et~al., 2026]{hazell2026national}
Hazell, J., Patterson, C., Sarsons, H., and Taska, B. (2026).
\newblock National wage setting.
\newblock {\em American Economic Review (forthcoming)}.

\bibitem[He and {le Maire}, 2025]{he.lemaire2025}
He, A.~X. and {le Maire}, D. (2025).
\newblock Managing inequality: Manager-specific wage premiums and selection in
  the managerial labor market.
\newblock {\em Review of Economics and Statistics (forthcoming)}.

\bibitem[Hertel-Fernandez, 2019]{hertel2019state}
Hertel-Fernandez, A. (2019).
\newblock {\em State Capture: How Conservative Activists, Big Businesses, and
  Wealthy Donors Reshaped the American States---and the Nation}.
\newblock Oxford University Press, New York.

\bibitem[Hjort et~al., 2026]{hjort2026across}
Hjort, J., Li, X., and Sarsons, H. (2026).
\newblock Across-country wage compression in multinationals.
\newblock {\em American Economic Review}, 116(1):52--87.

\bibitem[Lachowska et~al., 2022]{lachowska2022wage}
Lachowska, M., Mas, A., Saggio, R., and Woodbury, S.~A. (2022).
\newblock Wage posting or wage bargaining? a test using dual jobholders.
\newblock {\em Journal of Labor Economics}, 40(S1):S469--S493.

\bibitem[Levitt and List, 2007]{levitt.list2007}
Levitt, S.~D. and List, J.~A. (2007).
\newblock What do laboratory experiments measuring social preferences reveal
  about the real world?
\newblock {\em Journal of Economic Perspectives}, 21(2):153--174.

\bibitem[Mar{\'e}chal et~al., 2025]{marechal2025whose}
Mar{\'e}chal, M.~A., Cohn, A., Yusof, J., and Fisman, R. (2025).
\newblock Whose preferences matter for redistribution? cross-country evidence.
\newblock {\em Journal of Political Economy Microeconomics}, 3(1):1--24.

\bibitem[Mathisen, 2023]{mathisen2023}
Mathisen, R.~B. (2023).
\newblock Affluence and influence in a social democracy.
\newblock {\em American Political Science Review}, 117(2):751--758.

\bibitem[Meager, 2019]{meager2019}
Meager, R. (2019).
\newblock Understanding the average impact of microcredit expansions: A
  {Bayesian} hierarchical analysis of seven randomized experiments.
\newblock {\em American Economic Journal: Applied Economics}, 11(1):57--91.

\bibitem[Meager, 2022]{meager2022}
Meager, R. (2022).
\newblock Aggregating distributional treatment effects: A {Bayesian}
  hierarchical analysis of the microcredit literature.
\newblock {\em American Economic Review}, 112(6):1818--1847.

\bibitem[Mills, 1956]{mills1956power}
Mills, C.~W. (1956).
\newblock {\em The Power Elite}.
\newblock Oxford University Press, New York.

\bibitem[Mollerstrom et~al., 2015]{mollerstrom.etal2015}
Mollerstrom, J., Reme, B.-A., and S{\o}rensen, E.~{\O}. (2015).
\newblock Luck, choice and responsibility --- an experimental study of fairness
  views.
\newblock {\em Journal of Public Economics}, 131:33--40.

\bibitem[M{\"u}ller and Renes, 2021]{muller.renes2021}
M{\"u}ller, D. and Renes, S. (2021).
\newblock Fairness views and political preferences: Evidence from a large and
  heterogeneous sample.
\newblock {\em Social Choice and Welfare}, 56(4):679--711.

\bibitem[Nax et~al., 2021]{nax2021elites}
Nax, H.~H., Depoorter, B., Grech, P.~D., Halten, J., Hoeppner, S., Newton, J.,
  Pradelski, B. S.~R., Soos, A., Wehrli, S., Ikica, B., et~al. (2021).
\newblock Are elites really less fair-minded?
\newblock Legal Studies Research Paper Series 3766728, UC College of the Law
  San Francisco.

\bibitem[Norton and Ariely, 2011]{norton.ariely2011}
Norton, M.~I. and Ariely, D. (2011).
\newblock Building a better {America}---one wealth quintile at a time.
\newblock {\em Perspectives on Psychological Science}, 6(1):9--12.

\bibitem[Norton and Ariely, 2013]{norton.ariely2013}
Norton, M.~I. and Ariely, D. (2013).
\newblock American's desire for less wealth inequality does not depend on how
  you ask them.
\newblock {\em Judgment and Decision Making}, 8(3):393--394.

\bibitem[Page and Gilens, 2017]{page2017}
Page, B.~I. and Gilens, M. (2017).
\newblock {\em Democracy in America?: What Has Gone Wrong and What We Can Do
  About It}.
\newblock University of Chicago Press, Chicago, IL.

\bibitem[Page et~al., 2018]{page2018billionaires}
Page, B.~I., Seawright, J., and Lacombe, M.~J. (2018).
\newblock {\em Billionaires and Stealth Politics}.
\newblock University of Chicago Press, Chicago.

\bibitem[Preuss et~al., 2025]{preuss2025inequality}
Preuss, M., Reyes, G., Somerville, J., and Wu, J. (2025).
\newblock Inequality of opportunity and income redistribution.
\newblock {\em Journal of Political Economy Microeconomics}.

\bibitem[Reyes, 2025]{reyes2025}
Reyes, G. (2025).
\newblock Coarse wage-setting and behavioral firms.
\newblock {\em Review of Economics and Statistics (accepted)}.

\bibitem[Rigby and Wright, 2013]{rigby2013political}
Rigby, E. and Wright, G.~C. (2013).
\newblock Political parties and representation of the poor in the {American}
  states.
\newblock {\em American Journal of Political Science}, 57(3):552--565.

\bibitem[Saez and Zucman, 2022]{saez2022top}
Saez, E. and Zucman, G. (2022).
\newblock Top wealth in {America}: A reexamination.
\newblock Working Paper 30396, National Bureau of Economic Research.

\bibitem[Schakel, 2021]{schakel2021unequal}
Schakel, W. (2021).
\newblock Unequal policy responsiveness in the {Netherlands}.
\newblock {\em Socio-Economic Review}, 19(1):37--57.

\bibitem[Sundemo and L{\"o}fgren, 2025]{sundemo2025business}
Sundemo, M. and L{\"o}fgren, {\AA}. (2025).
\newblock Do business and economics studies erode prosocial values?
\newblock {\em Southern Economic Journal}, 92(2):504--526.

\bibitem[Vivalt, 2020]{vivalt2020}
Vivalt, E. (2020).
\newblock How much can we generalize from impact evaluations?
\newblock {\em Journal of the European Economic Association}, 18(6):3045--3089.

\bibitem[Wai et~al., 2024]{wai2024most}
Wai, J., Anderson, S.~M., Perina, K., Worrell, F.~C., and Chabris, C.~F.
  (2024).
\newblock The most successful and influential {Americans} come from a
  surprisingly narrow range of `elite' educational backgrounds.
\newblock {\em Humanities and Social Sciences Communications}, 11(1):1129.

\bibitem[Winters, 2011]{winters2011oligarchy}
Winters, J.~A. (2011).
\newblock {\em Oligarchy}.
\newblock Cambridge University Press, New York.

\bibitem[Winters and Page, 2009]{winters2009oligarchy}
Winters, J.~A. and Page, B.~I. (2009).
\newblock Oligarchy in the {United States}?
\newblock {\em Perspectives on Politics}, 7(4):731--751.

\bibitem[Yusof and Sartor, 2025]{yusof2025market}
Yusof, J. and Sartor, S. (2025).
\newblock Market luck: Skill-biased inequality and redistributive preferences.
\newblock Working Paper 475, University of Zurich, Department of Economics.

\end{thebibliography}

	\end{singlespace}

	\clearpage 
	\appendix
	\begin{center}
		\noindent {\LARGE \textbf{Appendix}}
	\end{center}

	\setcounter{table}{0}
	\setcounter{figure}{0}
	\setcounter{equation}{0}
	\renewcommand{\thetable}{A\arabic{table}}
	\renewcommand{\thefigure}{A\arabic{figure}}
	\renewcommand{\theequation}{A\arabic{equation}}

	\section{Appendix Figures and Tables}\label{app:figs}

	\begin{figure}[H]
		\caption{Flow of the Experiment}\label{fig:flow}
		\begin{center}
			\definecolor{workercolor}{RGB}{200,220,240}
			\definecolor{spectatorcolor}{RGB}{255,220,180}
			\definecolor{treatmentcolor}{RGB}{220,240,220}
			
			\tikzstyle{block} = [rectangle, draw, fill=workercolor, text width=16em, text centered, rounded corners, minimum height=3em, font=\small]
			\tikzstyle{treatmentblock} = [rectangle, draw, fill=treatmentcolor, text width=12em, text centered, rounded corners, minimum height=3em, font=\small]
			\tikzstyle{spectatorblock} = [rectangle, draw, fill=spectatorcolor, text width=16em, text centered, rounded corners, minimum height=3em, font=\small]
			\tikzstyle{subtext} = [font=\footnotesize\itshape]
			\tikzstyle{line} = [draw, -latex', thick]
			
			\begin{tikzpicture}[node distance = 1cm, auto, scale=0.6, transform shape]
				\node [block] (worker_recruitment) {Worker recruitment};
				\node [subtext, below=0.1cm of worker_recruitment] {Workers recruited from Prolific};
				
				\node [block, below=2.5cm of worker_recruitment, node distance=2.2cm] (production) {Production stage};
				\node [subtext, below=0.1cm of production] {Workers complete encryption task (5 minutes)};
				
				\node [block, below=2.5cm of production, node distance=2.2cm] (earnings) {Earnings stage};
				\node [subtext, below=0.1cm of earnings] {Workers paired, winner allocated \$6, loser \$0};

				\node [spectatorblock, below=3cm of earnings, node distance=2cm] (spectator) {Spectator recruitment};
				\node [subtext, below=0.1cm of spectator] {MBA students from Cornell University};
				
				\node [spectatorblock, below=2.5cm of spectator, node distance=2.2cm] (redistribution) {Redistribution stage};
				\node [subtext, below=0.1cm of redistribution] {Spectators make final earnings allocations};
				
				\node [treatmentblock, below left=2.5cm and 2cm of redistribution] (performance) {Performance treatment};
				\node [subtext, below=0.1cm of performance] {Winner determined by task performance};
				
				\node [treatmentblock, below=2.5cm of redistribution, node distance=4cm] (luck) {Luck treatment};
				\node [subtext, below=0.1cm of luck] {Winner determined by coin flip};
				
				\node [treatmentblock, below right=2.5cm and 2cm of redistribution] (efficiency) {Efficiency cost treatment};
				\node [subtext, below=0.1cm of efficiency] {Redistribution reduces total earnings};
				
				\node [spectatorblock, below=2.5cm of luck, node distance=2.5cm] (survey) {Exit survey};
				\node [subtext, below=0.1cm of survey] {Demographics, SES, career aspirations, social views};
				
				\path [line] (worker_recruitment) -- (production);
				\path [line] (production) -- (earnings);
				\path [line] (spectator) -- (redistribution);
				\path [line] (redistribution) -- (performance);
				\path [line] (redistribution) -- (luck);
				\path [line] (redistribution) -- (efficiency);
				\path [line] (performance) -- (survey);
				\path [line] (luck) -- (survey);
				\path [line] (efficiency) -- (survey);
				
				\node [draw, rectangle, fill=white, text width=8em, dashed, font=\footnotesize] at (-9,-17) {Randomized order of treatment conditions};
				\path [draw, ->, dashed] (-7,-17) -- (-4,-17);
				
				\node [draw, rectangle, fill=white, text width=12em, font=\small] at (7.5,-2) {
					\begin{tabular}{ll}
						\tikz\node [fill=workercolor, minimum size=1em] {}; & Worker stage \\[0.2cm]
						\tikz\node [fill=spectatorcolor, minimum size=1em] {}; & Spectator stage \\[0.2cm]
						\tikz\node [fill=treatmentcolor, minimum size=1em] {}; & Treatment conditions 
					\end{tabular}
				};
			\end{tikzpicture}
		\end{center}
		{\footnotesize \textit{Note:} This figure illustrates the three-stage experimental design. In the production stage, workers complete an encryption task. In the earnings stage, workers are paired and initial earnings are assigned based on either performance or luck. In the redistribution stage, MBA student spectators determine final earnings allocations across three treatment conditions presented in randomized order.}
	\end{figure}

	\newcommand{\rectangle}{\fboxsep0pt\fbox{\rule{3.5em}{0pt}\rule{0pt}{2ex}}} 
	\begin{figure}[H]
		\caption{Example of the Worker Encryption Task}\label{fig:encryptiontask}
		\begin{center}
			\footnotesize\sffamily
			\begin{tabular}{|c|c|c|c|c|c|c|c|c|c|c|c|c|}
				\hline
				\textbf{Q} & \textbf{X} & \textbf{D} & \textbf{A} & \textbf{C} & \textbf{V} & \textbf{U} & \textbf{R} & \textbf{P} & \textbf{W} & \textbf{L} & \textbf{Y} & \textbf{G} \\
				\hline
				754	& 579 & 860	& 708 & 344	& 725 & 950	& 314 & 532	& 595 & 654 & 838 & 327\\
				\hline
				\textbf{Z} & \textbf{F} & \textbf{M} & \textbf{N} & \textbf{T} & \textbf{B} & \textbf{K} & \textbf{O} & \textbf{H} & \textbf{S} & \textbf{E} & \textbf{I} & \textbf{J} \\
				\hline
				190	& 776 & 627	& 980 & 830 & 803 & 603	& 673 & 536 & 490 & 545 & 445 & 925\\
				\hline
			\end{tabular}
			
			\begin{tabular}{cccc}
				&&&\\	
				\multicolumn{4}{c}{Please translate the following word into code:}\\
				&&&\\[-1ex]
				\textbf{RPZ}: & \large{$\rectangle$} & \large{$\rectangle$} & \large{$\rectangle$}\\
			\end{tabular}
			
		\end{center}

		\singlespacing \justify \footnotesize \normalfont
		\textit{Notes:} This figure shows an example of an encryption completed by workers. For each three-letter ``word,'' workers receive a codebook that maps letters to three-digit numbers. After encrypting one word, a new word appears along with a new codebook. The words, codes, and the sequence of letters in the codebook are randomized for each new word. Feedback on the correctness of encryptions is not provided. Workers have five minutes to complete as many encryptions as possible. \par
		
	\end{figure}

	\clearpage
	\begin{figure}[H]\caption{Screenshots of Redistributive Decision Screens} \label{fig:screenshots}
		{\footnotesize
			\begin{centering}	
				\protect
				\begin{minipage}{.85\textwidth}
					\captionof*{figure}{Panel A. Luck condition}
					\fbox{\includegraphics[width=\linewidth]{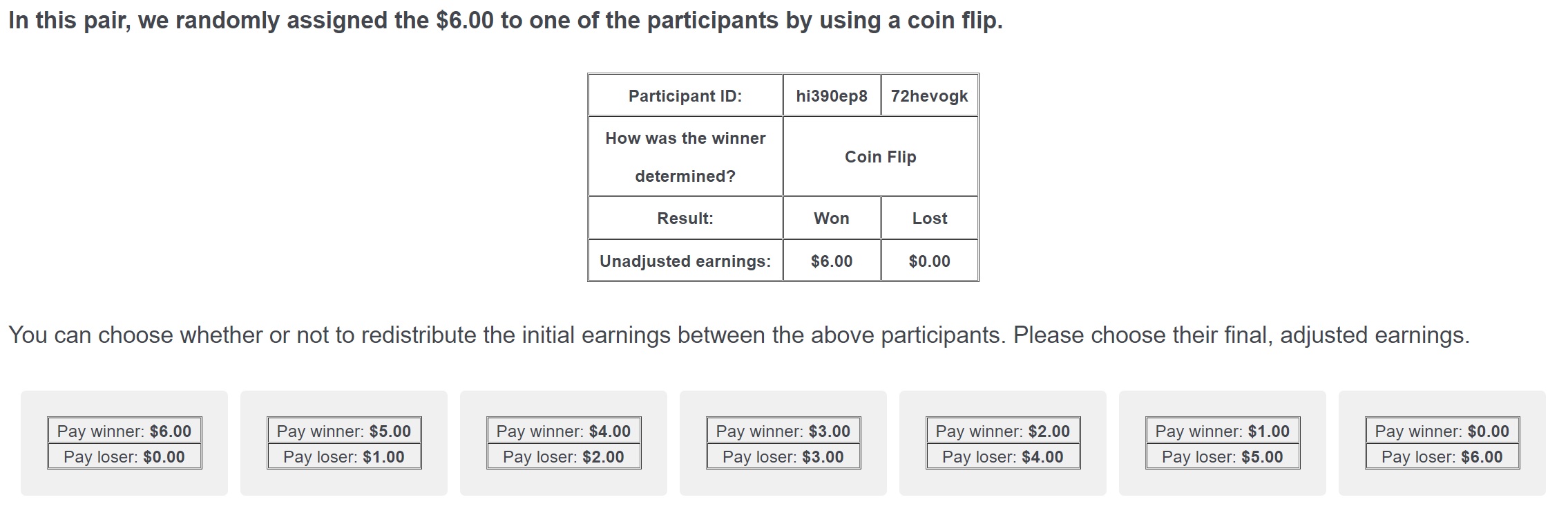}}\\
				\end{minipage} 
				\begin{minipage}{.85\textwidth}
					\captionof*{figure}{Panel B. Performance condition}				
					\fbox{\includegraphics[width=\linewidth]{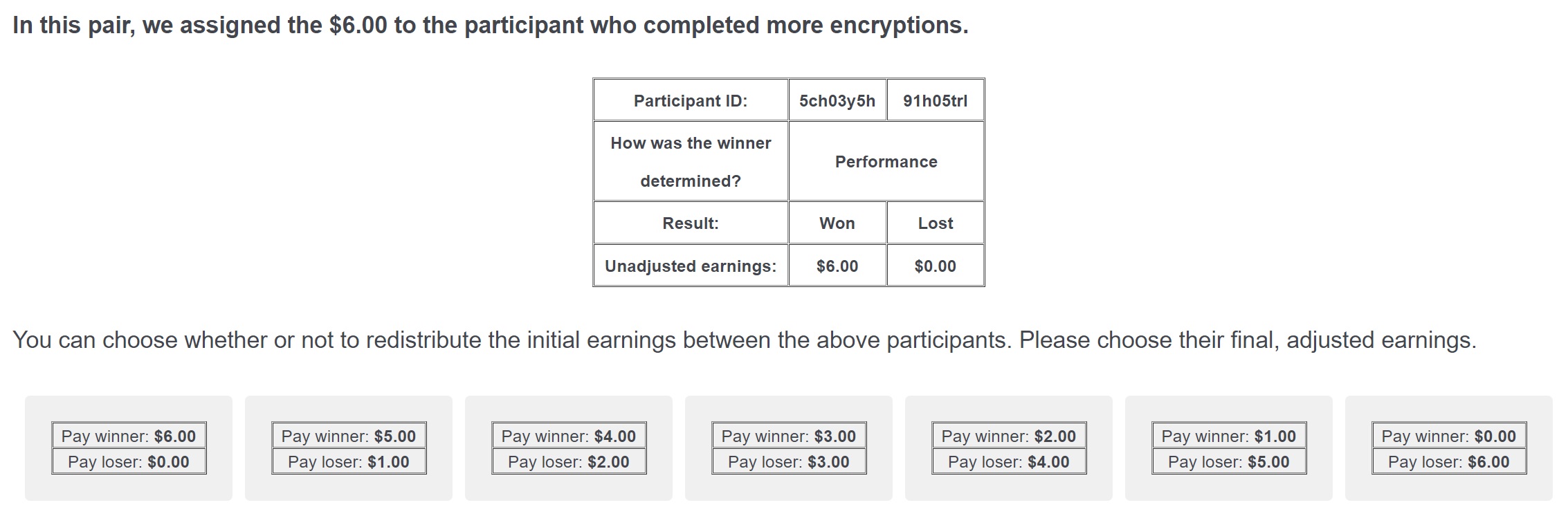}}\\
				\end{minipage}\hspace{1em}
				\begin{minipage}{.85\textwidth}
					\captionof*{figure}{Panel C. Efficiency-cost condition}				
					\fbox{\includegraphics[width=\linewidth]{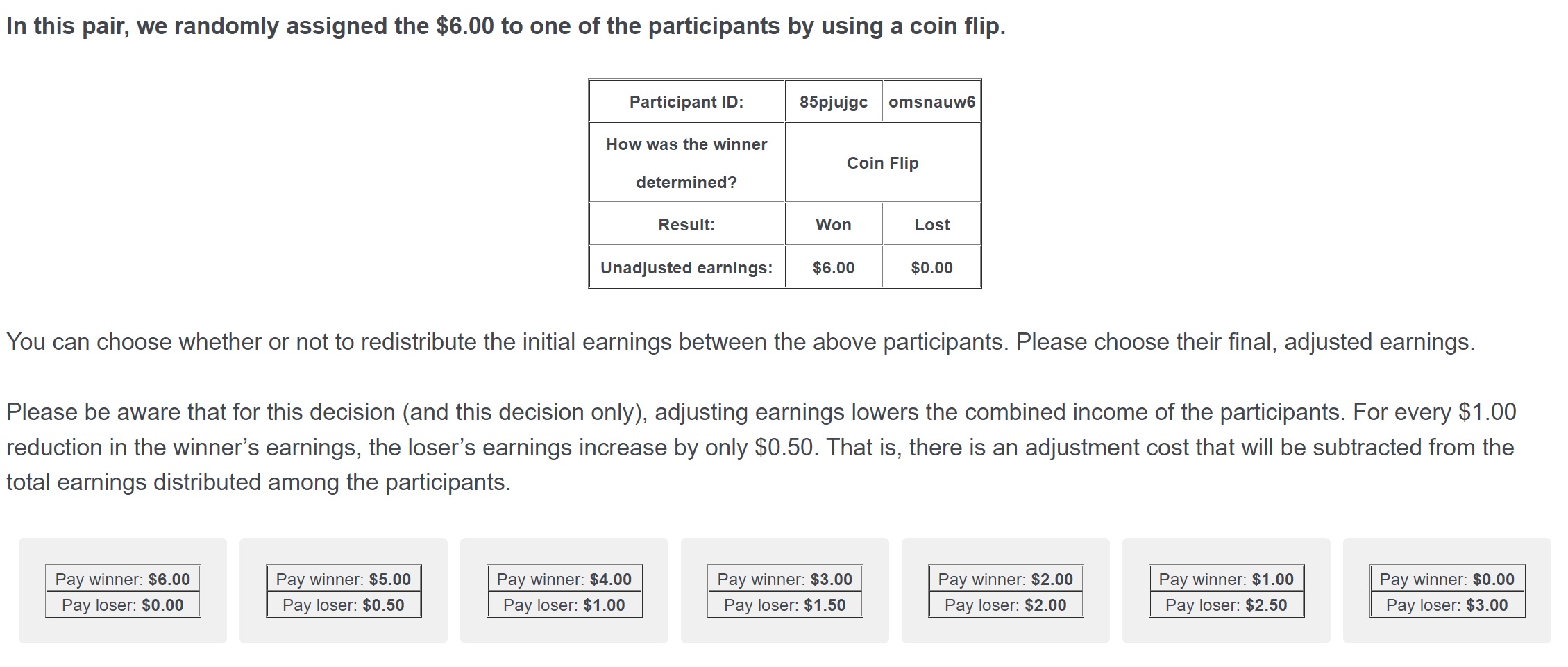}}
				\end{minipage}
				\par\end{centering}
			\singlespacing \justify \footnotesize
			\textit{Notes:} This figure shows screenshots of the redistribution screen shown to spectators for each worker pair condition. \par
		}
	\end{figure}

	\clearpage
	\begin{figure}[H]
		\caption{Fairness Ideals: Between- vs. Within-Subject Identification}\label{fig:ideals_btw_wth}
		\centering
		\includegraphics[width=.75\linewidth]{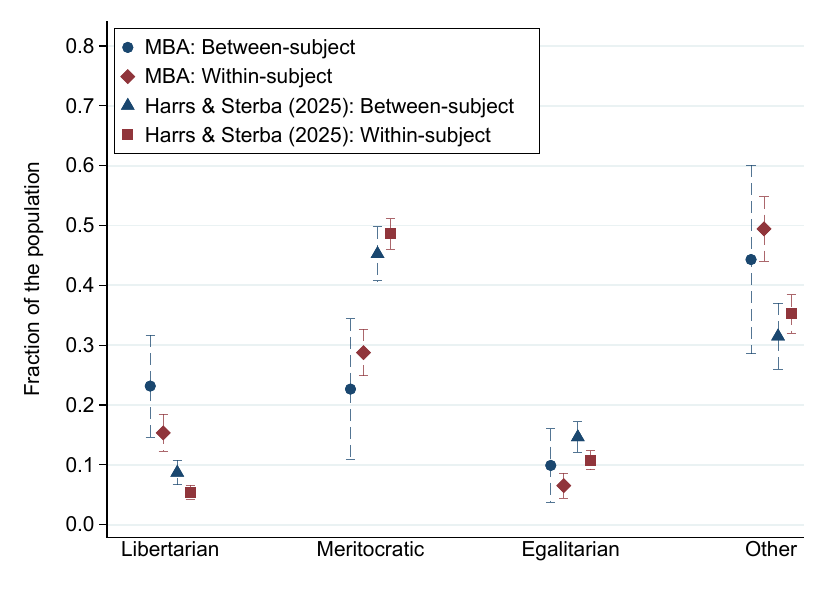}
		{\footnotesize
			\singlespacing \justify
			\textit{Notes:} This figure shows estimates of the fraction of libertarians, meritocrats, and egalitarians using two different research designs for MBA students and the \cite{harrs2025fairness} Prolific sample. See Section~\ref{app:framework} for the definitions and estimation of each fairness ideal. For MBA students, between-subject estimates (circles) use one redistributive choice per spectator and within-subject estimates (diamonds) use two choices. For \cite{harrs2025fairness}, between-subject estimates (triangles) use only the condition shown first, while within-subject estimates (squares) use both the luck and merit conditions per respondent. Vertical dashed lines represent 95 percent confidence intervals calculated using robust standard errors clustered at the spectator level. \par			
		}
	\end{figure}

	\begin{table}[H]\caption{Summary Statistics of the Sample by First Treatment Shown} \label{tab:summ_trt}
		{\footnotesize
			\begin{centering} 
				\protect
				\begin{tabular}{lccccc}
					\addlinespace \addlinespace \midrule			
					&     && \multicolumn{3}{c}{First condition shown}  \\ \cmidrule{4-6} 
					& All && Luck  & Performance & Efficiency \\
					& (1) && (2)   &    (3)      &     (4)     \\
					\midrule 	
					\multicolumn{5}{l}{\hspace{-1em} \textbf{Panel A. Demographic characteristics}} \\ 
					Age $\le$23 &0.031 & &0.043 &0.034 &0.013 \\
Age 24--27 &0.380 & &0.419 &0.315 &0.408 \\
Age 28--31 &0.453 & &0.387 &0.483 &0.500 \\
Age $\ge$32 &0.136 & &0.151 &0.169 &0.079 \\
Male &0.554 & &0.521 &0.589 &0.553 \\
Cohort 2023 &0.509 & &0.537 &0.451 &0.541 \\
Cohort 2024 &0.491 & &0.463 &0.549 &0.459 \\
Born in the USA &0.562 & &0.606 &0.545 &0.527 \\
 \midrule
					
					\multicolumn{5}{l}{\hspace{-1em} \textbf{Panel B. Financial situation while growing up}} \\ 
					Always had enough money for necessities and luxuries &0.198 & &0.191 &0.169 &0.240 \\
Always had enough for necessities and occasional luxuries &0.422 & &0.447 &0.438 &0.373 \\
Usually had enough for necessities, rarely for luxuries &0.279 & &0.277 &0.281 &0.280 \\
Sometimes did not have enough money for necessities &0.081 & &0.074 &0.090 &0.080 \\
Frequently did not have enough money for necessities &0.019 & &0.011 &0.022 &0.027 \\
 \midrule						
					
					\multicolumn{5}{l}{\hspace{-1em} \textbf{Panel C. Employment preferences}} \\ 
					Planning on getting a job in the USA &0.941 & &0.978 &0.899 &0.946 \\
Working long hours is sometimes necessary &0.811 & &0.806 &0.851 &0.770 \\
Work-life balance is more important than a high salary &0.524 & &0.505 &0.448 &0.635 \\
Wants career in private sector &0.839 & &0.806 &0.851 &0.865 \\
Wants career in a non-profit &0.173 & &0.204 &0.138 &0.176 \\
Wants to become an entrepreneur &0.484 & &0.430 &0.483 &0.554 \\
Aspire to be a manager &0.913 & &0.903 &0.920 &0.919 \\
 \midrule											
					
					\multicolumn{5}{l}{\hspace{-1em} \textbf{Panel D. Voting behavior and social views}} \\ 					
					Voted in the last elections &0.693 & &0.714 &0.701 &0.658 \\
Success is luck and connections &0.362 & &0.355 &0.414 &0.311 \\
Hard work brings a better life &0.815 & &0.828 &0.816 &0.797 \\
 \midrule		
					
					Passed attention check &0.933 & &0.968 &0.908 &0.919 \\
Minutes spent in survey &4.553 & &4.560 &4.640 &4.454 \\
Completed all three redistributive decisions &0.959 & &0.989 &0.989 &0.894 \\
Number of individuals &271 & &95 &91 &85 \\
Number of redistributive decisions &793 & &284 &271 &238 \\
 \midrule		
					
					F-statistic &-- & &1.042 &1.049 &1.034 \\
p-value F-statistic &-- & &0.410 &0.404 &0.417 \\
 \midrule \addlinespace \addlinespace

				\end{tabular}
				\par\end{centering}
			
			\begin{singlespace} \vspace{-.5cm}
				\noindent \justify  \textit{Notes:} This table shows summary statistics of MBA students in our sample. All variables are based on data self-reported by MBA students in the exit survey of our study. Employment preferences and social views are based on MBA students' agreement with several statements in a five-point Likert scale grid. For each statement, we define an indicator variable that equals one if the student selects ``strongly agree'' or ``agree,'' and zero otherwise. Column 1 reports overall summary statistics. Columns 2--4 split the sample by the first condition shown: luck (earnings randomly assigned), performance (earnings determined by performance on the encryption task), or efficiency cost (earnings randomly assigned with costly redistribution). \par
				
			\end{singlespace}	
			
		}
	\end{table}

	\begin{table}[H]{\footnotesize
			\begin{center}
				\caption{The Impact of the Order in Which Spectators Saw Each Condition} \label{tab:gini_order}
				\newcommand\w{2}
				\begin{tabular}{l@{}lR{\w cm}@{}L{0.5cm}R{\w cm}@{}L{0.5cm}R{\w cm}@{}L{0.5cm}R{\w cm}@{}L{0.5cm}}
					\midrule
					&& \multicolumn{8}{c}{Dependent variable: Gini coefficient} \\ \cmidrule{3-10}
					& \multicolumn{6}{c}{Separately by condition} &&  \\ 	\cmidrule{3-7} 		
					&& Luck && Performance && Efficiency && Pooled  \\
					&& (1) && (2) && (3) && (4)  \\
					\midrule
					Luck $\times$ Screen 2&&$-$0.014&&&&&&$-$0.014&\\
&&(0.059&)&&&&&(0.059&)\\
Luck $\times$ Screen 3&&$-$0.022&&&&&&$-$0.022&\\
&&(0.059&)&&&&&(0.060&)\\
Performance $\times$ Screen 2&&&&0.028&&&&0.028&\\
&&&&(0.050&)&&&(0.050&)\\
Performance $\times$ Screen 3&&&&0.087&\sym{*}&&&0.087&\sym{*}\\
&&&&(0.052&)&&&(0.052&)\\
Efficiency $\times$ Screen 2&&&&&&0.017&&0.017&\\
&&&&&&(0.057&)&(0.057&)\\
Efficiency $\times$ Screen 3&&&&&&$-$0.105&\sym{*}&$-$0.105&\sym{*}\\
&&&&&&(0.058&)&(0.058&)\\
Constant&&0.432&\sym{***}&0.623&\sym{***}&0.638&\sym{***}&0.565&\sym{***}\\
&&(0.040&)&(0.034&)&(0.041&)&(0.022&)\\
  \midrule
					\(N\) (individuals)&&261&&263&&269&&271&\\
 \midrule
					
				\end{tabular}
			\end{center}
			\begin{singlespace}  \vspace{-.5cm}
				\noindent \justify \textit{Notes:} This table displays estimates of order effects on the Gini coefficient. In columns 1--3, we regress the Gini on dummies for seeing a given condition on the second and third screens. The constant represents the average Gini coefficient for spectators who saw a given condition on the first screen. In column 4, we interact all conditions with the order dummies. Heteroskedasticity-robust standard errors clustered at the spectator level in parentheses. {*} $p<0.10$, {*}{*} $p<0.05$, {*}{*}{*} $p<0.01$. \par
			\end{singlespace} 	
		}
	\end{table}
	
	\clearpage
	\begin{table}[H]{\footnotesize
			\begin{center}
				\caption{Time Spent on Redistribution Screens by Spectators With Unclassified Ideals} \label{tab:other_time}
				\newcommand\w{1.95}
				\begin{tabular}{l@{}lR{\w cm}@{}L{0.5cm}R{\w cm}@{}L{0.5cm}R{\w cm}@{}L{0.5cm}R{\w cm}@{}L{0.5cm}}
					\midrule
					&& \multicolumn{8}{c}{Dependent variable: Time spent (seconds)} \\ \cmidrule{3-10}
					&&             && \multicolumn{6}{c}{By redistribution condition} \\ \cmidrule{5-10}
					&& All screens && Performance && Luck && Efficiency Cost \\  
					&& (1) && (2) && (3) && (4) \\  
					\midrule
					Other ideal&&0.382&&$-$1.280&&2.311&&$-$2.770&\\
&&(4.467&)&(1.983&)&(1.674&)&(2.974&)\\
Constant&&73.822&\sym{***}&21.266&\sym{***}&19.336&\sym{***}&35.338&\sym{***}\\
&&(3.489&)&(1.467&)&(1.225&)&(2.104&)\\
 \midrule
					\(N\) (individuals)&&260&&263&&261&&269&\\
 \midrule
				\end{tabular}
			\end{center}
			\begin{singlespace} \vspace{-.5cm}
				\noindent \justify \textit{Notes:} This table presents regression estimates of the relationship between having an unclassified fairness ideal and time spent on redistribution decision screens. The outcome is the total time spent across all redistribution screens. Standard errors clustered at the spectator level in parentheses. {*} $p<0.10$, {*}{*} $p<0.05$, {*}{*}{*} $p<0.01$. \par
			\end{singlespace}
		}
	\end{table}
	
	\clearpage
	\begin{table}[H]
		\begin{center}	
			\caption{Joint Distribution of Redistribution in Luck and Performance Conditions} \label{tab:joint}
			\begin{tabular}{ccccccccc}
				& & \multicolumn{7}{c}{Earnings Redistributed in Performance Condition} \\[0.5ex]
				\hline
				&   & \$0 & \$1 & \$2 & \$3 & \$4 & \$5 & \$6 \\
				\cline{2-9}
				\multirow{7}{*}{\begin{sideways}Luck Condition\end{sideways}}
				& \$0 & \cellcolor{green_lab!25}0.153 & 0.031 & 0.027 & 0.011 & 0.000 & 0.000 & 0.008 \\
& \$1 & 0.034 & 0.057 & 0.023 & 0.008 & 0.004 & 0.000 & 0.000 \\
& \$2 & 0.077 & 0.084 & 0.080 & 0.008 & 0.000 & 0.000 & 0.000 \\
& \$3 & \cellcolor{blue_lab!25}0.138 & \cellcolor{blue_lab!25}0.069 & \cellcolor{blue_lab!25}0.080 & \cellcolor{red_lab!25}0.065 & 0.000 & 0.000 & 0.004 \\
& \$4 & 0.000 & 0.004 & 0.015 & 0.004 & 0.000 & 0.000 & 0.000 \\
& \$5 & 0.000 & 0.000 & 0.000 & 0.000 & 0.000 & 0.000 & 0.000 \\
& \$6 & 0.000 & 0.000 & 0.000 & 0.004 & 0.004 & 0.004 & 0.004 \\

				\hline
			\end{tabular}
			\begin{tabular}{ll}
				\cellcolor{green_lab!25} & Libertarians \\
				\cellcolor{blue_lab!25} & Meritocrats \\
				\cellcolor{red_lab!25} & Egalitarians 
			\end{tabular}

		\end{center}
		
		\bigskip
		
		\begin{singlespace} {\footnotesize \vspace{-.5cm}
				\noindent \justify \textit{Notes:} This table shows the joint distribution of redistribution choices across luck and performance conditions, excluding the efficiency cost treatment. Each cell represents the proportion of spectators who chose to redistribute \$$x$ dollars in the luck condition (rows) and \$$y$ dollars in the performance condition (columns). Choices were elicited in \$0.50 increments and aggregated to \$1 bins for this table. For example, \getval{joint_mba_pct_00} percent of spectators chose to redistribute \$0 in both conditions, while \getval{joint_mba_pct_33} percent chose to redistribute \$3 (i.e., equalize earnings) in both conditions. \par
				
				The green shaded cell indicates spectators classified as libertarians (no redistribution in either condition), the blue shaded cells indicate spectators classified as meritocratic (earnings equalization in the luck condition and more redistribution in luck than performance), and the orange shaded cell indicates spectators classified as egalitarians (equal redistribution in both conditions). Spectators whose redistribution choices fall outside the shaded areas are classified as having ``other'' fairness ideals.
				
			}
		\end{singlespace} 	
		
	\end{table}

	\begin{table}[H]{\footnotesize
			\begin{center}
				\caption{Correlates of Fairness Ideals} \label{tab:ideals_correlates_bonf}
				\newcommand\w{1.8}
				\begin{tabular}{l@{}lR{\w cm}@{}L{0.45cm}R{\w cm}@{}L{0.45cm}R{\w cm}@{}L{0.45cm}R{\w cm}@{}L{0.45cm}R{\w cm}@{}L{0.45cm}}
					\midrule
					&& 	\multicolumn{10}{c}{Dependent Variable:}   \\\cmidrule{3-12} 
					&& Libertarian &&  Moderate  && Meritocrat  &&Egalitarian && Other  \\	
					&& (1)        && (2)          && (3)         && (4) && (5) \\	
					\midrule
					Male&&0.146&\sym{***}&$-$0.082&&$-$0.102&&$-$0.038&&0.075&\\
&&(0.042&)&(0.050&)&(0.057&)&(0.032&)&(0.057&)\\

					Aged 28 or over&&0.016&&0.031&&0.029&&$-$0.016&&$-$0.060&\\
&&(0.045&)&(0.050&)&(0.057&)&(0.032&)&(0.058&)\\

					Born USA&&0.064&&$-$0.027&&$-$0.026&&0.007&&$-$0.019&\\
&&(0.044&)&(0.051&)&(0.058&)&(0.031&)&(0.058&)\\

					Stay USA&&0.080&&0.002&&$-$0.260&&0.001&&0.178&\\
&&(0.069&)&(0.107&)&(0.133&)&(0.067&)&(0.093&)\\

					Always enough money&&0.030&&0.055&&$-$0.015&&$-$0.009&&$-$0.062&\\
&&(0.045&)&(0.050&)&(0.058&)&(0.032&)&(0.060&)\\

					Voted last elections&&0.007&&$-$0.044&&0.159&\sym{**}&0.041&&$-$0.163&\sym{*}\\
&&(0.048&)&(0.057&)&(0.056&)&(0.030&)&(0.065&)\\

					Success luck network&&0.015&&$-$0.109&&$-$0.048&&0.065&&0.076&\\
&&(0.048&)&(0.047&)&(0.059&)&(0.036&)&(0.061&)\\

					Hard work better life&&0.032&&0.075&&$-$0.034&&0.030&&$-$0.103&\\
&&(0.055&)&(0.056&)&(0.075&)&(0.035&)&(0.078&)\\

					Long hours necessary&&$-$0.042&&0.130&\sym{**}&$-$0.103&&0.031&&$-$0.016&\\
&&(0.062&)&(0.049&)&(0.077&)&(0.034&)&(0.074&)\\

					Work$-$life balance&&$-$0.038&&0.093&&$-$0.043&&0.033&&$-$0.044&\\
&&(0.046&)&(0.049&)&(0.057&)&(0.031&)&(0.058&)\\

					Career in priv. sector&&0.067&&$-$0.095&&0.086&&$-$0.037&&$-$0.021&\\
&&(0.053&)&(0.074&)&(0.072&)&(0.049&)&(0.079&)\\

					Career in a non$-$profit&&$-$0.021&&0.019&&$-$0.023&&0.056&&$-$0.032&\\
&&(0.058&)&(0.067&)&(0.074&)&(0.051&)&(0.075&)\\

					Wants entrepreneur&&0.049&&$-$0.020&&$-$0.029&&0.044&&$-$0.044&\\
&&(0.046&)&(0.049&)&(0.057&)&(0.032&)&(0.058&)\\

					Wants manager&&0.019&&0.008&&$-$0.079&&$-$0.026&&0.079&\\
&&(0.077&)&(0.087&)&(0.107&)&(0.064&)&(0.095&)\\

					\midrule
					Mean Dep. Var.&&0.181&&0.188&&0.277&&0.063&&0.292&\\
N (Spectators)&&271&&271&&271&&271&&271&\\

					\midrule
				\end{tabular}
			\end{center}
			\begin{singlespace}  \vspace{-.5cm}
				\noindent \justify \textit{Notes:} This table shows estimates of the correlates of having a given fairness ideal. The regression equation is: $$Y_{i} = \alpha + \gamma X_i + \nu_{i},$$ where $Y_i$ is a fairness ideal and $X_i$ is an individual-level characteristic. \par
				
				Each column shows the result for a different dependent variable. In column 1, the outcome equals one if the student does not redistribute earnings in either the luck or the performance environments. In column 2, the outcome equals one if the student is classified as a moderate (as defined in Section~\ref{sub:views_elites}). In column 3, the outcome equals one if the student equalizes earnings in the luck environment but gives strictly more to the winner in the performance environment. In column 4, the outcome equals one if the student divides earnings equally in the luck and performance environments. In column 5, the outcome equals one if the student has ``other'' fairness ideals but is not classified as a moderate.
				
				Heteroskedasticity-robust standard errors clustered at the spectator level in parentheses. Thresholds {*} $p<0.10$, {*}{*} $p<0.05$, {*}{*}{*} $p<0.01$ are Bonferroni corrected for five tests.
				
			\end{singlespace} 	
		}
	\end{table}

	\clearpage
	\setcounter{table}{0}
	\setcounter{figure}{0}
	\setcounter{equation}{0}	
	\renewcommand{\thetable}{B\arabic{table}}
	\renewcommand{\thefigure}{B\arabic{figure}}
	\renewcommand{\theequation}{B\arabic{equation}}
	
	\section{Empirical Appendix} \label{app:empirical}
	
	\subsection{Worker Sample} \label{app:worker}
	
	This appendix provides details on the characteristics of the worker sample.
	
	Appendix Table~\ref{tab:summ_worker} provides summary statistics on workers. The sample is diverse in terms of age, gender, and racial background. The average worker is approximately \getval{worker_age_mean} years old, with a standard deviation of \getval{worker_age_sd} years. About \getval{worker_pct_male} percent of workers are male, and \getval{worker_pct_born_usa} percent were born in the U.S. The majority identify as White (\getval{worker_pct_white} percent), followed by Black (\getval{worker_pct_black} percent), Asian (\getval{worker_pct_asian} percent), and individuals of mixed or other racial backgrounds.
		
\begin{figure}[H]\caption{Distribution of Tasks Completed in the Worker Task} \label{fig:hist_tasks}
		\centering
		\includegraphics[width=.75\linewidth]{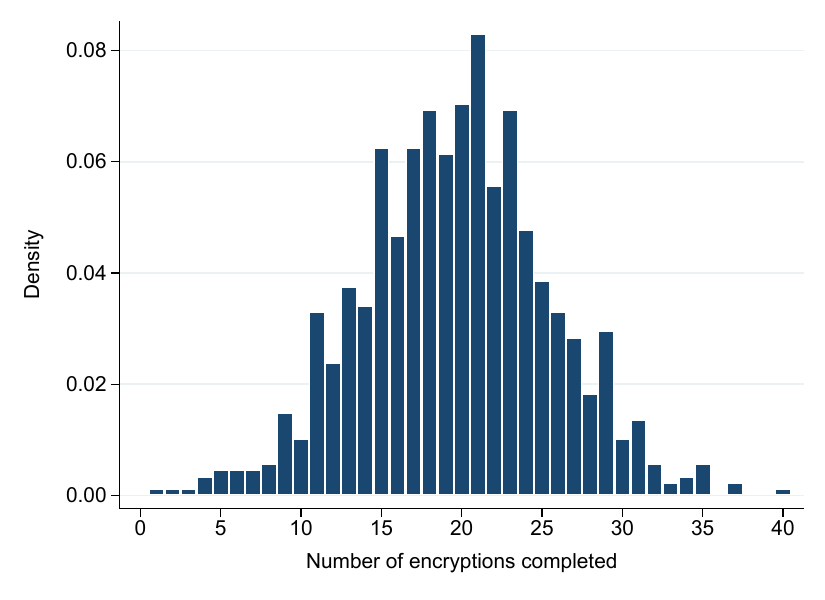}
		\singlespacing \justify \footnotesize
		\textit{Notes:} 
		This figure shows the distribution of the total number of correct three-letter encryptions completed by workers. The mean number of encryptions completed is \getval{worker_enc_mean_1dp} and the standard deviation is \getval{worker_enc_sd_1dp}.
	\end{figure}

	Appendix Figure~\ref{fig:hist_tasks} shows the distribution of encryption task performance. Workers completed an average of \getval{worker_enc_mean} encryptions, with a standard deviation of \getval{worker_enc_sd}. The average number of encryption attempts was \getval{worker_att_mean}. The distribution is approximately normal, with most workers completing between 10 and 30 encryptions.
	
	\begin{table}[H]\caption{Summary Statistics of the Worker Sample} \label{tab:summ_worker} \vspace{-.4cm}
		{\footnotesize
			\begin{centering} 
				\protect
				\begin{tabular}{lccc}
					\addlinespace \addlinespace \midrule			
					& Mean & SD  & N   \\
					& (1) & (2)  & (3)   \\
					\midrule 	
					
					\multicolumn{3}{l}{\hspace{-1em} \textbf{Panel A. Demographic characteristics}} \\ 
					Age &40.403 &13.829 &863 \\
Male &0.487 &0.500 &871 \\
White &0.697 &0.460 &855 \\
Black &0.116 &0.320 &855 \\
Asian &0.090 &0.286 &855 \\
Mixed race &0.065 &0.248 &855 \\
Other race &0.032 &0.175 &855 \\
Born in the USA &0.945 &0.229 &869 \\
Language is English &0.947 &0.223 &875 \\
 \midrule
					
					\multicolumn{3}{l}{\hspace{-1em} \textbf{Panel B. Employment status}} \\ 
					Is a student &0.136 &0.343 &589 \\
Works full-time &0.503 &0.501 &475 \\
Works part-time &0.162 &0.369 &475 \\
Unemployed &0.112 &0.315 &475 \\
 \midrule											
					
					\multicolumn{3}{l}{\hspace{-1em} \textbf{Panel C. Performance on the encryption task}} \\ 					
					Tasks attempted &20.919 &5.821 &880 \\
Tasks completed &19.670 &5.910 &880 \\
 \midrule		
					
				\end{tabular}
				\par\end{centering}
			
			\singlespacing\justify\footnotesize
		\textit{Notes:} This table shows summary statistics of our worker sample. Variables in Panels A and B are based on demographic and employment data provided by Prolific rather than collected through our survey. Sample sizes vary across variables as this information is only available for workers who have provided it to Prolific. Variables in Panel C are based on performance on the encryption task and are available for all workers. \par

		}
	\end{table}
	
	\clearpage

	\subsection{Descriptive Evidence} \label{app:descriptive}
	
	This appendix presents descriptive evidence on implemented inequality and redistribution choices across experimental conditions.
	
\begin{figure}[H]\caption{Histogram of the Gini Coefficient and Earnings Redistributed} \label{fig:gini}
		{\footnotesize
			\begin{centering}	
				\protect
				\begin{minipage}{.48\textwidth}
					\captionof*{figure}{Panel A. Gini coefficient}
					\includegraphics[width=\linewidth]{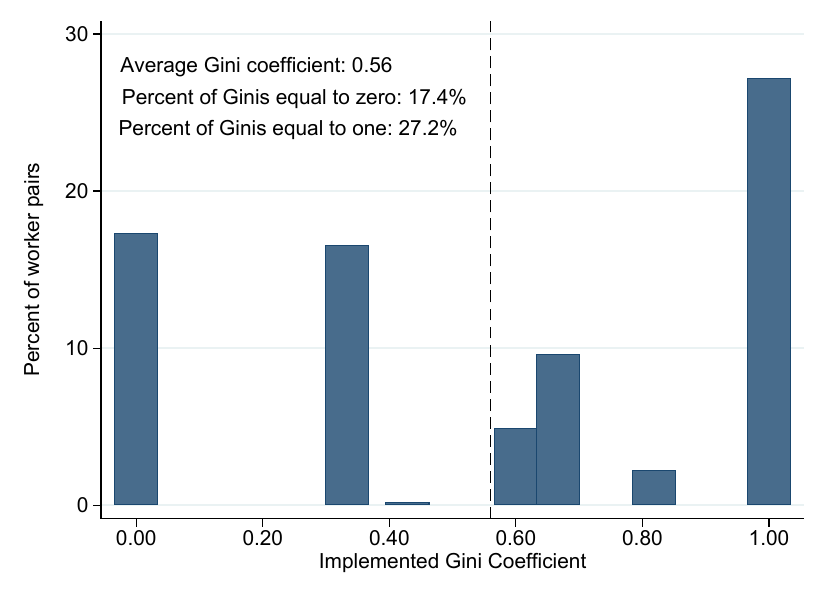}
				\end{minipage}\hspace{1em}
				\begin{minipage}{.48\textwidth}
					\captionof*{figure}{Panel B. Earnings redistributed}				
					\includegraphics[width=\linewidth]{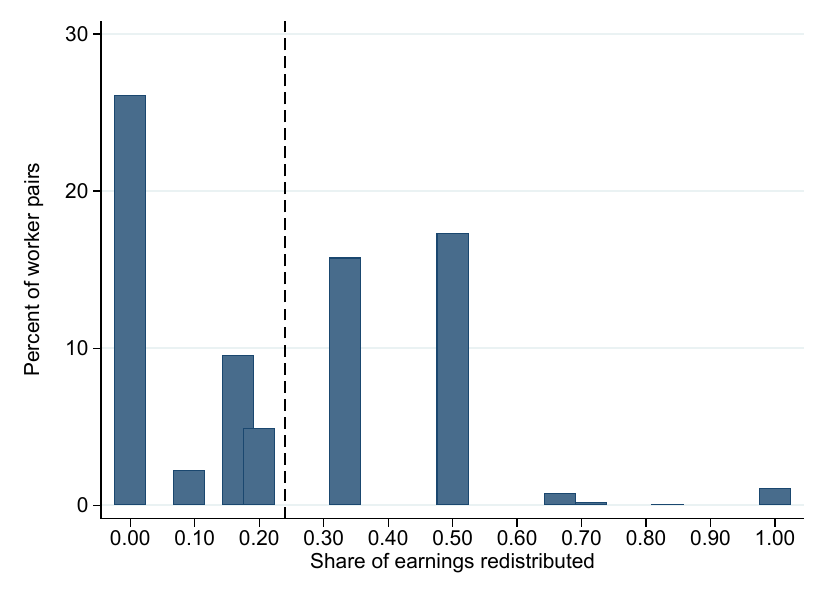}
				\end{minipage}
				\par\end{centering}
			\singlespacing \justify \footnotesize
			\textit{Notes:} This figure shows a histogram of the implemented Gini coefficients (Panel A) and share of earnings redistributed (Panel B). To produce this figure, we use data on all spectator redistributive decisions. \par
		}
	\end{figure}

	Appendix Figure~\ref{fig:gini} shows a histogram of the implemented Gini coefficients (Panel A) and share of earnings redistributed (Panel B). Implemented inequality varies substantially across spectators. On average across worker pairs, spectators implemented a Gini coefficient of $\getval{desc_gini_overall}$ and redistributed $\$\getval{desc_redist_dollars}$ of the winner's earnings to the loser. The two modal Gini coefficients are one (perfect inequality, $\getval{desc_pct_perfineq}$ percent of worker pairs) and zero (perfect equality, $\getval{desc_pct_perfeq}$ percent).
	
\begin{figure}[H]\caption{Average Gini and Earnings Redistributed by Condition} \label{fig:gini_cond}
		{\footnotesize
			\begin{centering}	
				\protect
				\begin{minipage}{.48\textwidth}
					\captionof*{figure}{Panel A. Gini coefficient}
					\includegraphics[width=\linewidth]{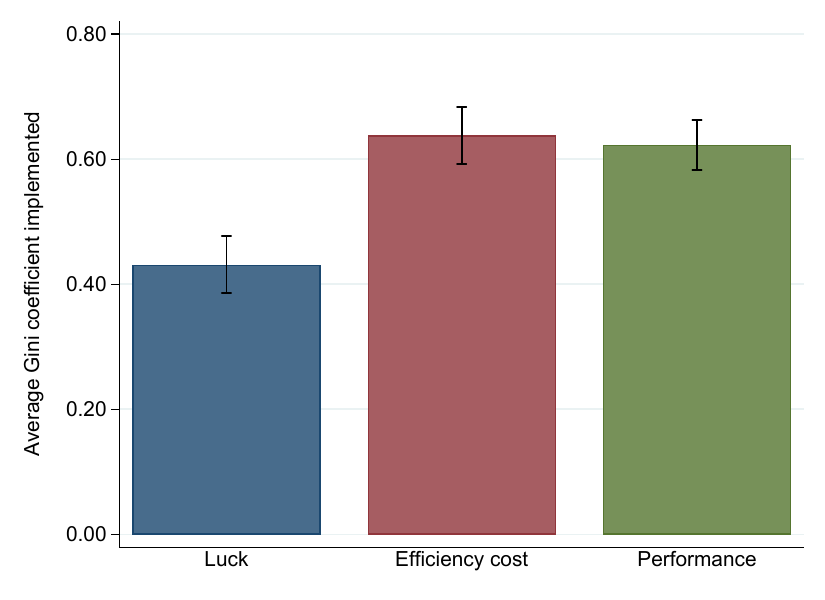}
				\end{minipage}\hspace{1em}
				\begin{minipage}{.48\textwidth}
					\captionof*{figure}{Panel B. Earnings redistributed}				
					\includegraphics[width=\linewidth]{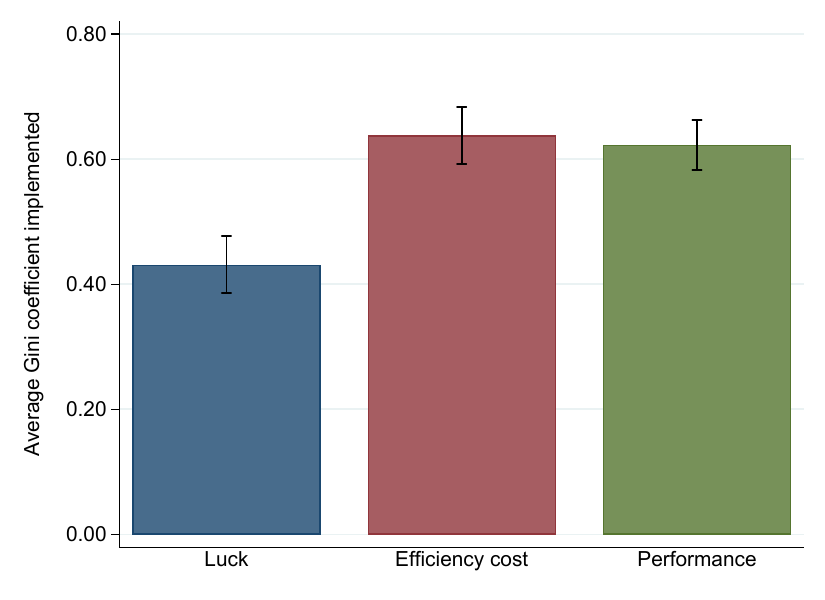}
				\end{minipage}
				\par\end{centering}
			\singlespacing \justify \footnotesize
			\textit{Notes:} This figure shows the average implemented Gini coefficients (Panel A) and share of earnings redistributed (Panel B) across each condition. To produce this figure, we only use the first redistributive decision of each spectator. \par
		}
	\end{figure}

	Appendix Figure~\ref{fig:gini_cond} presents the average Gini coefficient (Panel A) and the share of earnings redistributed (Panel B) across different experimental conditions. Panel A shows that implemented inequality varies significantly by condition. In the luck treatment, where earnings are randomly assigned, the average Gini coefficient is $\getval{gini_mean_luck}$. In contrast, inequality is substantially higher in the performance and efficiency cost conditions. The efficiency cost condition generates the highest inequality, with an average Gini coefficient of $\getval{gini_mean_eff}$. Panel B illustrates the fraction of earnings redistributed across conditions. Spectators redistribute the highest share of earnings in the luck condition, while redistribution decreases sharply in the performance and efficiency cost conditions.

	\subsection{Robustness Checks} \label{sec:robustness}

	This appendix presents several robustness checks of our main results. We examine the sensitivity of our findings to different sample restrictions and alternative outcome measures. All robustness checks follow the specification in equation \eqref{eq:gini} and present results using both within-subject and between-subject variation.
	
	First, we verify that our results are not driven by inattentive participants. Appendix Table~\ref{tab:rob_att} reproduces our main analysis excluding spectators who failed an embedded attention check in our survey. The attention check asked participants to select a specific response option within a matrix of survey questions. The estimates remain virtually unchanged after excluding inattentive respondents.
	
	Second, we examine whether our results are affected by participants who rushed through the experiment. Appendix Table~\ref{tab:rob_tim} presents estimates excluding spectators who completed the survey in less than \getval{rush_cutoff_min} minutes, corresponding to the bottom five percent of the completion time distribution. The magnitude and statistical significance of our main effects remain stable in this restricted sample.
	
	Third, we assess the robustness of our findings to potentially inconsistent response patterns. Appendix Table~\ref{tab:rob_los} shows results after excluding spectators who allocated strictly more earnings to the loser than to the winner in at least one worker pair---a pattern that could indicate confusion about the task. Our main conclusions about the source-of-inequality effect and efficiency concerns remain unchanged in this restricted sample.
	
	Fourth, we examine whether our results are driven by spectators whose redistributive choices do not conform to standard fairness ideals. Appendix Table~\ref{tab:rob_oth} presents estimates excluding spectators classified as having ``other'' fairness ideals---those who do not fit the egalitarian, libertarian, or meritocratic classifications. The performance condition continues to increase the Gini coefficient by \getval{rob_oth_perf} points in the between-subject specification without controls, and the efficiency-cost condition increases it by \getval{rob_oth_eff} points (both $p < 0.01$).
	
	Fifth, we examine whether our results are driven by spectators who may have misunderstood the efficiency cost treatment. Appendix Table~\ref{tab:rob_eff} presents estimates excluding spectators who redistributed more earnings when there was an efficiency cost than when there was no efficiency cost---a pattern inconsistent with standard economic theory. This restriction excludes \getval{eff_excl_n} spectators (\getval{eff_excl_pct} percent of our sample). Despite this exclusion, our main findings remain robust. The performance condition continues to increase the Gini coefficient by \getval{rob_eff_perf} points in the within-subject specification without controls (column 1), nearly identical to our baseline estimate of \getval{gini_b1}. The efficiency-cost condition increases the Gini by \getval{rob_eff_eff} points, larger than our baseline estimate of \getval{gini_b2}. The baseline inequality level increases slightly from \getval{gini_b0} to \getval{rob_eff_cons}.
	
	Sixth, we examine whether our results differ systematically across the two MBA cohorts in our sample. Appendix Table~\ref{tab:rob_coh} presents estimates from equation \eqref{eq:gini} estimated separately for MBA students from the 2023 and 2024 cohorts. The effect of the performance condition is stable across cohorts, ranging from \getval{rob_coh23_perf} to \getval{rob_coh24_perf} in the within-subject specifications without controls. Similarly, the efficiency-cost condition effects are consistent across cohorts, with estimates of \getval{rob_coh23_eff} and \getval{rob_coh24_eff} for the 2023 and 2024 cohorts, respectively. The baseline levels of inequality, captured by the constants, are also similar across cohorts (\getval{rob_coh23_cons} and \getval{rob_coh24_cons}).
	
	Seventh, we account for potential order effects by including round fixed effects in our estimation. Appendix Table~\ref{tab:rob_round} presents estimates from our main specification with additional controls for the order in which spectators encountered each treatment condition. The results remain consistent with our baseline findings. When controlling for round fixed effects, the performance condition increases the Gini coefficient by $\getval{rob_round_perf}$ points in the within-subject specification without additional controls, nearly identical to our main estimate. Similarly, the efficiency cost condition increases the Gini coefficient by $\getval{rob_round_eff}$ points, matching our baseline result. This consistency indicates that our findings are not driven by the sequential nature of the experimental design or by potential learning effects across redistributive decisions.
	
	Finally, we verify that our results are not sensitive to our choice of dependent variable. Appendix Table~\ref{tab:rob_red} presents estimates using the net-of-efficiency-cost share of earnings redistributed as the outcome variable instead of the Gini coefficient. This alternative specification directly measures redistribution behavior while accounting for efficiency costs. The results remain qualitatively similar, confirming robustness to alternative outcome measures.
	
	Across all these robustness checks, the key patterns in our data---MBA students' greater tolerance for inequality, weaker sensitivity to the source of inequality than the general population, and stronger response to efficiency costs---remain stable and statistically significant.

	\begin{table}[H]{\footnotesize
			\begin{center}
				\caption{Robustness Check of Main Effects to Excluding Inattentive Spectators} \label{tab:rob_att}
				\newcommand\w{1.65}
				\begin{tabular}{l@{}lR{\w cm}@{}L{0.5cm}R{\w cm}@{}L{0.5cm}R{\w cm}@{}L{0.5cm}R{\w cm}@{}L{0.5cm}}
					\midrule
					&& \multicolumn{8}{c}{Dependent variable: Gini coefficient} \\ \cmidrule{3-10}
					& \multicolumn{4}{c}{All redistributive decisions} && \multicolumn{4}{c}{First decision only} \\ 	\cmidrule{3-5} \cmidrule{7-10}				
					&& (1) && (2) && (3) && (4)  \\
					\midrule
					Performance condition&&0.242&\sym{***}&0.242&\sym{***}&0.186&\sym{***}&0.186&\sym{***}\\
&&(0.030&)&(0.036&)&(0.054&)&(0.057&)\\
Efficiency condition&&0.186&\sym{***}&0.186&\sym{***}&0.194&\sym{***}&0.202&\sym{***}\\
&&(0.025&)&(0.030&)&(0.062&)&(0.063&)\\
Constant&&0.415&\sym{***}&0.415&\sym{***}&0.430&\sym{***}&0.437&\sym{***}\\
&&(0.026&)&(0.019&)&(0.040&)&(0.073&)\\
  \midrule
					Additional controls? 				&& No && Yes && No  && Yes \\
					\(N\) (redistributive decisions)&&711&&711&&237&&237&\\

					\(N\) (individuals)&&237&&237&&237&&237&\\
 \midrule
					
				\end{tabular}
			\end{center}
			\begin{singlespace}  \vspace{-.5cm}
				\noindent \justify \textit{Notes:} This table is analogous to Table~\ref{tab:gini}, but excludes students who failed the attention check. See notes to Table~\ref{tab:gini} for details. {*} $p<0.10$, {*}{*} $p<0.05$, {*}{*}{*} $p<0.01$. \par
			\end{singlespace} 	
		}
	\end{table}

	\begin{table}[H]{\footnotesize
			\begin{center}
				\caption{Robustness Check of Main Effects to Excluding Spectators Who Rushed Through the Experiment} \label{tab:rob_tim}
				\newcommand\w{1.65}
				\begin{tabular}{l@{}lR{\w cm}@{}L{0.5cm}R{\w cm}@{}L{0.5cm}R{\w cm}@{}L{0.5cm}R{\w cm}@{}L{0.5cm}}
					\midrule
					&& \multicolumn{8}{c}{Dependent variable: Gini coefficient} \\ \cmidrule{3-10}
					& \multicolumn{4}{c}{All redistributive decisions} && \multicolumn{4}{c}{First decision only} \\ 	\cmidrule{3-5} \cmidrule{7-10}				
					&& (1) && (2) && (3) && (4)  \\
					\midrule
					Performance condition&&0.249&\sym{***}&0.248&\sym{***}&0.206&\sym{***}&0.202&\sym{***}\\
&&(0.029&)&(0.036&)&(0.053&)&(0.055&)\\
Efficiency condition&&0.181&\sym{***}&0.180&\sym{***}&0.186&\sym{***}&0.194&\sym{***}\\
&&(0.024&)&(0.030&)&(0.059&)&(0.061&)\\
Constant&&0.421&\sym{***}&0.421&\sym{***}&0.430&\sym{***}&0.430&\sym{***}\\
&&(0.025&)&(0.019&)&(0.040&)&(0.073&)\\
  \midrule
					Additional controls? 				&& No && Yes && No  && Yes \\
					\(N\) (redistributive decisions)&&762&&762&&257&&257&\\

					\(N\) (individuals)&&257&&257&&257&&257&\\
 \midrule
					
				\end{tabular}
			\end{center}
			\begin{singlespace}  \vspace{-.5cm}
				\noindent \justify \textit{Notes:} This table is analogous to Table~\ref{tab:gini}, but excludes students who completed the survey in less than \getval{rush_cutoff_min} minutes (corresponding to the bottom five percent of the total completion time distribution). See notes to Table~\ref{tab:gini} for details. {*} $p<0.10$, {*}{*} $p<0.05$, {*}{*}{*} $p<0.01$. \par
			\end{singlespace} 	
		}
	\end{table}

	\begin{table}[H]{\footnotesize
			\begin{center}
				\caption{Robustness Check of Main Effects to Excluding Spectators Who Allocated Strictly More Earnings to the Loser} \label{tab:rob_los}
				\newcommand\w{1.65}
				\begin{tabular}{l@{}lR{\w cm}@{}L{0.5cm}R{\w cm}@{}L{0.5cm}R{\w cm}@{}L{0.5cm}R{\w cm}@{}L{0.5cm}}
					\midrule
					&& \multicolumn{8}{c}{Dependent variable: Gini coefficient} \\ \cmidrule{3-10}
					& \multicolumn{4}{c}{All redistributive decisions} && \multicolumn{4}{c}{First decision only} \\ 	\cmidrule{3-5} \cmidrule{7-10}				
					&& (1) && (2) && (3) && (4)  \\
					\midrule
					Performance condition&&0.257&\sym{***}&0.259&\sym{***}&0.212&\sym{***}&0.214&\sym{***}\\
&&(0.029&)&(0.036&)&(0.053&)&(0.056&)\\
Efficiency condition&&0.186&\sym{***}&0.186&\sym{***}&0.221&\sym{***}&0.236&\sym{***}\\
&&(0.025&)&(0.030&)&(0.059&)&(0.062&)\\
Constant&&0.410&\sym{***}&0.409&\sym{***}&0.402&\sym{***}&0.423&\sym{***}\\
&&(0.025&)&(0.019&)&(0.040&)&(0.072&)\\
  \midrule
					Additional controls? 				&& No && Yes && No  && Yes \\
					\(N\) (redistributive decisions)&&744&&744&&254&&254&\\

					\(N\) (individuals)&&254&&254&&254&&254&\\
 \midrule
					
				\end{tabular}
			\end{center}
			\begin{singlespace}  \vspace{-.5cm}
				\noindent \justify \textit{Notes:} This table is analogous to Table~\ref{tab:gini}, but excludes students who allocated strictly more earnings to the loser than to the winner in at least one worker pair. See notes to Table~\ref{tab:gini} for details. {*} $p<0.10$, {*}{*} $p<0.05$, {*}{*}{*} $p<0.01$. \par
			\end{singlespace} 	
		}
	\end{table}

	\begin{table}[H]{\footnotesize
			\begin{center}
				\caption{Robustness Check of Main Effects to Excluding Spectators with Non-Standard Fairness Ideals} \label{tab:rob_oth}
				\newcommand\w{1.65}
				\begin{tabular}{l@{}lR{\w cm}@{}L{0.5cm}R{\w cm}@{}L{0.5cm}R{\w cm}@{}L{0.5cm}R{\w cm}@{}L{0.5cm}}
					\midrule
					&& \multicolumn{8}{c}{Dependent variable: Gini coefficient} \\ \cmidrule{3-10}
					& \multicolumn{4}{c}{All redistributive decisions} && \multicolumn{4}{c}{First decision only} \\ 	\cmidrule{3-5} \cmidrule{7-10}				
					&& (1) && (2) && (3) && (4)  \\
					\midrule
					Performance condition&&0.419&\sym{***}&0.418&\sym{***}&0.296&\sym{***}&0.325&\sym{***}\\
&&(0.037&)&(0.046&)&(0.089&)&(0.092&)\\
Efficiency condition&&0.253&\sym{***}&0.248&\sym{***}&0.199&\sym{**}&0.174&\sym{*}\\
&&(0.038&)&(0.047&)&(0.098&)&(0.098&)\\
Constant&&0.303&\sym{***}&0.305&\sym{***}&0.378&\sym{***}&0.439&\sym{***}\\
&&(0.040&)&(0.026&)&(0.073&)&(0.117&)\\
  \midrule
					Additional controls? 				&& No && Yes && No  && Yes \\
					\(N\) (redistributive decisions)&&405&&405&&141&&141&\\

					\(N\) (individuals)&&141&&141&&141&&141&\\
 \midrule
					
				\end{tabular}
			\end{center}
			\begin{singlespace}  \vspace{-.5cm}
				\noindent \justify \textit{Notes:} This table is analogous to Table~\ref{tab:gini}, but excludes spectators classified as having ``other'' fairness ideals (i.e., those who do not fit the standard egalitarian, libertarian, or meritocratic classifications). See notes to Table~\ref{tab:gini} for details. {*} $p<0.10$, {*}{*} $p<0.05$, {*}{*}{*} $p<0.01$. \par
			\end{singlespace} 	
		}
	\end{table}

	\begin{table}[H]{\footnotesize
			\begin{center}
				\caption{Robustness Check of Main Effects to Excluding Spectators Who Redistributed More Under Efficiency Costs} \label{tab:rob_eff}
				\newcommand\w{1.65}
				\begin{tabular}{l@{}lR{\w cm}@{}L{0.5cm}R{\w cm}@{}L{0.5cm}R{\w cm}@{}L{0.5cm}R{\w cm}@{}L{0.5cm}}
					\midrule
					&& \multicolumn{8}{c}{Dependent variable: Gini coefficient} \\ \cmidrule{3-10}
					& \multicolumn{4}{c}{All redistributive decisions} && \multicolumn{4}{c}{First decision only} \\ 	\cmidrule{3-5} \cmidrule{7-10}				
					&& (1) && (2) && (3) && (4)  \\
					\midrule
					Performance condition&&0.234&\sym{***}&0.237&\sym{***}&0.187&\sym{***}&0.187&\sym{***}\\
&&(0.034&)&(0.041&)&(0.060&)&(0.062&)\\
Efficiency condition&&0.243&\sym{***}&0.245&\sym{***}&0.250&\sym{***}&0.252&\sym{***}\\
&&(0.027&)&(0.033&)&(0.061&)&(0.066&)\\
Constant&&0.447&\sym{***}&0.445&\sym{***}&0.456&\sym{***}&0.437&\sym{***}\\
&&(0.028&)&(0.021&)&(0.043&)&(0.081&)\\
  \midrule
					Additional controls? 				&& No && Yes && No  && Yes \\
					\(N\) (redistributive decisions)&&655&&655&&225&&225&\\

					\(N\) (individuals)&&225&&225&&225&&225&\\
 \midrule
					
				\end{tabular}
			\end{center}
			\begin{singlespace}  \vspace{-.5cm}
				\noindent \justify \textit{Notes:} This table is analogous to Table~\ref{tab:gini}, but excludes spectators who redistributed more earnings when there was an efficiency cost than when there was no efficiency cost. This exclusion addresses concerns that some spectators may have misunderstood the efficiency cost treatment. See notes to Table~\ref{tab:gini} for details. {*} $p<0.10$, {*}{*} $p<0.05$, {*}{*}{*} $p<0.01$. \par
			\end{singlespace} 	
		}
	\end{table}

	\begin{table}[H]{\footnotesize
			\begin{center}
				\caption{Robustness Check of Main Effects to Estimating the Regression Separately for Each Cohort} \label{tab:rob_coh}
				\newcommand\w{1.65}
				\begin{tabular}{l@{}lR{\w cm}@{}L{0.5cm}R{\w cm}@{}L{0.5cm}R{\w cm}@{}L{0.5cm}R{\w cm}@{}L{0.5cm}}
					\midrule
					&& \multicolumn{8}{c}{Dependent variable: Gini coefficient} \\ \cmidrule{3-10}
					& \multicolumn{4}{c}{All redistributive decisions} && \multicolumn{4}{c}{First decision only} \\ 	\cmidrule{3-5} \cmidrule{7-10}				
					&& (1) && (2) && (3) && (4)  \\
					\midrule
					\multicolumn{3}{l}{\hspace{-1em} \textbf{Panel A. 2023 Cohort}} \\ \addlinespace
					Performance condition&&0.235&\sym{***}&0.234&\sym{***}&0.180&\sym{**}&0.140&\sym{*}\\
&&(0.042&)&(0.052&)&(0.075&)&(0.079&)\\
Efficiency condition&&0.182&\sym{***}&0.184&\sym{***}&0.195&\sym{**}&0.187&\sym{**}\\
&&(0.034&)&(0.042&)&(0.078&)&(0.086&)\\
Constant&&0.419&\sym{***}&0.419&\sym{***}&0.438&\sym{***}&0.439&\sym{***}\\
&&(0.035&)&(0.027&)&(0.053&)&(0.114&)\\
  \midrule
					Additional controls? 				&& No && Yes && No  && Yes \\
					\(N\) (redistributive decisions)&&403&&403&&138&&138&\\

					\(N\) (individuals)&&138&&138&&138&&138&\\

					\midrule
					\multicolumn{3}{l}{\hspace{-1em} \textbf{Panel B. 2024 Cohort}} \\ \addlinespace
					Performance condition&&0.245&\sym{***}&0.251&\sym{***}&0.202&\sym{***}&0.218&\sym{***}\\
&&(0.039&)&(0.048&)&(0.075&)&(0.078&)\\
Efficiency condition&&0.195&\sym{***}&0.191&\sym{***}&0.220&\sym{***}&0.246&\sym{***}\\
&&(0.034&)&(0.041&)&(0.084&)&(0.088&)\\
Constant&&0.421&\sym{***}&0.421&\sym{***}&0.424&\sym{***}&0.422&\sym{***}\\
&&(0.035&)&(0.025&)&(0.060&)&(0.095&)\\
  \midrule
					Additional controls? 				&& No && Yes && No  && Yes \\
					\(N\) (redistributive decisions)&&390&&390&&133&&133&\\

					\(N\) (individuals)&&133&&133&&133&&133&\\
 \midrule				
				\end{tabular}
			\end{center}
			\begin{singlespace}  \vspace{-.5cm}
				\noindent \justify \textit{Notes:} This table is analogous to Table~\ref{tab:gini}, but estimated separately for MBA students of the 2023 cohort (Panel A) and 2024 cohort (Panel B). See notes to Table~\ref{tab:gini} for details. {*} $p<0.10$, {*}{*} $p<0.05$, {*}{*}{*} $p<0.01$. \par
			\end{singlespace} 	
		}
	\end{table}

	\begin{table}[H]{\footnotesize
			\begin{center}
				\caption{Robustness Check of Main Effects to Including Round Fixed Effects} \label{tab:rob_round}
				\newcommand\w{1.65}
				\begin{tabular}{l@{}lR{\w cm}@{}L{0.5cm}R{\w cm}@{}L{0.5cm}R{\w cm}@{}L{0.5cm}R{\w cm}@{}L{0.5cm}}
					\midrule
					&& \multicolumn{8}{c}{Dependent variable: Gini coefficient} \\ \cmidrule{3-10}
					& \multicolumn{4}{c}{All redistributive decisions} && \multicolumn{4}{c}{First decision only} \\ 	\cmidrule{3-5} \cmidrule{7-10}				
					&& (1) && (2) && (3) && (4)  \\
					\midrule
					Performance condition&&0.240&\sym{***}&0.242&\sym{***}&0.191&\sym{***}&0.184&\sym{***}\\
&&(0.029&)&(0.029&)&(0.052&)&(0.054&)\\
Efficiency condition&&0.188&\sym{***}&0.187&\sym{***}&0.206&\sym{***}&0.211&\sym{***}\\
&&(0.024&)&(0.024&)&(0.057&)&(0.060&)\\
Constant&&0.420&\sym{***}&0.420&\sym{***}&0.432&\sym{***}&0.425&\sym{***}\\
&&(0.024&)&(0.015&)&(0.040&)&(0.071&)\\
  \midrule
					Round fixed effects? 				&& Yes && Yes && Yes  && Yes \\				
					Additional controls? 				&& No  && Yes && No   && Yes \\
					\(N\) (redistributive decisions)&&793&&784&&271&&271&\\

					\(N\) (individuals)&&271&&262&&271&&271&\\
 \midrule
					
				\end{tabular}
			\end{center}
			\begin{singlespace}  \vspace{-.5cm}
				\noindent \justify \textit{Notes:} This table is analogous to Table~\ref{tab:gini}, but the regression equation also controls for round fixed effects (only relevant for the columns 1 and 2). See notes to Table~\ref{tab:gini} for details. {*} $p<0.10$, {*}{*} $p<0.05$, {*}{*}{*} $p<0.01$. \par
			\end{singlespace} 	
		}
	\end{table}

	\begin{table}[H]{\footnotesize
			\begin{center}
				\caption{Robustness Check of Main Effects to Using the Share of Earnings Redistributed as the Outcome} \label{tab:rob_red}
				\newcommand\w{1.65}
				\begin{tabular}{l@{}lR{\w cm}@{}L{0.5cm}R{\w cm}@{}L{0.5cm}R{\w cm}@{}L{0.5cm}R{\w cm}@{}L{0.5cm}}
					\midrule
					&& \multicolumn{8}{c}{Dependent variable: Share of earnings redistributed} \\ \cmidrule{3-10}
					& \multicolumn{4}{c}{All redistributive decisions} && \multicolumn{4}{c}{First decision only} \\ 	\cmidrule{3-5} \cmidrule{7-10}				
					&& (1) && (2) && (3) && (4)  \\
					\midrule
					Performance condition&&$-$0.123&\sym{***}&$-$0.124&\sym{***}&$-$0.116&\sym{***}&$-$0.121&\sym{***}\\
&&(0.016&)&(0.019&)&(0.031&)&(0.031&)\\
Efficiency condition&&$-$0.103&\sym{***}&$-$0.106&\sym{***}&$-$0.125&\sym{***}&$-$0.145&\sym{***}\\
&&(0.014&)&(0.017&)&(0.033&)&(0.033&)\\
Constant&&0.313&\sym{***}&0.314&\sym{***}&0.323&\sym{***}&0.305&\sym{***}\\
&&(0.014&)&(0.010&)&(0.024&)&(0.037&)\\
  \midrule
					Additional controls? 				&& No && Yes && No  && Yes \\
					\(N\) (redistributive decisions)&&793&&793&&271&&271&\\

					\(N\) (individuals)&&271&&271&&271&&271&\\
 \midrule
					
				\end{tabular}
			\end{center}
			\begin{singlespace}  \vspace{-.5cm}
				\noindent \justify \textit{Notes:} This table is analogous to Table~\ref{tab:gini}, but the dependent variable is the net-of-efficiency-cost share of earnings redistributed. See notes to Table~\ref{tab:gini} for details. {*} $p<0.10$, {*}{*} $p<0.05$, {*}{*}{*} $p<0.01$. \par
			\end{singlespace} 	
		}
	\end{table}

\clearpage
\setcounter{table}{0}
\setcounter{figure}{0}
\setcounter{equation}{0}
\renewcommand{\thetable}{C\arabic{table}}
\renewcommand{\thefigure}{C\arabic{figure}}
\renewcommand{\theequation}{C\arabic{equation}}

\section{Comparison of Benchmark Studies} \label{app:benchmark}

This appendix compares experimental designs, recruitment strategies, and sample characteristics across the five studies (Tables~\ref{tab:comparison}--\ref{tab:ideals}). Appendix Table~\ref{tab:summ_benchmark} summarizes the key design features and spectator demographics.

\subsection{Benchmark Studies}

\textit{\citet{almas.etal2020}.} \citet{almas.etal2020} conducted a spectator experiment across the United States and Norway using nationally representative samples recruited through Norstat and its U.S.\ collaborator Research Now. Workers completed assignments on Amazon Mechanical Turk. Each spectator was randomized into one of three conditions---luck, merit, or efficiency cost---and made a single redistribution decision (between-subject design), choosing how to divide \$6 between a winning and a losing worker. In the luck and merit treatments, spectators chose from a discrete menu in \$1 increments; in the efficiency treatment, redistribution incurred a 50 percent cost. We use only the U.S.\ sample for our comparison.

\textit{\citet{cohn.etal2023}.} \citet{cohn.etal2023} recruited a large sample of U.S.\ respondents through YouGov between December 2016 and April 2017, oversampling households in the top 5 percent by income or wealth (annual household income above \$250,000 or gross liquid assets of at least \$1 million). Workers completed a real-effort task (checking and correcting digitized identification entries) on Amazon Mechanical Turk. Each spectator made a single redistribution decision (between-subject design), choosing how to divide the winner's \$6 earnings using a discrete menu in \$1 increments. The study includes three treatments---luck, mixed, and merit---but does not include an efficiency cost treatment.

\textit{\citet{preuss2025inequality} (SCE).} \citet{preuss2025inequality} embedded a spectator experiment within the New York Federal Reserve's Survey of Consumer Expectations (SCE), yielding a nationally representative U.S.\ sample. Workers completed an encryption task on Amazon Mechanical Turk. Each spectator made 12 redistribution decisions (within-subject design), allocating \$5 between workers in \$0.50 increments. Decisions varied in the probability that earnings were determined by a coin flip versus worker performance; we use only the pure luck and pure performance conditions.

\textit{\citet{harrs2025fairness}.} \citet{harrs2025fairness} recruited a broadly representative U.S.\ sample of spectators through Prolific. Workers were recruited separately on Amazon Mechanical Turk. Each spectator made two redistribution decisions in randomized order---one in a luck condition and one in a merit condition (within-subject design). The total earnings at stake were \$4 and spectators can transfer amounts in \$0.10 increments---lower stakes and finer increments than the other studies (\$5--\$6 and \$0.50--\$1, respectively).

\subsection{Design Differences}

All five studies share the impartial-spectator paradigm \citep{cappelen.etal2013} in which spectators allocate income between other workers. The workers whose income spectators redistribute are U.S.-based online workers, recruited from Prolific in our study and from Amazon Mechanical Turk in the benchmark studies. Nonetheless, some differences between the cited studies exist. 

First, our MBA spectators were recruited during mandatory class sessions, reducing the self-selection concerns that arise in online experiments \citep{levitt.list2007}. The benchmark studies use nationally representative panels (Norstat/Research Now for \citeauthor{almas.etal2020}; YouGov for \citeauthor{cohn.etal2023}; the NY Fed SCE for \citeauthor{preuss2025inequality}) or online platforms (Prolific for \citeauthor{harrs2025fairness}).

Second, the number of decisions per spectator differs across studies. Our design and those of \citet{preuss2025inequality} and \citet{harrs2025fairness} are within-subject, meaning each spectator makes multiple redistribution decisions across conditions. In contrast, \citet{almas.etal2020} and \citet{cohn.etal2023} use between-subject designs in which each spectator makes a single decision. When comparing across studies in our main analysis, we use between-subject estimates---restricting to spectators' first decisions in within-subject designs---to ensure comparability.

Third, the transfer mechanisms and stakes differ. \citet{almas.etal2020} and \citet{cohn.etal2023} use discrete transfer menus in \$1 increments from a \$6 total. \citet{preuss2025inequality} uses \$0.50 increments from a \$5 total. \citet{harrs2025fairness} uses finer increments (\$0.10) and lower stakes (\$4). Despite these differences, all studies allow us to compute a standardized Gini coefficient as the outcome measure.

Fourth, the treatment set differs. Our experiment and \citet{almas.etal2020} include an efficiency cost treatment alongside the luck and performance conditions. \citet{cohn.etal2023} includes luck, mixed, and merit treatments but no efficiency cost. The remaining benchmark studies include only luck and performance (or merit) conditions.

\begin{table}[H]{\footnotesize
		\begin{center}
			\caption{Summary of Study Designs and Spectator Demographics} \label{tab:summ_benchmark}
			\begin{tabular}{p{4cm}C{2.1cm}C{2.1cm}C{2.1cm}C{2.1cm}C{2.1cm}}
				\midrule
				& This study & \cite{almas.etal2020} & \cite{cohn.etal2023} & \cite{preuss2025inequality} & \cite{harrs2025fairness} \\
				\midrule
				\multicolumn{6}{l}{\hspace{-1em} \textbf{Panel A. Study design}} \\\addlinespace
				Year of data collection & 2023--2024 & 2014 & 2016--2017 & 2021 & 2020--2022 \\
				Spectator recruitment & In class & Research Now & YouGov & NY Fed SCE & Prolific \\
				Worker recruitment & Prolific & MTurk & MTurk & MTurk & MTurk \\
				Stakes & \$6 & \$6 & \$6 & \$5 & \$4 \\
				Decisions per spectator & 3 & 1 & 1 & 12 & 2 \\
				Transfer increments & \$0.50 & \$1 & \$1 & \$0.50 & \$0.10 \\
				Treatments & L, P, E & L, P, E & L, Mx, P & L, P & L, P \\
				Design & Within & Between & Between & Within & Within \\
				\midrule
				\multicolumn{6}{l}{\hspace{-1em} \textbf{Panel B. Spectator demographics}} \\\addlinespace
				Male (\%) & 55.4 & 48.9 & 44.0 & 50.3 & 48.2 \\
Mean age (years) & 28.3 & 44.5 & 47.9 & 49.5 & 43.5 \\
College educated (\%) & 100.0 & 58.2 & 44.8 & 62.4 & 69.9 \\
HH income $\geq$ \$100k (\%) & --- & 24.5 & 18.9 & 25.4 & 27.0 \\
White (\%) & --- & --- & 74.6 & 82.2 & 75.5 \\
$N$ & 271 & 1,000 & 882 & 197 & 1,476 \\

				\midrule
			\end{tabular}
		\end{center}
		\begin{singlespace}  \vspace{-.5cm}
			\noindent \justify \textit{Notes:} This table compares study designs and spectator demographics across the five studies used in our benchmark analysis. Panel A describes key design features; ``---'' indicates the variable is not available in that dataset. Panel B reports spectator demographics. For \cite{cohn.etal2023}, demographics in Panel B are population-weighted to account for the oversampling of households in the top 5\% by income or wealth; household income $\geq$\$100k is computed from self-reported family annual income categories. For \cite{almas.etal2020}, annual household income is computed from monthly income. All incentivized studies implement one randomly selected spectator decision. L = Luck, P = Performance/Merit, Mx = Mixed, E = Efficiency cost. \par
		\end{singlespace}
	}
\end{table}

\subsection{Pooling the Benchmark Studies} \label{app:pooling}

We summarize four benchmark estimates from comparable U.S. impartial-spectator designs (\citep{almas.etal2020, cohn.etal2023, preuss2025inequality, harrs2025fairness}) with a single pooled estimate, computed using a \citet{dersimonian1986meta} random-effects meta-analysis. The \citeauthor{dersimonian1986meta} estimator is a widely used conventional random-effects estimator in the medical and social sciences, and random-effects or hierarchical pooling of study-level estimates is also common in economics---for example, in \citet{bellavance.etal2009}'s meta-analysis of the value of a statistical life and in the hierarchical analyses of \citet{meager2019, meager2022} and \citet{vivalt2020}.

We combine $K$ study-level estimates $\hat\theta_i$, each with squared standard error $s_i^2$. Let the inverse-variance (fixed-effect) weights be $w_i = 1/s_i^2$, and let $Q = \sum_{i} w_i (\hat\theta_i - \bar\theta_{\text{FE}})^2$, where $\bar\theta_{\text{FE}} = \sum_i w_i \hat\theta_i / \sum_i w_i$, denote Cochran's heterogeneity statistic. The \citeauthor{dersimonian1986meta} method-of-moments estimator of the between-study variance $\tau^2$ is
\begin{align}
	\hat\tau^2 = \max\left\{0,\ \frac{Q - (K-1)}{\sum_i w_i - \sum_i w_i^2 / \sum_i w_i}\right\}.
\end{align}
The pooled estimate weights each study by the inverse of its total variance, $w_i^{*} = 1/(s_i^2 + \hat\tau^2)$:
\begin{align}
	\hat\theta_{\text{RE}} = \frac{\sum_i w_i^{*} \hat\theta_i}{\sum_i w_i^{*}}, \qquad \widehat{\mathrm{Var}}(\hat\theta_{\text{RE}}) = \frac{1}{\sum_i w_i^{*}},
\end{align}
and the 95 percent confidence intervals we report for the pooled estimate (e.g., Figure~\ref{fig:gini_comparison} and the ``Pooled'' column of Table~\ref{tab:comparison}) are $\hat\theta_{\text{RE}} \pm 1.96\,\widehat{\mathrm{Var}}(\hat\theta_{\text{RE}})^{1/2}$. The estimator nests inverse-variance fixed effects as the special case $\hat\tau^2 = 0$, and the share of cross-study variation reflecting genuine heterogeneity rather than sampling error is $I^2 = \max\{0, (Q-(K-1))/Q\}$.

\subsection{Comparison with \citet{fisman.etal2015a}} \label{app:fisman_comparison}

\citet{fisman.etal2015a} study elites' efficiency preferences using a modified dictator game in which subjects divide an endowment between themselves and another participant. To compare efficiency concerns across designs, we estimate the analogue of their efficiency parameter $\rho$ in our data. Specifically, \citet{fisman.etal2015a} assume that dictators maximize a CES utility function:
\begin{align}
	\left(\alpha \, y_S^\rho + (1-\alpha) \, y_O^\rho\right)^{1/\rho},
\end{align}
where $y_S$ is the income the dictator assigns to themselves, $y_O$ is the income assigned to the other participant, $\alpha$ captures the weight on own payoff, and $\rho$ governs the equality--efficiency tradeoff. For our impartial spectators, we assume the analogous function with equal weights ($\alpha = 1/2$), since spectators have no stake in the allocation:
\begin{align}
	\left(\tfrac{1}{2} \, y_H^\rho + \tfrac{1}{2} \, y_L^\rho\right)^{1/\rho},
\end{align}
where $y_H$ and $y_L$ denote the earnings assigned to the winner and loser, respectively. In the efficiency-cost condition, the spectator's budget constraint is $y_H + 2 y_L = M$, where the price of redistribution is $2$ (each dollar taken from the winner generates only 50 cents for the loser). The first-order condition yields $(y_L / y_H)^{\rho - 1} = 2$, which we invert to recover the implied $\rho$ for each spectator:
\begin{align}
	\rho_i = 1 - \frac{\ln(2)}{\ln(y_H / y_L)}.
\end{align}
Spectators who do not redistribute ($y_L = 0$) are assigned $\rho = 1$ (utilitarian); those who fully equalize ($y_H = y_L$) are assigned $\rho = -20$, matching the effective floor in the \citet{fisman.etal2015a} replication data. Since $\rho$ is unbounded below and can take arbitrarily negative values, we compare medians to limit the influence of outliers.

A key challenge is that $\rho$ has different interpretations across designs. In our setting, $\rho$ measures how quickly spectators become more tolerant of inequality as redistribution becomes costlier. In \citet{fisman.etal2015a}, by contrast, dictators allocate their own income, so $\rho$ measures how quickly they become more tolerant of advantageous inequality as the cost of redistribution rises. Because 78 percent of YLS students are estimated to prioritize their own income over others' (that is, have $\alpha > 0.6$), the efficiency parameter in their setting partly reflects self-interest rather than a pure preference for efficiency over equality.

We present two comparisons (Appendix Table~\ref{tab:fisman_rho}). The first comparison uses the full YLS sample ($N = \getval{fisman_n_yls_all}$): the median $\rho$ is $\getval{fisman_rho_yls_all}$ for YLS students and $\getval{fisman_rho_mba}$ for MBA students---statistically indistinguishable ($p = \getval{fisman_p_yls_all}$, rank-sum test). This comparison includes all subjects but is harder to interpret because the dictator design conflates efficiency preferences with selfishness for subjects with high $\hat{\alpha}$. The second comparison restricts the \citet{fisman.etal2015a} data to the 15 percent of YLS students they classify as ``fair-minded'' ($0.45 \le \hat{\alpha} \le 0.55$, $N = \getval{fisman_n_yls_fair}$), who assign roughly equal weight to their own and the other participant's payoff. This removes the selfishness contamination, yielding a median $\rho$ of $\getval{fisman_rho_yls_fair}$---close to our MBA estimate of $\getval{fisman_rho_mba}$.

Both comparisons point to the same conclusion: MBA students' efficiency preferences are similar to those of YLS elites. Both elite groups are substantially more efficiency-seeking than the general population, whose fair-minded subsample has a median $\rho$ of $\getval{fisman_rho_genpop_fair}$ (Table~\ref{tab:fisman_rho}, Panel B). Our Dyson replication confirms this pattern, with Dyson students exhibiting a median $\rho$ of $\getval{fisman_rho_dyson}$---identical to MBA students. However, only \getval{fisman_n_yls_fair} of the \getval{fisman_n_yls_all} YLS students are classified as fair-minded. For the remaining 85 percent, the estimated $\rho$ confounds efficiency preferences with self-interest, so the \citet{fisman.etal2015a} data cannot tell us how the majority of elites would trade off efficiency against inequality between others. Our impartial-spectator design closes this gap by measuring efficiency preferences for the full sample, free of self-interest confounds.

\begin{table}[H]
	\caption{Efficiency Preferences: MBA Spectators vs.\ \citet{fisman.etal2015a}}\label{tab:fisman_rho}
	{\footnotesize
		\begin{center}
			\begin{tabular}{l ccccc}
				\midrule
				& Median $\rho$ & Mean $\rho$ & p25 & p75 & $N$ \\
				\midrule
				\multicolumn{6}{l}{\hspace{-1em} \textbf{Panel A. Spectator design (self-interest removed)}} \\
				\addlinespace
				MBA students & 0.500 & -3.834 & 0.000 & 1.000 & 266 \\
Dyson students & 0.500 & -3.665 & 0.000 & 1.000 & 167 \\
Alm\r{a}s et al.\ --- U.S. sample & -20.000 & -11.327 & -20.000 & 1.000 & 332 \\

				\midrule
				\multicolumn{6}{l}{\hspace{-1em} \textbf{Panel B. Dictator design, fair-minded subjects ($0.45 \le \hat{\alpha} \le 0.55$)}} \\
				\addlinespace
				Fisman et al.\ --- YLS students & 0.533 & -1.521 & 0.024 & 0.904 & 30 \\
Fisman et al.\ --- General population & 0.276 & -1.314 & -0.114 & 0.466 & 227 \\

				\midrule
				\multicolumn{6}{l}{\hspace{-1em} \textbf{Panel C. Dictator design, all subjects}} \\
				\addlinespace
				Fisman et al.\ --- YLS students & 0.699 & -0.182 & 0.214 & 0.826 & 208 \\
Fisman et al.\ --- General population & 0.034 & -1.243 & -0.731 & 0.477 & 717 \\

				\midrule
			\end{tabular}
		\end{center}
		\begin{singlespace} \vspace{-.5cm}
			\noindent \justify \textit{Notes:} This table compares the CES efficiency parameter $\rho$ across studies. Higher $\rho$ indicates stronger efficiency preferences ($\rho = 1$: utilitarian; $\rho = 0$: Cobb--Douglas; $\rho \to -\infty$: Rawlsian). Panel~A reports implied $\rho$ for our spectator samples, recovered from efficiency-cost treatment choices using $\rho_i = 1 - \ln(2)/\ln(y_H/y_L)$. Panel~B restricts the \cite{fisman.etal2015a} sample to ``fair-minded'' subjects, for whom the selfishness--efficiency confound is minimal. Panel~C shows all \cite{fisman.etal2015a} subjects. The spectator design in Panel~A removes the self-interest confound by construction. \par
		\end{singlespace}
	}
\end{table}

\clearpage
\section{Replication: Dyson Business Students} \label{app:dyson}

\setcounter{table}{0}
\setcounter{figure}{0}
\setcounter{equation}{0}
\renewcommand{\thetable}{D\arabic{table}}
\renewcommand{\thefigure}{D\arabic{figure}}
\renewcommand{\theequation}{D\arabic{equation}}

This appendix presents our replication with undergraduate business students from Cornell University's Dyson School of Applied Economics and Management.

\subsection{Sample Characteristics}

Appendix Table~\ref{tab:dyson_summ_spectators} presents summary statistics of the Dyson sample. The sample consists primarily of young students (mean age \getval{dyson_age_mean} years), with a roughly equal gender split (\getval{dyson_pct_male} percent male). The majority were born in the U.S.\ (\getval{dyson_pct_born_usa} percent) and report relatively high socioeconomic status during childhood---\getval{dyson_pct_comfort} percent had enough money for necessities and at least occasional luxuries while growing up. Most aspire to managerial positions (\getval{dyson_pct_manager} percent) and private-sector careers (\getval{dyson_pct_private} percent), and about half plan to
pursue an MBA (\getval{dyson_pct_mba} percent). Only \getval{dyson_pct_voted} percent report voting in recent elections, likely reflecting their young age.

\subsection{Implemented Inequality}

Appendix Table~\ref{tab:dyson_gini} presents estimates of equation \eqref{eq:gini} for the Dyson sample. When worker earnings are randomly assigned, Dyson students implement a Gini coefficient of \getval{dyson_gini_b0} (column 1), similar to the MBA estimate. The performance condition increases the implemented Gini coefficient by \getval{dyson_gini_b1} Gini points ($p<0.01$), or \getval{dyson_gini_b1_pct} percent of the baseline Gini. The efficiency cost condition increases the Gini coefficient by \getval{dyson_gini_b2} Gini points ($p<0.01$), or \getval{dyson_gini_b2_pct} percent of the baseline. These effects are robust to including spectator fixed effects (column 2) and using only spectators' first choices (columns 3 and 4).

\subsection{Distribution of Fairness Ideals}

Appendix Tables \ref{tab:dyson_joint} and \ref{tab:dyson_ideals} present the distribution of fairness ideals in the Dyson sample. Using the between-subject design, we find that \getval{dyson_egal_btw} percent are egalitarians, \getval{dyson_lib_btw} percent are libertarians, \getval{dyson_merit_btw} percent are meritocrats, and \getval{dyson_other_btw} percent have other fairness ideals. The within-subject design yields similar estimates. Compared to MBA students, Dyson students show lower rates of egalitarianism (\getval{dyson_egal_btw} vs.\ \getval{egal_share_btw} percent), higher rates of meritocracy (\getval{dyson_merit_btw} vs.\ \getval{merit_share_btw} percent), and similar rates of libertarianism (\getval{dyson_lib_btw} vs.\ \getval{lib_share_btw} percent). Like MBA students, they are less likely to conform to standard fairness ideals than the general population. About a quarter of Dyson students (\getval{dyson_moderate_share} percent) behave as ``moderates.''

\begin{table}[H]\caption{Summary Statistics of the Dyson Sample} \label{tab:dyson_summ_spectators}
	{\footnotesize
		\begin{centering} 
			\protect
			\begin{tabular}{lccc}
				\addlinespace \addlinespace \midrule			
				& Mean & SD  & N   \\
				& (1) & (2)  & (3)   \\
				\midrule 	
				
				\multicolumn{3}{l}{\hspace{-1em} \textbf{Panel A. Demographic characteristics}} \\ 
				Average age &19.534 &0.689 &161 \\
Male &0.543 &0.500 &162 \\
Born in the USA &0.747 &0.436 &162 \\
 \midrule
				
				\multicolumn{3}{l}{\hspace{-1em} \textbf{Panel B. Financial situation while growing up}} \\ 
				Always had enough money for necessities and luxuries &0.244 &0.431 &156 \\
Always had enough for necessities and occasional luxuries &0.474 &0.501 &156 \\
Usually had enough for necessities, rarely for luxuries &0.231 &0.423 &156 \\
Sometimes did not have enough money for necessities &0.038 &0.193 &156 \\
Frequently did not have enough money for necessities &0.013 &0.113 &156 \\
 \midrule						
				
				\multicolumn{3}{l}{\hspace{-1em} \textbf{Panel C. Employment preferences}} \\ 
				Working long hours is sometimes necessary &0.776 &0.418 &161 \\
Wants career in a non-profit &0.199 &0.400 &161 \\
Wants to become an entrepreneur &0.516 &0.501 &161 \\
Wants career in private sector &0.727 &0.447 &161 \\
Work-life balance is more important than a high salary &0.596 &0.492 &161 \\
Aspire to be a manager &0.807 &0.396 &161 \\
Wants to do an MBA &0.516 &0.501 &161 \\
 \midrule											
				
				\multicolumn{3}{l}{\hspace{-1em} \textbf{Panel D. Attitudes toward pay determination}} \\ 
				How much responsibility should decide pay &0.994 &0.079 &161 \\
How well someone works should decide pay &1.000 &0.000 &161 \\
How hard someone works should decide pay &0.957 &0.205 &161 \\
 \midrule											
				
				\multicolumn{3}{l}{\hspace{-1em} \textbf{Panel E. Voting behavior and social views}} \\ 					
				Voted in the last elections &0.204 &0.404 &152 \\
Success is luck and connections &0.280 &0.450 &161 \\
Hard work brings a better life &0.863 &0.345 &161 \\
 \midrule		
				
				Completed all three redistributive decisions &0.970 &0.171 &167 \\
Passed attention check &0.938 &0.242 &161 \\
Minutes spent in survey &5.674 &2.849 &167 \\
 \midrule	
				
			\end{tabular}
			\par\end{centering}
		
		\singlespacing\justify\footnotesize
		
		\textit{Notes:} This table shows summary statistics of Dyson students in our sample. All variables are based on data self-reported by Dyson students in the exit survey of our study. Employment preferences and social views are based on Dyson students' agreement with several statements in a five-point Likert scale grid. For each statement, we define an indicator variable that equals one if the student selects ``strongly agree'' or ``agree,'' and zero if the student selects ``neither agree nor disagree,'' ``disagree,'' or ``strongly disagree.''	For the attitudes toward pay determination, we define an indicator variable that equals one if the student selects ``essential,'' ``very important,'' or ``fairly important'' and zero if the student selects ``not very important'' or ``not important at all.'' \par
		
	}
\end{table}

\clearpage
\begin{table}[H]{\footnotesize
		\begin{center}
			\caption{Implemented Gini Coefficient Across Environments (Dyson Sample)} \label{tab:dyson_gini}
			\newcommand\w{1.65}
			\begin{tabular}{l@{}lR{\w cm}@{}L{0.5cm}R{\w cm}@{}L{0.5cm}R{\w cm}@{}L{0.5cm}R{\w cm}@{}L{0.5cm}}
				\midrule
				&& \multicolumn{8}{c}{Dependent variable: Gini coefficient} \\ \cmidrule{3-10}
				& \multicolumn{4}{c}{All redistributive decisions} && \multicolumn{4}{c}{First decision only} \\ 	\cmidrule{3-5} \cmidrule{7-10}				
				&& (1) && (2) && (3) && (4)  \\
				\midrule
				Performance condition&&0.311&\sym{***}&0.304&\sym{***}&0.334&\sym{***}&0.330&\sym{***}\\
&&(0.036&)&(0.044&)&(0.071&)&(0.072&)\\
Efficiency condition&&0.198&\sym{***}&0.194&\sym{***}&0.157&\sym{**}&0.152&\sym{*}\\
&&(0.028&)&(0.035&)&(0.078&)&(0.080&)\\
Constant&&0.416&\sym{***}&0.420&\sym{***}&0.413&\sym{***}&0.441&\sym{***}\\
&&(0.031&)&(0.023&)&(0.060&)&(0.087&)\\
  \midrule
				Additional controls? 				&& No && Yes && No  && Yes \\
				\(N\) (redistributive decisions)&&494&&494&&167&&167&\\

				\(N\) (individuals)&&167&&167&&167&&167&\\
 \midrule
				
			\end{tabular}
		\end{center}
		\begin{singlespace}  \vspace{-.5cm}
			\noindent \justify \textit{Notes:} This table displays estimates of $\beta_0$, $\beta_1$, and $\beta_2$ from equation \eqref{eq:gini} estimated on the Dyson sample. The omitted treatment category is the luck condition. The specifications in columns 2 and 4 include additional controls. In column 2, we include spectator fixed effects. In column 4, we control for gender, parental financial situation while growing up, and a dummy for being born in the U.S. Heteroskedasticity-robust standard errors clustered at the spectator level in parentheses. {*} $p<0.10$, {*}{*} $p<0.05$, {*}{*}{*} $p<0.01$. \par
		\end{singlespace} 	
	}
\end{table}

\begin{table}[H]
\begin{center}	
	\caption{Joint Distribution of Redistribution in Luck and Performance Treatments (Dyson Sample)} \label{tab:dyson_joint}
	\begin{tabular}{ccccccccc}
		& & \multicolumn{7}{c}{Earnings Redistributed in Performance Condition} \\[0.5ex]
		\hline
		&   & \$0 & \$1 & \$2 & \$3 & \$4 & \$5 & \$6 \\
		\cline{2-9}
		\multirow{7}{*}{\begin{sideways}Luck Condition\end{sideways}}  
		& \$0 & \cellcolor{green_lab!25}0.160 & 0.061 & 0.018 & 0.000 & 0.000 & 0.000 & 0.000 \\
& \$1 & 0.061 & 0.037 & 0.025 & 0.006 & 0.006 & 0.000 & 0.000 \\
& \$2 & 0.117 & 0.067 & 0.049 & 0.006 & 0.000 & 0.000 & 0.000 \\
& \$3 & \cellcolor{blue_lab!25}0.166 & \cellcolor{blue_lab!25}0.067 & \cellcolor{blue_lab!25}0.074 & \cellcolor{red_lab!25}0.061 & 0.000 & 0.000 & 0.000 \\
& \$4 & 0.000 & 0.000 & 0.000 & 0.000 & 0.000 & 0.000 & 0.006 \\
& \$5 & 0.000 & 0.006 & 0.000 & 0.000 & 0.000 & 0.000 & 0.000 \\
& \$6 & 0.000 & 0.000 & 0.000 & 0.006 & 0.000 & 0.000 & 0.000 \\

		\hline
	\end{tabular}
	\begin{tabular}{ll}
		\cellcolor{green_lab!25} & Libertarians \\
		\cellcolor{blue_lab!25} & Meritocrats \\
		\cellcolor{red_lab!25} & Egalitarians 
	\end{tabular}
	
\end{center}
\begin{singlespace}  \vspace{-.5cm}
	\justify \textit{Notes:} This table shows the joint distribution of redistribution choices across luck and performance conditions in the Dyson sample, excluding the efficiency cost treatment. See notes to Table~\ref{tab:joint} for details. \par
	
\end{singlespace}

\end{table}

\begin{table}[H]\caption{Distribution of Fairness Ideals: Elites vs. Average Citizens \\ (Including Dyson Sample)} \label{tab:dyson_ideals}
{\footnotesize 
	\begin{center} 
		\protect
		\begin{tabular}{lcccc}
			\midrule 
			& \multicolumn{4}{c}{Percentage of...} \\ \cmidrule{2-5}
			& Libertarians & Meritocrats & Egalitarians & Other ideals  \\
			& (1) & (2) & (3) & (4) \\\hline \addlinespace
			
			\multicolumn{5}{l}{\hspace{-1em} \textbf{Panel A. MBA and undergraduate business students}} \\\addlinespace
			
			MBA students (spectators' first choices) &23.16 &22.65 &9.89 &44.30 \\
 &(4.35) &(5.99) &(3.15) &(8.05) \\
 
			MBA students (repeated-measures design) &15.33 &28.74 &6.51 &49.43 \\
 &(1.58) &(1.98) &(1.08) &(2.76) \\

			Dyson students (spectators' first choices) &23.91 &37.60 &1.85 &36.63 \\
 &(6.36) &(7.78) &(1.85) &(10.22) \\
 
			Dyson students (repeated-measures design) &15.95 &30.67 &6.13 &47.24 \\
 &(2.03) &(2.56) &(1.33) &(3.53) \\

			\midrule
			\multicolumn{5}{l}{\hspace{-1em} \textbf{Panel B. Average American}} \\ \addlinespace

			SCE -- Average American &39.77 &36.67 &8.57 &14.99 \\
 &(5.25) &(6.31) &(2.75) &(8.66) \\

			\cite{almas.etal2020} -- Average American &29.43 &37.54 &15.32 &17.72 \\
 &(2.50) &(3.43) &(1.98) &(4.69) \\

			\cite{cohn.etal2023} -- Average American &12.80 &60.33 &17.45 &9.42 \\
 &(1.89) &(3.22) &(2.19) &(4.33) \\

			\cite{harrs2025fairness} -- Average American &8.68 &45.24 &14.61 &31.46 \\
 &(1.04) &(2.30) &(1.30) &(2.84) \\

			\midrule
			\multicolumn{5}{l}{\hspace{-1em} \textbf{Panel C. Higher-income samples}} \\\addlinespace
			SCE -- USA income over 100k &44.44 &40.32 &3.23 &12.01 \\
 &(12.05) &(13.27) &(3.23) &(18.22) \\

			\cite{almas.etal2020} -- USA income over 100k &22.67 &41.92 &14.08 &21.33 \\
 &(4.87) &(7.11) &(4.16) &(9.57) \\

			\cite{cohn.etal2023} -- USA top 5 percent &25.15 &59.14 &8.97 &6.75 \\
 &(3.33) &(4.37) &(2.38) &(5.99) \\

			\cite{harrs2025fairness} -- USA income over 100k &8.96 &41.44 &7.61 &41.99 \\
 &(2.02) &(4.42) &(1.89) &(5.21) \\

			\midrule
			\multicolumn{5}{l}{\hspace{-1em} \textbf{Panel D. Low-income samples}} \\\addlinespace
			SCE -- USA income below 100k &38.57 &35.14 &10.81 &15.48 \\
 &(5.86) &(7.32) &(3.63) &(10.06) \\

			\cite{almas.etal2020} -- USA income below 100k &31.40 &36.33 &15.65 &16.63 \\
 &(2.89) &(3.92) &(2.25) &(5.37) \\

			\cite{cohn.etal2023} -- USA bottom 95 percent &12.06 &60.54 &17.83 &9.57 \\
 &(2.75) &(4.62) &(3.06) &(6.19) \\

			\cite{harrs2025fairness} -- USA income below 100k &8.58 &46.71 &17.16 &27.55 \\
 &(1.21) &(2.69) &(1.62) &(3.36) \\

			\midrule
			
		\end{tabular}
	\end{center}
	\begin{singlespace}  \vspace{-.5cm}
		\justify \textit{Notes:} This table shows estimates of the fraction of egalitarians, libertarians, and meritocrats in various studies and samples. See notes to Table~\ref{tab:ideals} for details. \par
		
	\end{singlespace}
}
\end{table}

\clearpage
\setcounter{table}{0}
\setcounter{figure}{0}
\setcounter{equation}{0}	
\renewcommand{\thetable}{E\arabic{table}}
\renewcommand{\thefigure}{E\arabic{figure}}
\renewcommand{\theequation}{E\arabic{equation}}

\section{Empirical Framework}\label{app:framework}

This appendix presents a statistical framework of income redistribution that mirrors our experimental setup. We use the framework to define each type of fairness ideal and the identification assumptions.

\subsection{Defining Fairness Ideals} \label{sub:def}

We begin by formalizing how spectators make redistribution decisions based on their fairness ideals. Consider a spectator who observes the initial earnings distribution $(w_H, w_L) \in {\rm I\!R}_+ \times {\rm I\!R}_+$ of two workers who competed for a fixed prize, where $w_H > w_L$ represents the earnings of the worker who won the competition (the ``high-earnings'' worker) and $w_L$ the earnings of the worker who lost (the ``low-earnings'' worker).

We study an environment in which worker earnings are allocated based on some combination of performance and luck. Let $z_p \in [0, 1]$ denote the probability that the outcomes of a worker pair $p$ are determined by chance. $z_p$ can be interpreted as the role that luck plays in determining income inequality. When $z_p = 0$, the initial earnings are determined by workers' performance. The worker with a better performance earns $w_H$, and the other worker earns $w_L$; thus, inequality reflects differences in performance. When $z_p = 1$, earnings are randomly assigned to workers, and inequality therefore reflects luck alone. Spectators observe $w_H$, $w_L$, and $z_p$ but do not observe worker performance.

Let $Y_{ip}$ be the earnings redistributed from the high- to the low-earnings worker in worker pair $p$ by spectator $i$. We assume that a spectator's choice of how much earnings to redistribute, $Y_{ip}$, is guided by their \textit{fairness ideal}, that is, their preferences for what constitutes a fair earnings allocation given the role played by luck in determining worker earnings, $z_p$. We model fairness ideals as a mapping from initial distributions to final earnings distributions. Following the literature, we focus on three main types of fairness ideals:	

\begin{definition}[Fairness ideals]\label{def:ideals} 
	
	\begin{itemize}
		\item[]
		\item Spectator $i$ is an \emph{egalitarian} if $Y_{ip}(z_p) = \frac{w_H - w_L }{2}$ for all $z_p$.
		\item Spectator $i$ is a \emph{libertarian} if $Y_{ip}(z_p) = 0$ for all  $z_p$.
		\item Spectator $i$ is a \emph{meritocrat} if $Y_{ip}(z_p)$ is strictly increasing in $z_p$ and $Y_{ip}(1) = \frac{w_H - w_L }{2}$.
	\end{itemize}
	
\end{definition}

Egalitarian spectators equalize the workers' earnings regardless of how the earnings were generated. Libertarian spectators leave the initial earnings unchanged regardless of how earnings differences arose. Meritocratic spectators condition their redistributive decisions on the source of inequality. Specifically, as the role of chance in determining earnings $z_p$ increases, they redistribute more earnings and, when inequality is entirely luck-based (i.e., there is no merit to the earnings allocation), they equalize earnings. 

The aggregate level of redistribution in a population of spectators, $\E[Y_{ip}]$, links directly to the distribution
of fairness ideals. Let $\alpha_E$, $\alpha_L$, and $\alpha_M$ be the proportions of egalitarians, libertarians, and meritocrats in a population. By the law of iterated expectations 
\begin{align} \label{eq:mean_red}
	\mathbb{E}[Y_{ip}] &= \alpha_E \frac{w_H - w_L }{2} + \alpha_M Y_{ip}^M + (1-\alpha_E-\alpha_L-\alpha_M) Y_{ip}^O, 
\end{align}	
where $Y_{ip}^M \equiv \E[Y_{ip} | \text{Meritocrat}]$ and $Y_{ip}^O \equiv \E[Y_{ip} | \text{Other ideal}]$ are the average earnings redistributed conditional on being a meritocrat and having other fairness ideals, respectively (the libertarian term is omitted because libertarians never redistribute). 

Equation \eqref{eq:mean_red} highlights two reasons why elites may have different redistributive choices than non-elites. First, the distribution of fairness ideals among elites (the $\alpha$'s) may differ from those among non-elites. Second, $Y_{ip}^M$ and $Y_{ip}^O$ may differ for elites and non-elites. In other words, elites who follow a meritocratic fairness ideal or fall outside the three classifications above may redistribute different amounts than the corresponding non-elites.

\subsection{Identification of Fairness Ideals}

\subsubsection{Research design.}

Our first research design relies on comparisons \textit{across} subjects, using only each spectator's first redistributive decision and excluding those who saw the efficiency cost environment first. We identify the fraction of egalitarians by the fraction of individuals who divide earnings equally when the winner is determined by performance. We identify the fraction of libertarians by the fraction of individuals who redistribute no earnings when the winner is determined by luck. Finally, we identify the fraction of meritocrats by the difference between (i) the fraction of individuals who allocate more to the winner when worker outcomes are determined by performance and (ii) the fraction of individuals who allocate more to the winner when worker outcomes are determined by luck. We refer to the remainder of the population as having ``other'' fairness ideals.

Our second research design relies on \textit{within}-subject comparisons across worker pairs, using all redistributive decisions except those from the efficiency cost environment. We classify a spectator as an egalitarian if they divide earnings equally in both the luck and performance environments, as a libertarian if they do not redistribute earnings in either environment, and as a meritocrat if they equalize earnings in the luck environment but give strictly more to the winner in the performance environment. This within-subject definition of meritocrats coincides with that of \citet{harrs2025fairness}. We refer to individuals who do not fit any of these classifications as having ``other'' fairness ideals.	

\subsubsection{Identification assumptions.}

For the between-subject design, we follow the standard assumptions in the literature \citep[e.g.,][]{almas.etal2020, cohn.etal2023}, which we formalize through the lens of our statistical framework. For the within-subject design, we extend these assumptions to account for multiple observations per spectator.

To identify egalitarians in the between-subject design, our identification assumption is that individuals who divide earnings equally when worker earnings are assigned based on performance would have also divided earnings equally under any other environment. Formally, if $Y_{ip}(z_p) = \frac{w_H - w_L }{2}$ for $z_p = 0$, then $Y_{ip}(z_p) = \frac{w_H - w_L }{2}$ for all $z_p$. In the within-subject design, our identification assumption is that individuals who divide equally in the luck \textit{and} performance environments would have divided equally under any other environment. Formally, if $Y_{ip}(z_p) = \frac{w_H - w_L }{2}$ for $z_p = 0$ \textit{and} $Y_{ip'}(z_{p'}) = \frac{w_H - w_L }{2}$ for $z_{p'} = 1$, then $Y_{ip}(z_p) = \frac{w_H - w_L }{2}$ for all $z_p$.

To identify libertarians in the between-subject design, our identification assumption is that individuals who do not redistribute earnings in the luck environment would not redistribute earnings under any other environment. Formally, if $Y_{ip}(z_p) = 0$ for $z_p = 1$, then $Y_{ip}(z_p) = 0$ for all $z_p$. In the within-subject design,  our identification assumption is that individuals who do not redistribute earnings in either the luck or performance environments would not redistribute earnings under any other environment. Formally, if $Y_{ip}(z_p) = 0$ for $z_p = 0$ \textit{and} $Y_{ip'}(z_{p'}) = 0$ for $z_{p'} = 1$, then $Y_{ip}(z_p) = 0$ for all $z_p$.

To identify meritocrats in the between-subject design, we follow \citet{almas.etal2020}. The share of spectators who allocate more to the winner in the performance condition equals the combined share of libertarians (who never redistribute) and meritocrats (who reward performance differences), while the share who allocate more to the winner in the luck condition equals the share of libertarians alone (since meritocrats equalize under luck). The difference between these two shares therefore identifies the meritocrat fraction. In the within-subject design, our identification assumption is that those who equalize earnings in the luck environment but give strictly more to the winner in the performance environment would strictly redistribute more earnings as the role of luck increases. Formally, if $Y_{ip}(0) < Y_{ip}(1) = \frac{w_H - w_L}{2}$, then $Y_{ip}(z_p) > Y_{ip}(z'_p)$ whenever $z_p > z'_p$.

Both the between- and within-subject designs have advantages and disadvantages for identification. To understand these, recall that a fairness ideal is a mapping from circumstances to choices. Reconstructing this mapping requires observing an infinite number of counterfactual choices. The research designs ``extrapolate'' spectator behavior based on a limited set of observed choices: one in the between-subject design and two (sequential) choices in the within-subject design. The between-subject design uses a single choice, so it avoids assumptions about sequential-decision bias but is more susceptible to measurement error and choice noise. In contrast, the within-subject design, using two choices, mitigates the measurement error concerns by incorporating more information about spectators' behavior, at the cost of assuming that the first choice does not influence the second.

\subsection{A Statistical Model of Redistribution and Inequality Source} \label{app:pot_outcomes}

Recall $Y_{ip}$ represents the earnings redistributed from the high- to the low-earnings worker in worker pair $p$ by spectator $i$. We model $Y_{ip}$ as a function of the role played by luck in determining worker earnings, $z_p$.\footnote{We abstract from modeling $Y_{ip}$ as a function of $w_H$ and $w_L$ for simplicity. In our experiment, $w_H$ and $w_L$ are constant across worker pairs, and the only variable that changes across worker pairs is $z_p$.} For the remainder of this subsection, we restrict attention to $z_p \in \{0, 1\}$ and consider the two environments studied by most experimental work: worker earnings are determined by performance ($z_{p} = 0$) or by chance ($z_{p} = 1$). Let $Y_{ip}(0)$ be the amount redistributed if earnings are determined by performance and $Y_{ip}(1)$ the amount redistributed if outcomes are determined by chance. These two redistribution levels denote potential outcomes for different levels of luck, but only one of the two outcomes is observed for a given worker pair. The observed redistribution, $Y_{ip}(z_p)$, can be written in terms of these potential outcomes as
\begin{align} \label{eq:pot_out}
	Y_{ip}(z_{p}) = Y_{ip}(0) + \hspace{-.4cm} \underbrace{\Big(Y_{ip}(1) -  Y_{ip}(0) \Big)}_{\text{``Source-of-inequality effect''}  (\beta_i)}\hspace{-.5cm} z_{p},
\end{align}
where $Y_{ip}(1) -  Y_{ip}(0) \equiv \beta_i$ measures the effect of changing the income-generating process from performance to chance, or ``source-of-inequality effect,'' for short. 

Suppose that we observed spectator choices for two worker pairs, $p$ and $p'$, with earnings in pair $p$ determined by performance ($z_p = 0$) and earnings in pair $p'$ randomly assigned ($z_{p'} = 1$). One could then compare average redistribution in the random-assignment pair, $\E[Y_{ip'} | z_{p'} = 1]$, with average redistribution in the performance pair, $\E[Y_{ip} | z_{p} = 0]$. This comparison can be written as
\begin{align} \label{eq:pot_out_decomp}
	\E[Y_{ip'} | z_{p'} = 1] - \E[Y_{ip} | z_{p} = 0]  &= \underbrace{\E[Y_{ip'}(1) | z_{p'} = 1] -  \E[Y_{ip'}(0) | z_{p'} = 1]}_{\text{Term 1: Source-of-inequality effect}} \notag \\ &+ \underbrace{\E[Y_{ip'}(0) | z_{p'} = 1] -  \E[Y_{ip}(0) | z_{p} = 0]}_{\text{Term 2: Sequential-decision bias}}.
\end{align}

Equation \eqref{eq:pot_out_decomp} shows that comparing the redistributive behavior of spectators for different worker pairs yields the sum of two terms. The first one is the effect of the source of inequality on earnings redistributed. The second term is a potential bias that arises when comparing sequential decisions. This term captures mechanisms such as anchoring effects, contrast effects, and learning effects. To address this potential bias, most literature uses a between-subject design in which spectators make only one decision. We provide both between-subject estimates (using only the first decision) and within-subject estimates (using the decisions in the luck and performance conditions) for comparison.

In addition to measuring the source-of-inequality effect, we also examine how introducing an efficiency cost affects redistribution ($Y_{ip}$). We analyze this using the same framework, holding constant the source of inequality and re-interpreting $z_p$ as an indicator for redistribution having a cost. In this case, $\beta_i$ in equation \eqref{eq:pot_out} and Term 1 in equation \eqref{eq:pot_out_decomp} measure the responsiveness of redistribution to the efficiency cost of redistribution.

\clearpage
\section{Narratives of Inequality by Fairness Ideal} \label{app:quotes}

\setcounter{table}{0}
\setcounter{figure}{0}
\setcounter{equation}{0}	
\renewcommand{\thetable}{F\arabic{table}}
\renewcommand{\thefigure}{F\arabic{figure}}
\renewcommand{\theequation}{F\arabic{equation}}

This appendix classifies MBA students' open-ended responses about the drivers of income inequality in the United States using large language models (LLMs). We describe the classification methodology, report the distribution of responses by fairness ideals, and provide qualitative evidence.

\subsection{Classification Methodology}

We classify the \getval{oeNcoded} open-ended responses collected from the second MBA cohort to the question: \textit{``What do you believe is the main driver of income inequality in the United States?''} Responses are typically short (most are 1--10 words), so many are ambiguous between adjacent categories.

We developed the codebook inductively, providing all responses to ChatGPT~5.2 Pro and instructing it to derive 6--8
categories from the data without a predefined scheme. The procedure yielded eight categories (see Appendix Table~\ref{tab:oe_codebook}): education (C01), unequal opportunities (C02), discrimination (C03), government policy (C04), corporate/elite power (C05), historical legacy (C06), behavioral/cultural (C07), and other/non-substantive (C08).

Each response was then classified using OpenAI's gpt-5-mini with structured outputs, producing both a multi-label coding (all relevant categories) and a single-best primary label. To establish robustness, we re-ran the classification with three additional models from two providers (gpt-5-nano, claude-sonnet-4-6, claude-haiku-4-5). Agreement across all four models is \getval{oeAgreeAllRate} percent, and most disagreements involve substantively adjacent categories or short, ambiguous responses (see Appendix Tables~\ref{tab:oe_agreement}--\ref{tab:oe_disagree}).

\subsection{Overall Distribution and Cross-Tabulation}

Appendix Table~\ref{tab:oe_overall} reports the multi-label distribution across all \getval{oeNcoded} responses. Because the classifier assigns every relevant category to each response, percentages sum to more than 100 percent. Education and skill formation is the most frequently cited driver (\getval{oeMlEduP} percent), followed by unequal opportunities (\getval{oeMlOppP} percent), government policy (\getval{oeMlPolP} percent), and discrimination (\getval{oeMlDisP} percent). Overall, \getval{oeMlMultiP} percent of responses mention two or more categories.

\begin{table}[H]
	\caption{Overall Distribution of Inequality Drivers, Multi-Label}\label{tab:oe_overall}
	\vspace{-6pt}
	{\small
		\begin{tabular*}{\textwidth}{@{\extracolsep{\fill}}cl cc}
			\midrule
			Code & Category & $N$ & \% \\
			\midrule
			C01 & Education and skill formation & 37 & 31.9 \\
C02 & Unequal opportunities, mobility barriers, starting conditions & 28 & 24.1 \\
C03 & Discrimination and identity-based bias & 19 & 16.4 \\
C04 & Government policy, politics, institutional governance & 20 & 17.2 \\
C05 & Corporate/elite power, wealth concentration & 13 & 11.2 \\
C06 & Historical legacy, broad systemic/structural inequity & 13 & 11.2 \\
C07 & Behavioral, cultural, and ideological explanations & 12 & 10.3 \\
C08 & No clear driver / rejects premise / non-substantive & 3 & 2.6 \\

			\midrule
		\end{tabular*}
	}

	\begin{minipage}{\textwidth}
		{\footnotesize\textit{Notes:} Multi-label distribution of \getval{oeNcoded} MBA students' open-ended responses across eight LLM-identified categories. Each response may be coded into multiple categories; percentages sum to more than 100\%. Classifications are based on gpt-5-mini. \par}
	\end{minipage}
\end{table}

\subsection{Qualitative Evidence by Fairness Ideal}

\begin{figure}[H]
	\caption{Inequality Drivers by Fairness Ideal}\label{fig:ineq_drivers}
	\centering
	\includegraphics[width=.75\linewidth]{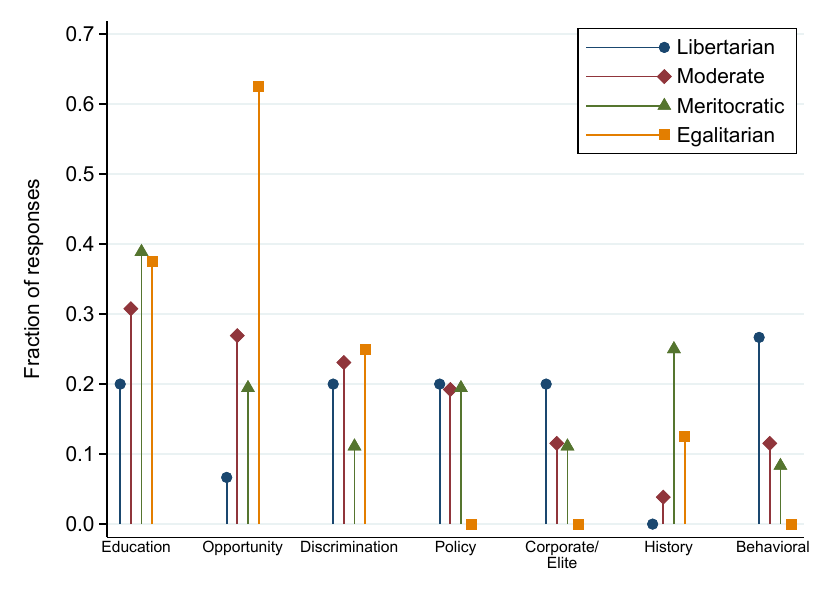}
	{
		\footnotesize \singlespacing \justify

		\textit{Notes:} Each marker shows the fraction of respondents within a fairness ideal whose response mentions the given category. Because responses may be coded into multiple categories, fractions can sum to more than one within each fairness type. Classifications are based on gpt-5-mini. \par
		
	}
\end{figure}

Appendix Table~\ref{tab:oe_crosstab} reports the multi-label category distribution by fairness ideal for the \getval{oeNclassified} classified respondents, and Appendix Figure~\ref{fig:ineq_drivers} visualizes these shares. The qualitative evidence below illustrates each type's distinctive profile with representative quotes.

\begin{table}[H]
	\caption{Inequality Drivers by Fairness Ideal, Multi-Label (\%)}\label{tab:oe_crosstab}
	\vspace{-6pt}
	{\small
		\begin{tabular*}{\textwidth}{@{\extracolsep{\fill}}l cccc}
			\midrule
			& Libertarian & Moderate & Meritocratic & Egalitarian \\
			\midrule
			C01: Education & 20.0 & 30.8 & 38.9 & 37.5 \\
C02: Opportunity & 6.7 & 26.9 & 19.4 & 62.5 \\
C03: Discrimination & 20.0 & 23.1 & 11.1 & 25.0 \\
C04: Policy & 20.0 & 19.2 & 19.4 & 0.0 \\
C05: Corporate/elite & 20.0 & 11.5 & 11.1 & 0.0 \\
C06: History & 0.0 & 3.8 & 25.0 & 12.5 \\
C07: Behavioral & 26.7 & 11.5 & 8.3 & 0.0 \\
C08: Non-substantive & 6.7 & 0.0 & 0.0 & 12.5 \\

			\midrule
			$N$ & \getval{oeNlib} & \getval{oeNmod} & \getval{oeNmerit} & \getval{oeNegal} \\
			\midrule
		\end{tabular*}
	}
	\begin{minipage}{\textwidth}
		{\footnotesize\textit{Notes:} Each cell reports the share of respondents within a fairness ideal whose response mentions the given category. Because responses may be coded into multiple categories, columns sum to more than 100 percent. The \getval{oeNother} respondents not classified as libertarian, moderate, meritocratic, or egalitarian are omitted. \par}
	\end{minipage}
\end{table}

\subsubsection{Libertarian Narratives.}

Libertarians are the only fairness type whose modal category is behavioral and cultural explanations (\getval{oeMlLibBehP} percent). They attribute inequality to personal choices, work ethic, and cultural differences---responses such as ``willingness to work hard,'' ``poor choices,'' and ``cultural differences.'' This individualistic framing is consistent with their preference not to redistribute: if inequality reflects personal agency, there is less justification for corrective transfers.

Beyond the behavioral emphasis, libertarian responses are spread evenly across education, discrimination, corporate power, and policy (each cited by \getval{oeMlLibEduP} percent). They are the least likely of any type to cite barriers such as unequal opportunities. Some express outright skepticism about inequality claims: ``Women demanding more money than what Men make, there is no such thing as income inequality in the U.S.\ unless you are brainwashed by false statistics, causation without correlation.''

\subsubsection{Moderate Narratives.}

Moderates identify multiple distinct mechanisms behind inequality. Their top categories are education (\getval{oeMlModEduP} percent), unequal opportunities (\getval{oeMlModOppP} percent), discrimination (\getval{oeMlModDisP} percent), and policy (\getval{oeMlModPolP} percent). \getval{oeModBarrierP} percent of moderates cite at least one barrier category, compared to only \getval{oeLibBarrierP} percent of libertarians.

Many point to starting conditions: ``Circumstance,'' ``Socioeconomic status at birth,'' ``How you grew up,'' and ``access to opportunities.'' This focus on initial endowments aligns with their experimental pattern of redistributing more when inequality stems from luck while still allowing some luck-based inequality to persist. Others emphasize educational factors---``difference in accessibility of education'' and ``intellect and education inequality''---without the systemic framing of egalitarians.

The most distinctive feature of the moderate profile is what it \textit{excludes}: near-zero historical or systemic framing (\getval{oeMlModHisP} percent), in stark contrast to meritocrats (\getval{oeMlMeritHisP} percent). Identity factors appear through mentions of ``white supremacy,'' ``racism, sexism, greed,'' and ``race,'' but typically as proximate mechanisms rather than deep structural forces. This pattern suggests that moderates are not well characterized as soft meritocrats or soft libertarians, but instead hold a distinct worldview.

\subsubsection{Meritocratic Narratives.}

Meritocrats concentrate on education (\getval{oeMlMeritEduP} percent), the highest share of any type. They describe ``access to high quality and relevant education,'' ``lack of equitable access to quality education,'' and education being ``too expensive for more people to access.'' This framing positions unequal educational access---rather than education per se---as the primary barrier to mobility, consistent with a worldview that values merit but acknowledges an uneven playing field.

Meritocrats also cite historical and systemic factors at the highest rate of any type (\getval{oeMlMeritHisP} percent). One student references ``historical issues that continue to affect current socio-economic stratification''; another cites ``history (residual injustice and cultural difference) exacerbated by a financial system that gives the `further ahead' exponential resources to get ahead.'' This willingness to acknowledge structural forces---while still operating within a framework that values merit-based outcomes---distinguishes meritocrats from libertarians, who never invoke historical framing.

\subsubsection{Egalitarian Narratives.}

Egalitarians emphasize barriers: \getval{oeMlEgalOppP} percent cite unequal opportunities and \getval{oeMlEgalDisP} percent cite discrimination---the highest barrier rates of any type. Rather than viewing education as a meritocratic sorting mechanism, egalitarians frame it as reinforcing existing disparities: ``Access to education (cost of studying/school/etc)'' and ``barriers to opportunity/education.'' Identity-based discrimination features prominently, with one student citing ``sexism and racism'' as primary drivers and another pointing to ``historical systems around identity and location.''

This framing---inequality as the product of institutional barriers rather than individual failings---is consistent with egalitarians' strong preference for redistribution. No egalitarian cites behavioral or cultural explanations, the sharpest contrast with libertarians (\getval{oeMlLibBehP} percent).

\subsection{Classification Robustness and Codebook}

This section provides details on the codebook and inter-model agreement.

\begin{table}[H]
	\caption{Inter-Model Agreement Rates}\label{tab:oe_agreement}
	\vspace{-6pt}
	{\small
		\begin{tabular*}{\textwidth}{@{\extracolsep{\fill}}l cc}
			\midrule
			Comparison & Agree / $N$ & Agreement (\%) \\
			\midrule
			\multicolumn{3}{@{}l}{\textit{Within-provider (OpenAI)}} \\
			\quad gpt-5-mini vs.\ gpt-5-nano & \getval{oeAgreeMiniNanoN} / \getval{oeNcoded} & \getval{oeAgreeMiniNanoRate} \\[3pt]
			\multicolumn{3}{@{}l}{\textit{Within-provider (Anthropic)}} \\
			\quad claude-sonnet-4-6 vs.\ claude-haiku-4-5 & \getval{oeAgreeSonnetHaikuN} / \getval{oeNcoded} & \getval{oeAgreeSonnetHaikuRate} \\[3pt]
			\multicolumn{3}{@{}l}{\textit{Cross-provider}} \\
			\quad gpt-5-mini vs.\ claude-sonnet-4-6 & \getval{oeAgreeMiniSonnetN} / \getval{oeNcoded} & \getval{oeAgreeMiniSonnetRate} \\
			\quad gpt-5-mini vs.\ claude-haiku-4-5 & \getval{oeAgreeMiniHaikuN} / \getval{oeNcoded} & \getval{oeAgreeMiniHaikuRate} \\
			\quad gpt-5-nano vs.\ claude-sonnet-4-6 & \getval{oeAgreeNanoSonnetN} / \getval{oeNcoded} & \getval{oeAgreeNanoSonnetRate} \\
			\quad gpt-5-nano vs.\ claude-haiku-4-5 & \getval{oeAgreeNanoHaikuN} / \getval{oeNcoded} & \getval{oeAgreeNanoHaikuRate} \\
			\midrule
			\quad All four models agree & \getval{oeAgreeAllN} / \getval{oeNcoded} & \getval{oeAgreeAllRate} \\
			\midrule
		\end{tabular*}
	}
	\vspace{-3pt}
	\begin{minipage}{\textwidth}
		{\footnotesize\textit{Notes:} Each row reports the share of \getval{oeNcoded} responses where the listed models assign the same single-best primary category. Cross-provider comparisons (OpenAI vs.\ Anthropic) are especially informative as the models differ in training data, architecture, and fine-tuning. \par}
	\end{minipage}
\end{table}

\begin{table}[H]
	\caption{Codebook for Open-Ended Inequality Response Classification}\label{tab:oe_codebook}
	\vspace{-6pt}
	{\scriptsize
		\begin{tabular*}{\textwidth}{@{}c @{\extracolsep{\fill}} p{2.8cm} p{4.5cm} p{4.5cm}@{}}
			\midrule
			Code & Category & Include if & Exclude if \\
			\midrule
			C01 & Education and skill formation & Mentions education (K--12, college) as key driver; emphasizes affordability/cost of schooling; points to skill gaps, credential requirements, or knowledge deficits & Focuses on unequal opportunity in general without education being central $\to$ C02; focuses on race/gender discrimination $\to$ C03 \\[6pt]
			C02 & Unequal opportunities, mobility barriers, starting conditions & Mentions lack of opportunities, ``uneven playing field''; emphasizes parental income, upbringing, geography; highlights connections/social capital & Core argument is education access/quality $\to$ C01; identity-based discrimination $\to$ C03; specific government lever $\to$ C04 \\[6pt]
			C03 & Discrimination and identity-based bias & References race/racism, gender/sexism, white supremacy, or discrimination; mentions bias tied to identity & Favoritism via networks without identity framing $\to$ C02; historical injustice without present-day discrimination $\to$ C06 \\[6pt]
			C04 & Government policy, politics, institutional governance & Mentions tax policy, regulation, laws, ``the government''; references politics, money in politics, or named political leaders; mentions social welfare/healthcare policy & Blames corporations/elites without emphasizing government levers $\to$ C05; only broad ``systemic'' language $\to$ C06 \\[6pt]
			C05 & Corporate/elite power, wealth concentration & References economic elite, corporations, or wealth concentration; mentions ``greed'' as elite capture; points to executive pay dynamics or capitalism & Focuses primarily on taxes/regulation/politics $\to$ C04; on starting conditions without elite capture $\to$ C02 \\[6pt]
			C06 & Historical legacy, broad systemic/structural inequity & Mentions history, residual injustice, long-run stratification; uses ``systemic challenges,'' ``social inequity,'' ``structural inequity'' & Specifies a primary mechanism fitting another category $\to$ code that; gives concrete opportunity pathway without historical framing $\to$ C02 \\[6pt]
			C07 & Behavioral, cultural, ideological & Mentions work ethic, willingness to work hard, poor choices, confidence, luck; mentions culture/cultural differences; frames inequality as driven by beliefs/perceptions & Emphasizes structural barriers to education $\to$ C01; opportunity constraints $\to$ C02; identity discrimination $\to$ C03 \\[6pt]
			C08 & No clear driver / rejects premise & Expresses uncertainty or no answer; denies inequality is real; says ``no one factor'' without naming any & Any specific driver is named (even if hedged) $\to$ code relevant category \\
			\midrule
		\end{tabular*}
	}
	\begin{minipage}{\textwidth}
		{\footnotesize\textit{Notes:} Codebook developed inductively by providing all \getval{oeNcoded} responses to ChatGPT~5.2 Pro with instructions to derive categories from the data without predefined schemes. Categories are defined to be mutually exclusive at the definition level; individual responses may receive multiple labels. \par}
	\end{minipage}
\end{table}

\begin{landscape}
	\begin{table}[htbp]
		\caption{Responses with Inter-Model Disagreement}\label{tab:oe_disagree}
		\vspace{-6pt}
		{\scriptsize
			\begin{tabular*}{\linewidth}{@{\extracolsep{\fill}} p{13cm} cccc @{}}
				\midrule
				Response text & \shortstack{gpt-5\\mini} & \shortstack{gpt-5\\nano} & \shortstack{claude\\sonnet} & \shortstack{claude\\haiku} \\
				\midrule
				Barriers to opportunity/education & C02 & C02 & \textbf{C01} & \textbf{C01} \\
				Current laws and inequality of race and gender in the decision room & \textbf{C04} & C03 & C03 & C03 \\
				Intellect and education inequality & \textbf{C07} & C01 & C01 & C01 \\
				Lack of centralized knowledge and access to resources & \textbf{C01} & C02 & C02 & C02 \\
				Lack of environment that encourages competition & \textbf{C05} & \textbf{C02} & C07 & C07 \\
				Lack of parental guidance and inadequate education & \textbf{C02} & C01 & C01 & C01 \\
				Laws, access to opportunities & C04 & C04 & \textbf{C02} & C04 \\
				Leftist ideas & C07 & C07 & C07 & \textbf{C08} \\
				Opportunity and education & C02 & C02 & C02 & \textbf{C01} \\
				People think investing is gambling. Basic knowledge about markets is absent & C01 & C01 & \textbf{C07} & \textbf{C07} \\
				Structural inequity, unequal opportunities & C06 & C06 & \textbf{C02} & C06 \\
				Systemic bias & C03 & C03 & \textbf{C06} & \textbf{C06} \\
				Systemic issues in regards to race, gender and overall socioeconomic situations & C03 & C03 & C03 & \textbf{C06} \\
				systemic poverty and capitalism & C06 & C06 & \textbf{C05} & \textbf{C05} \\
				Transparency and Opportunity & \textbf{C04} & C02 & C02 & C02 \\
				Trump & C04 & \textbf{C08} & C04 & \textbf{C08} \\
				Wage stagnation relative to inflation & C05 & \textbf{C06} & C05 & \textbf{C04} \\
				Women demanding more money than what Men make, there is no such thing as income inequality\dots & C03 & C03 & \textbf{C08} & \textbf{C08} \\
				Wrong people at the wrong place, playing a perception game on general population & C05 & \textbf{C02} & C05 & \textbf{C07} \\
				Wrongly enforced initiative for diversity initiatives & C04 & C04 & \textbf{C03} & C04 \\
				\midrule
			\end{tabular*}
		}
		\vspace{-3pt}
		\begin{minipage}{\linewidth}
			{\footnotesize\textit{Notes:} Table lists all \getval{oeNdisagree} responses (of \getval{oeNcoded}) where at least one of four models assigns a different primary category. Responses are sorted alphabetically. \textbf{Bold} entries indicate the model(s) that deviate from the modal classification. Of the \getval{oeNdisagree} disagreements, \getval{oeNsplitThreeOne} are 3-vs-1 splits and \getval{oeNsplitTwoTwo} are 2-vs-2 (or 2-vs-1-vs-1) splits. Category codes: C01 = Education; C02 = Opportunities; C03 = Discrimination; C04 = Policy; C05 = Corporate/elite power; C06 = Systemic; C07 = Behavioral/cultural; C08 = Non-substantive. See Appendix Table~\ref{tab:oe_codebook} for full definitions. \par}
		\end{minipage}
	\end{table}
\end{landscape}

%
%
%

\end{document}